\documentclass[longbibliography,aps,prr,twocolumn,superscriptaddress,showkeys,amsfonts,amssymb,amsmath,floatfix]{revtex4-2}
\usepackage{inputenc}
\usepackage{amsmath,amsbsy,amsfonts,revsymb}
\usepackage{graphicx}% Include figure files %%% \usepackage[draft]{graphicx}
\usepackage{bm}% bold math
\usepackage[dvipsnames]{xcolor}
\usepackage[multi-part-units=single]{siunitx}
\usepackage{bm}
\usepackage{mathtools} 
\usepackage{enumitem}
\usepackage{multirow}
\usepackage{upgreek}
\usepackage{placeins}
\usepackage[colorlinks=true,linkcolor=blue,citecolor=blue,urlcolor=blue]{hyperref}
\usepackage{soul}
\usepackage{balance}

\DeclarePairedDelimiter\dpa{(}{)} %Parenthèse adaptataives
\DeclarePairedDelimiter\dba{\langle}{\rangle}
\DeclarePairedDelimiter\dca{[}{]}

\DeclarePairedDelimiterX\Bra[1]{\langle}{\rvert}{#1\,}
\DeclarePairedDelimiter\ket{\lvert}{\rangle}
\DeclarePairedDelimiterX\Ket[1]{\lvert}{\rangle}{\,#1}
\DeclarePairedDelimiter\abs{\lvert}{\rvert}
\DeclarePairedDelimiterX\Abs[1]{\lvert}{\rvert}{\,#1\,}
\DeclarePairedDelimiterX\qsp[2]{\langle}{\rangle}{#1\delimsize\vert\mathopen{}#2}
\DeclarePairedDelimiterX\Qsp[2]{\langle}{\rangle}{#1\,\delimsize\vert\,\mathopen{}#2}
\DeclarePairedDelimiterX\mel[3]{\langle}{\rangle}{#1\delimsize\vert\mathopen{}#2\delimsize\vert\mathopen{}#3}
\DeclarePairedDelimiterX\Mel[3]{\langle}{\rangle}{#1\,\delimsize\vert\,\mathopen{}#2\,\delimsize\vert\,\mathopen{}#3}

\newcommand\rev[1]{\textcolor{black}{#1}}
\newcommand\fig[1]{\hyperref[{#1}]{\ref*{#1}}}
\newcommand\sfig[2]{\hyperref[{#1}]{\ref*{#1}{(#2)}}}
\newcommand\ssfig[3]{\hyperref[{#1}]{\ref*{#1}{(#2)}}$-$\hyperref[{#1}]{\ref*{#1}{(#3)}}}
\newcommand\TbIr{Tb$_{2}$Ir$_{2}$O$_{7}$}
\newcommand\TN{$T_{\mrm{N}}$}

\newcommand\mrm[1]{\mathrm{#1}}
\newcommand\wh[1]{\widehat{#1}}
\DeclareSIUnit \mub {\ensuremath{\upmu_{B}}}
\DeclareSIUnit\angstrom{\text{Å}}

\begin{document}
	
\title{Origin of the magnetic ground state of Tb$_{2}$Ir$_{2}$O$_{7}$}

\author {Y. Alexanian}\email{yann.alexanian@neel.cnrs.fr}
\affiliation{Universit\'e Grenoble Alpes, CNRS, Institut N\'eel, 38000 Grenoble, France}
\affiliation{Institut Laue Langevin, 38000 Grenoble, France}
\author{E. Lhotel}
\affiliation{Universit\'e Grenoble Alpes, CNRS, Institut N\'eel, 38000 Grenoble, France}
\author{J. Robert}
\affiliation{Universit\'e Grenoble Alpes, CNRS, Institut N\'eel, 38000 Grenoble, France}
\author{S. Petit}
\affiliation{Laboratoire Léon Brillouin, CEA, CNRS, Université Paris-Saclay, 91191 Gif-sur-Yvette, France}
\author{E. Lefrançois}
\affiliation{Universit\'e Grenoble Alpes, CNRS, Institut N\'eel, 38000 Grenoble, France}
\author{P. Lejay}
\affiliation{Universit\'e Grenoble Alpes, CNRS, Institut N\'eel, 38000 Grenoble, France}
\author{A. Hadj-Azzem}
\affiliation{Universit\'e Grenoble Alpes, CNRS, Institut N\'eel, 38000 Grenoble, France}
\author {F. Damay}
\affiliation{Laboratoire Léon Brillouin, CEA, CNRS, Université Paris-Saclay, 91191 Gif-sur-Yvette, France}
\author {J. Ollivier}
\affiliation{Institut Laue Langevin, 38000 Grenoble, France}
\author {B. F{\aa}k}
\affiliation{Institut Laue Langevin, 38000 Grenoble, France}
\author {S. De Brion}
\affiliation{Universit\'e Grenoble Alpes, CNRS, Institut N\'eel, 38000 Grenoble, France}
\author {R. Ballou}
\affiliation{Universit\'e Grenoble Alpes, CNRS, Institut N\'eel, 38000 Grenoble, France}
\author{V. Simonet}\email{virginie.simonet@neel.cnrs.fr}
\affiliation{Universit\'e Grenoble Alpes, CNRS, Institut N\'eel, 38000 Grenoble, France}

\date{\today}

\begin{abstract}

Magnetic-rare-earth pyrochlore iridates exhibit a rich variety of unconventional phases, driven by the complex interactions within and between the rare-earth and the iridium sublattices. In this study, we investigate the peculiar magnetic state of \TbIr, where a component of the Tb$^{3+}$ moment orders perpendicular to its local Ising anisotropy axis. By means of neutron diffraction and inelastic neutron scattering down to dilution temperatures, complemented by specific heat measurements, we show that this intriguing magnetic state is fully established at $\SI{1.5}{\kelvin}$ and we characterize its excitation spectrum across a broad range of energies. Our calculations reveal that bilinear interactions between Tb$^{3+}$ ions subjected to the Ir molecular field capture several key features of the experiments, but need to be supplemented to fully reproduce the observed behavior.

\end{abstract}

\maketitle

\section{Introduction}

Materials that combine strong spin–orbit entanglement with electron correlations offer a rich playground for the emergence of original quantum states of matter such as exotic spin liquids and topological phases \cite{Pesin2010,WitczakKrempa2014}. Among them, iridium oxides have attracted considerable attention due to their spin–orbit-entangled $j_{\mathrm{eff}} = 1/2$ ground state \cite{Kim2008,Kim2009}. The resulting pseudo-spin degrees of freedom interact via highly anisotropic interactions, which has led to the prediction of a wide range of unconventional electronic and magnetic ground states across the iridate family \cite{Rau2016,Schaffer2016}. A prominent example is offered by the pyrochlore iridates $R_{2}$Ir$_{2}$O$_{7}$, where the Ir$^{4+}$ and $R^{3+}$ (rare earth) ions form two interpenetrating pyrochlore lattices of corner sharing tetrahedra \cite{Subramanian1983}. Most members of this series exhibit a metal–insulator transition at $T_{\mrm{N}} \simeq \SIrange{120}{150}{\kelvin}$ (except $R = \mrm{Nd}$  with $T_{\mrm{N}} \simeq \SI{30}{\kelvin}$ and $R = \mrm{Pr}$ which remains metallic) \cite{Yanagishima2001,Matsuhira2011}. This is accompanied by long-range magnetic ordering of the Ir sublattice into an all-in–all-out (AIAO) configuration where all magnetic moments on a given tetrahedron point either toward or inward \cite{Disseler2012,Sagayama2013,Disseler2014,Lefrancois2015,Guo2016}. This AIAO order breaks time-reversal symmetry while preserving inversion, a combination allowing the emergence of magnetic topological phases such as magnetic Weyl semi-metals or axionic insulators \cite{Yang2010,Wan2011,Go2012,Varnava2018}. Although a direct observation of the band topology remains elusive, indirect experimental signatures have already been reported \cite{Machida2007,Machida2009,Sushkov2015,Ueda2018}.

Beyond their putative topological nature, pyrochlore iridates also exhibit unconventional low-energy magnetic behavior when $R^{3+}$ is magnetic. A first key ingredient is the Ir$^{4+}$ molecular field acting along the local $\dba*{111}$ anisotropy axis of the $R^{3+}$ ions. For magnetic ions with Heisenberg or Ising anisotropy parallel to this field ($R = \mrm{Ho}$, Dy, Nd, Gd), an AIAO magnetic order of the rare-earth moments is induced below $T_{\mrm{N}}$ \cite{Tomiyasu2012,Guo2016,Lefrancois2017,Lefrancois2019,Cathelin2020}. In contrast, for a planar XY anisotropy ($R = \mrm{Er}$, Yb), no such ordering is observed. Then, at lower temperatures, when rare-earth interactions become significant, a variety of unexpected phases has been reported. \rev{In $R = \mrm{Ho}$ and $\mrm{Dy}$ compounds, a fragmented magnetic phase appears, where half of each magnetic moment takes part to an AIAO ordered structure while the other half participates to a disordered Coulomb phase} \cite{BrooksBartlett2014,Lefrancois2017,Cathelin2020}. In the Er-based compound, only short-range correlations persist down to at least $\SI{70}{\milli\kelvin}$ \cite{Lefrancois2015}, whereas the Yb system develops a ferromagnetic ground state with a strongly reduced ordered moment \cite{Jacobsen2020}. In the Gd case, despite the nearly isotropic nature of Gd$^{3+}$ spins, correlations of spin components perpendicular to the Ir$^{4+}$ molecular field develop at lower temperature than the induced AIAO order \cite{Lefrancois2019}.

Finally, \TbIr\ is far from the least remarkable case. As in other pyrochlore iridates, an AIAO ordering is induced on the Tb$^{3+}$ ions below $T_{\mrm{N}}\approx\SI{130}{\kelvin}$. Further cooling to $T < \SI{10}{\kelvin}$ leads to the emergence of an antiferromagnetic ordering of the component perpendicular to the local $\dba*{111}$ anisotropy axis in the so-called $\Gamma_5$ representation \cite{Guo2017}. This is particularly intriguing, as such a configuration competes with the predominantly Ising-like nature of the Tb$^{3+}$ ions, further reinforced by the molecular field from the Ir sublattice. This unconventional behavior resonates with that of other Tb-based pyrochlores, whose magnetic ground states are highly sensitive to weak interaction terms \cite{Rau2019}. \rev{This gives rise to a remarkable diversity of states, from ordered spin-ice in Tb$_{2}$Sn$_{2}$O$_{7}$ \cite{Mirebeau2005,DalmasdeReotier2006}, competing dipolar-quadrupolar correlations impeding long range order in Tb$_{2}$Ge$_{2}$O$_{7}$ \cite{Hallas2020}, and to potential disorder-induced Coulomb phases in Tb$_{2}$Hf$_{2}$O$_{7}$ \cite{Sibille2017,Anand2018} and Tb$_{2}$ScNbO$_{7}$ \cite{Alexanian2023b}}. Tb$_{2}$Ti$_{2}$O$_{7}$ is a case in point with a particularly rich phase diagram, combining spin-liquid physics and complex dipolar and quadrupolar orders, whose microscopic origin remains intensely debated \cite{Gardner1999,Fennell2012,Taniguchi2013,Guitteny2013,Fritsch2013,Fennell2014,Kermarrec2015,Takatsu2016,Kadowaki2022,Alexanian2023,Roll2024}. The coexistence of magnetic orders in \TbIr\ therefore stands as a particularly striking case that calls for deeper investigation. Clarifying its origin may also shed light on the subtle competing interactions that govern the broad family of Tb-based pyrochlores.

In this study, we report neutron diffraction and inelastic neutron scattering measurements from room temperature down to dilution temperatures, as well as specific heat measurements at low temperatures on polycrystalline \TbIr. Our data confirm the presence of the $\Gamma_{5}$ magnetic order below $\SI{10}{\kelvin}$, and show that it remains unchanged below $\SI{1.5}{\kelvin}$. We then characterize the crystal field excitation spectrum in detail and probe the low-energy magnetic excitations. They exhibit dispersive behavior consistent with the role of interactions in promoting $\Gamma_{5}$ order and associated collective excitations. We model all these data using a Hamiltonian including crystal field effects together with bilinear Tb-Tb interactions and an effective Ir molecular field acting on the Tb. Our calculations reproduce the global behavior of this material including the existence of a $\Gamma_{5}$ magnetic order. At variance with experiment, they predict however a lower ordering temperature, which shows the need to consider additional ingredients and a more complex role of the iridium.

\section{Methods}

\subsection{Experimental methods}
\label{sec:exp_details}

\rev{The present study was performed on a polycrystalline Tb$_{2}$Ir$_{2}$O$_{7}$ sample synthesized by solid-state reaction using a flux method described in Ref.~\cite{Lefrancois2015}. Its structure and quality was checked by X-ray diffraction at room temperature. The lattice parameter and the 48f oxygen positional parameter were found to be $a=\SI{10.2368(5)}{\angstrom}$ and $x=\num{0.35(2)}$, respectively. A Tb$_{2}$O$_{3}$ impurity phase of less than $\SI{3}{\percent}$ was observed in addition to the main Tb$_{2}$Ir$_{2}$O$_{7}$ phase, which appeared of good quality. No antisite disorder or non-stoichiometry was detected.}

Specific heat measurements were performed on a Quantum Design Physical Property Measurement System (PPMS). Samples were pressed into pellets, stuck to the PPMS puck with apiezon grease. The addenda contribution was measured apart and removed from the total specific heat measured with the samples.

Powder neutron diffraction was carried out on the G4.1 diffractometer at the Laboratoire Léon Brillouin (LLB) using an incident wavelength $\lambda_{\mathrm{i}} = \SI{2.426}{\angstrom}$, over a temperature range from $\SI{130}{\kelvin}$ down to $\SI{26}{\milli\kelvin}$. Between $\SI{1.5}{\kelvin}$ and $\SI{26}{\milli\kelvin}$, the diffractometer was equipped with a Cryoconcept-France HD dilution refrigerator. \rev{To mitigate neutron absorption by iridium and ensure thermalization, the sample was placed in an annular copper sample holder optimized to maximize neutron transmission and enclosed in a helium-filled copper box.}  Simultaneous Rietveld refinements of the nuclear and magnetic structures were performed using the \textsc{FullProf} Suite. 

\rev{Powder inelastic neutron scattering was carried out on three time-of-flight spectrometers at the Institut Laue-Langevin (ILL) using the same annular sample holder}. Data from room temperature down to $\SI{1.5}{\kelvin}$ were collected on IN4c ($\lambda_{\mathrm{i}} = \SI{0.8}{\angstrom}, \SI{1.2}{\angstrom}$) and IN5 ($\lambda_{\mathrm{i}} = \SI{4.8}{\angstrom}$). Measurements between $\SI{1.5}{\kelvin}$ and $\SI{45}{\milli\kelvin}$ were obtained on IN6 ($\lambda_{\mathrm{i}} = \SI{5.1}{\angstrom}$) equipped with a dilution fridge.

\subsection{Numerical methods}
\label{sec:num_details}

Numerical calculation of neutron scattering functions, ordered magnetic moment, and magnetic specific heat were carried out using a model including the crystal-field Hamiltonian, the interactions among the Tb$^{3+}$ total angular momenta $J$, and an effective magnetic field oriented along the local $z$ axis accounting for the influence of the Ir magnetic moments.

First, the crystal-field parameters were determined using a reverse Monte Carlo procedure with a simpler model neglecting Tb-Tb and Tb-Ir interactions. A cost functional defined as the sum of squared differences between the calculated and experimental high-energy ($E \ge \SI{8}{\milli\electronvolt}$) inelastic neutron spectra was minimized using a Metropolis algorithm with simulated annealing. This procedure typically involved $\num{10000}$ steps to ensure convergence.

Tb$–$Tb and Tb$–$Ir interactions were subsequently incorporated at the mean-field level through a self-consistent iterative scheme. The corresponding interaction parameters were refined by comparing the calculated and experimental ordered magnetic moments and magnetic specific heat, while keeping the crystal-field parameters fixed to their previously determined values. Convergence was achieved when no further decrease in the free energy was observed between successive iterations. To ensure robustness and to avoid trapping in local minima, the procedure was repeated from multiple random initial configurations.

The low-energy magnetic scattering function was finally obtained from the dynamical magnetic susceptibility computed within the random phase approximation (RPA), using the crystal-field and interaction parameters established in the preceding mean-field step. The computational framework has been described in detail elsewhere, see e.g. Ref. \cite{Roll2024}.

\section{Experimental results}

\subsection{Neutron diffraction and specific heat measurements: Magnetic order}

\begin{figure}
\includegraphics[width=\columnwidth]{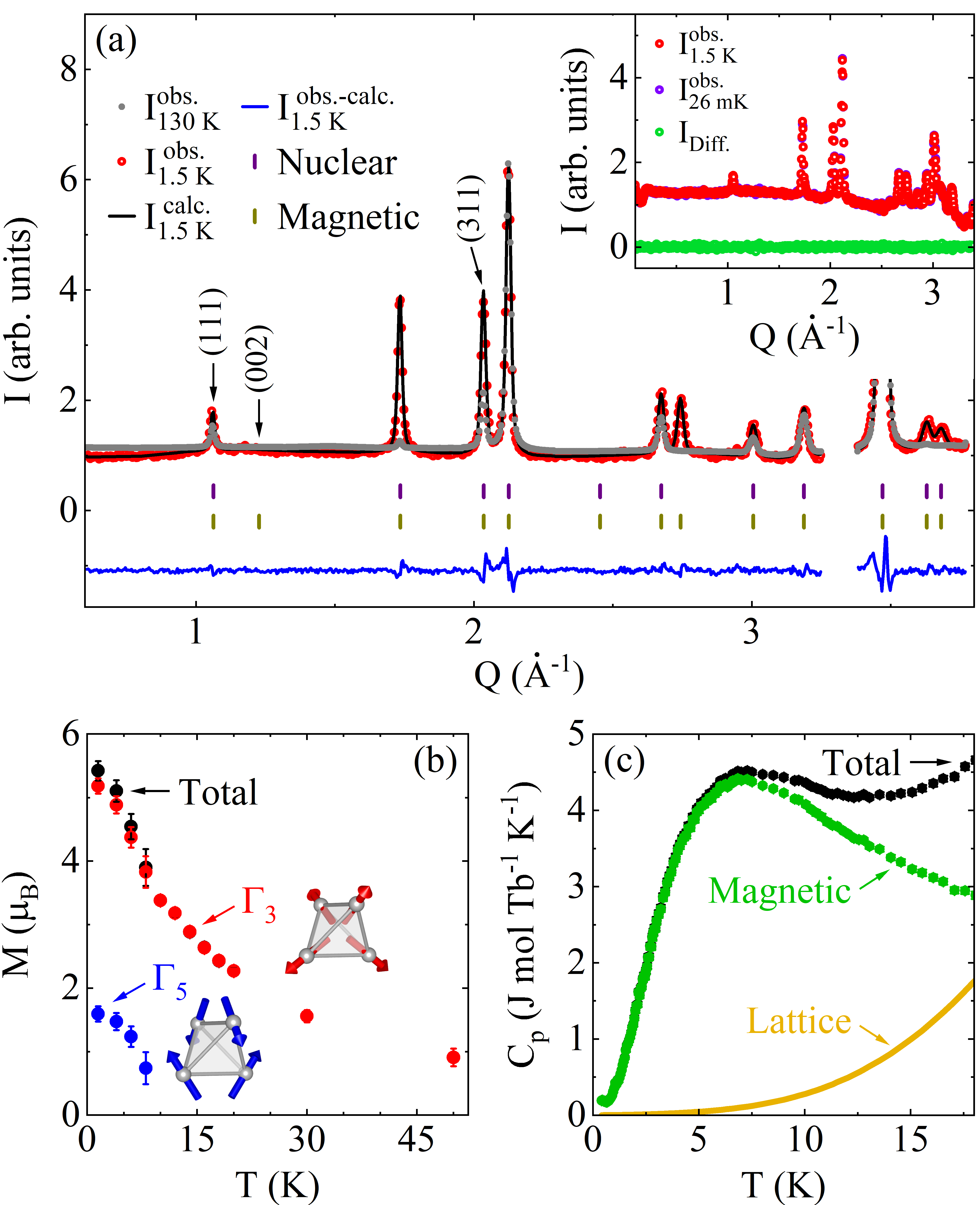}
\caption{%
(a) Powder neutron diffractograms measured in G4.1 at $\SI{130}{\kelvin}$ (grey dots) and $\SI{1.5}{\kelvin}$ (red dots), along with the corresponding Rietveld refinement (black line). \rev{The excluded region corresponds to a reflection from the minor Tb$_{2}$O$_{3}$ impurity phase.} The difference between the two latter (blue line) illustrates the quality of the refinement (Bragg R-factor $R_{\mrm{B}} = 1.51$, RF-factor $R_{\mrm{F}} = 1.92$ and Magnetic R-factor $R_{\mrm{M}} = 1.69$). Nuclear and magnetic Bragg peak positions are indicated by purple and gold vertical ticks, respectively. Inset: Measurements at $\SI{1.5}{\kelvin}$ (red dots) and $\SI{26}{\milli\kelvin}$ (purple dots) using a dilution refrigerator. Their difference (green) shows no significant evolution. 
(b) Temperature evolution of the refined Tb$^{3+}$ total ordered magnetic moment (black dots), its AIAO component (red dots), and its $\Gamma_{5}$ component (blue dots). 
(c) Temperature dependence of the specific heat (black dots). The magnetic contribution (green dots) is obtained by subtracting the lattice part (orange line), estimated from the scaled heat capacity of the non-magnetic rare-earth ion analogue Eu$_{2}$Ir$_{2}$O$_{7}$.}
\label{fig:1}
\end{figure}

Fig.~\fig{fig:1} presents the characterization of the \TbIr\ magnetic order. Powder neutron diffractograms recorded at $T = \SI{130}{\kelvin}$ (close to the Néel temperature \TN\ of the Ir sublattice), $\SI{1.5}{\kelvin}$, and $\SI{26}{\milli\kelvin}$ are shown in Fig.~\sfig{fig:1}{a}. \rev{Magnetic Bragg peaks rise below $\SI{50}{\kelvin}$ \cite{Guo2017} and can all be indexed} with a $\bm{k} = \bm{0}$ propagation vector, indicating the onset of long-range magnetic order. The representation of such order in the Tb (Wyckoff 16c) and Ir (Wyckoff 16d) sites can be decomposed as
\begin{equation}
\Gamma = \Gamma_{3} \oplus \Gamma_{5} \oplus \Gamma_{7} \oplus 2\Gamma_{9},
\end{equation}
where the $\Gamma_n$ denote irreducible representations, each defining a distinct set of magnetic moment basis vectors. To track the temperature evolution of the Tb$^{3+}$ magnetic structure, \rev{we performed successive Rietveld refinements at and below $\SI{50}{\kelvin}$}. Due to the weak Ir$^{4+}$ moment, we constrained it to a constant AIAO value of $\SI{0.36}{\mub}$ based on preliminary refinements, significantly improving the fits. Above $\SI{10}{\kelvin}$, the data are well described by a pure $\Gamma_3$ (AIAO) magnetic structure of the Tb$^{3+}$ ions, with moments aligned along their local $\dba*{111}$ Ising axes. 

\rev{Below $\SI{10}{\kelvin}$, additional intensity emerges at the $\dpa*{111}$ position, incompatible with a purely AIAO state, while the absence of magnetic intensity at the $\dpa*{002}$ and $\dpa*{311}$ positions rule out $\Gamma_7$ and $\Gamma_9$ orders. We therefore allow in the refinement a small antiferromagnetic $\Gamma_5$ component perpendicular to the Tb$^{3+}$ ions local $\dba*{111}$ axes (see sketch in Fig.~\sfig{fig:1}{b}), in line with previous studies \cite{Guo2017}. Since $\Gamma_5$ is a two-dimensional irreducible representation and neutron powder diffraction cannot resolve the mixing of basis vectors, only a single one was included ($\Psi_2$ in the notation of Ref. \cite{Yan2017}).} Our measurements performed at dilution temperatures reveal no further changes in the magnetic structure down to at least $T = \SI{26}{\milli\kelvin}$. This indicates that the low-temperature state is already well established at $T=\SI{1.5}{\kelvin}$, where we found an ordered moment of $\SI{5.42\pm0.15}{\mub}$ composed of a dominant $\Gamma_{3}$ contribution of $\SI{5.18\pm0.12}{\mub}$ and a smaller $\Gamma_{5}$ contribution of $\SI{1.59\pm0.12}{\mub}$.

In Fig.~\sfig{fig:1}{b}, we show the temperature dependence of the total Tb$^{3+}$ magnetic moment, along with its $\Gamma_{3}$ and $\Gamma_{5}$ components. The $\Gamma_{3}$ contribution exhibits a temperature evolution characteristic of an induced magnetic order, in this case driven by the molecular field produced by the six Ir$^{4+}$ neighbours of each Tb$^{3+}$, \rev{themselves arranged in an $\Gamma_{3}$ order from $T_\mrm{N} \sim \SI{130}{\kelvin}$. Note that in this scenario, the $\Gamma_{3}$ magnetic moment of the Tb$^{3+}$ ions likely grows from $T_\mrm{N}$, although it can only be resolved around $\SI{50}{\kelvin}$ and below, see Ref.~\cite{Guo2017} for details.} The evolution of the $\Gamma_{5}$ component is harder to track since we cannot exclude unresolved weak moments above $T = \SI{8}{\kelvin}$. 

To probe the existence and nature of a possible transition associated with the $\Gamma_{5}$ component, we performed specific heat measurements, see Fig.~\sfig{fig:1}{c}. The magnetic contribution was extracted by subtracting the rescaled heat capacity of the isostructural non-magnetic rare-earth ion compound Eu$_{2}$Ir$_{2}$O$_{7}$, following the procedure described in Ref.~\cite{Hardy2003}. \rev{Only a broad anomaly is observed around $T = \SI{7}{\kelvin}$, which shifts to higher temperature and becomes weaker under applied magnetic field (see Appendix~\ref{app:AddCpData})}. Note that our previous study on the same sample has shown a bump in the magnetic susceptibility around $\SI{10}{\kelvin}$, coinciding with the appearance of a metamagnetic transition at $\SI{1.8}{\tesla}$ in the magnetic isotherms \cite{Lefrancois2015}. These observations point to the onset of the $\Gamma_5$ magnetic order associated with the Tb-Tb interactions, although the usual signature of a second order phase transition as a sharp peak in the specific heat is missing.

\subsection{Thermal inelastic neutron scattering : High energy spectra}
\label{sec:inelastic}

\begin{figure}
\includegraphics[width=\columnwidth]{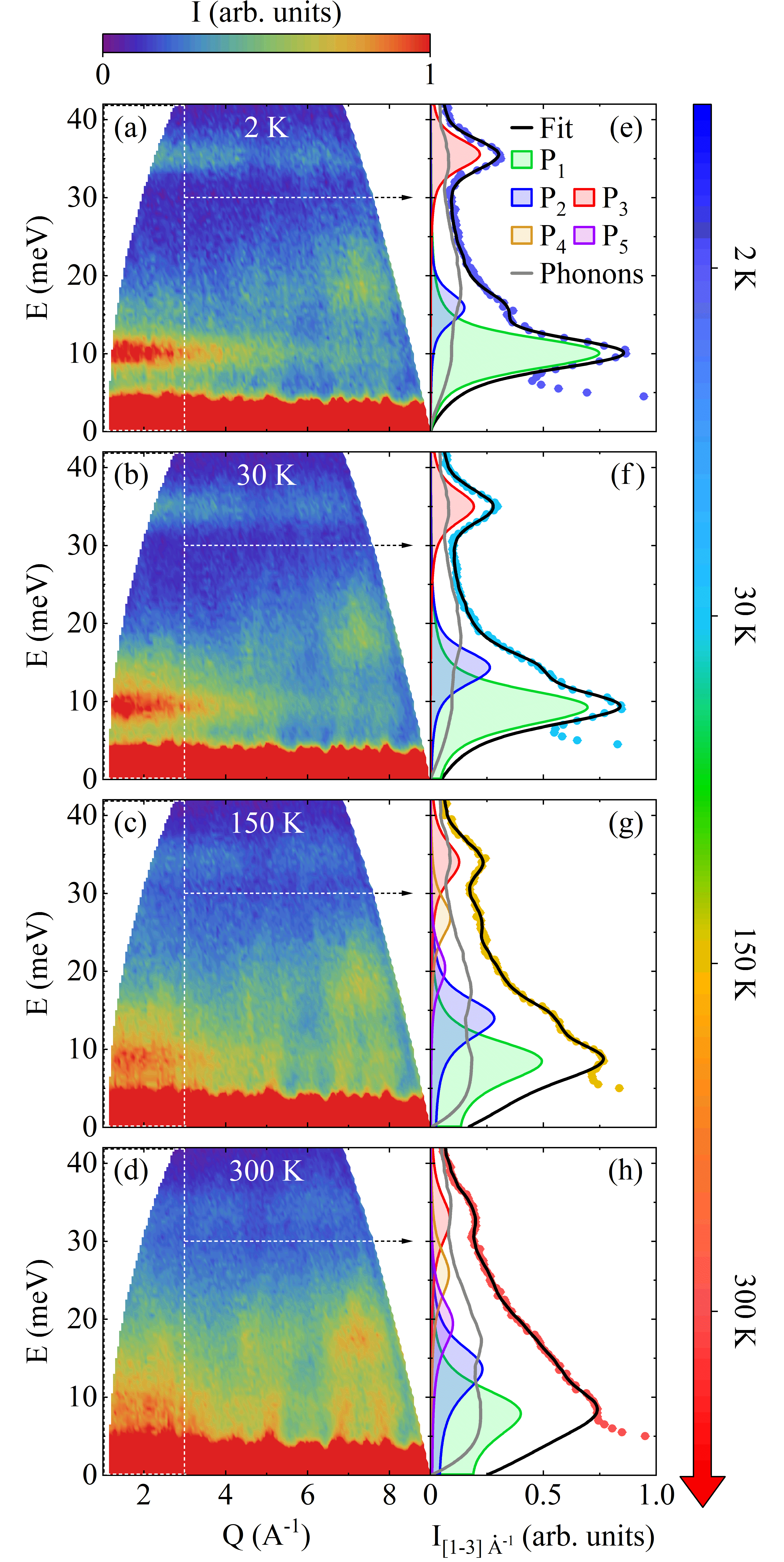}
\caption{(a-d) Scattering function intensity maps measured on IN4c at $T = 2$, $30$, $150$, and $\SI{300}{\kelvin}$, respectively. (e-h) Corresponding momentum-integrated intensities over the range $Q = \SIrange{1}{3}{\angstrom^{-1}}$. Colored dots represent the experimental data, the thick black line is the fit, and the thinner colored lines indicate the different contributions to the fit.}
\label{fig:2}
\end{figure}

The scattering function intensity maps $S(Q,E)$, measured at $T = \qtylist{2;30;150;300}{\kelvin}$ are presented in Figs.~\ssfig{fig:2}{a}{d}. At 2 K, we observe a pronounced signal around $E = \SI{10}{\milli\electronvolt}$, which decreases with increasing $Q$, indicating a magnetic origin. An inspection of the momentum-integrated intensities over $Q = \SIrange{1}{3}{\angstrom^{-1}}$ (hereafter denoted $I_{\left[1-3\right]\,\si{\angstrom^{-1}}}(E)$, see Figs.~\ssfig{fig:2}{e}{h}) reveals an additional weaker excitation at $E \simeq \SI{16}{\milli\electronvolt}$ (confirmed by anti-Stokes measurements on the IN5 spectrometer, see Appendix~\ref{app:AddNeutronsData}). Given their non-dispersive nature, we attribute these two excitations to crystal field levels. In contrast, the scattering observed at large $Q$ values in the $S(Q,E)$ maps, particularly near $E = \SI{20}{\milli\electronvolt}$, is characteristic of phonons. This is further supported by the increased intensity at elevated temperatures, where the acoustic phonon branches become clearly visible. Finally, at higher energy, a nearly $Q$-independent signal is visible around $E = \SI{36}{\milli\electronvolt}$, suggesting a mixed magnetic and phononic origin (dominant at low and high $Q$, respectively). This interpretation is reinforced by data collected with higher incident energy (see Appendix~\ref{app:AddNeutronsData}), which further reveal no additional excitation up to $E = \SI{95}{\milli\electronvolt}$.

We next fitted $I_{\left[1-3\right]\,\si{\angstrom^{-1}}}(E)$ with a phononic $I_{\mrm{ph}}(E)$ and a crystal field $I_{\mrm{cf}}(E)$ contribution at all temperatures. The phononic part was obtained from the high-temperature ($T\ge\SI{150}{\kelvin}$) high-$Q$ integrated intensities $I_{\left[6-8\right]\,\si{\angstrom^{-1}}}(E)$, adjusted by the ratio of the detailed balance factor at lower temperatures and rescaled to account for the reduced phonon intensity at lower transferred momentum. The crystal field part was modeled as a sum of Voigt profiles P$_k$ representing (cluster of) energy levels, each defined as a Lorentzian (with common full width at half maximum $2\gamma$) convolved with a Gaussian of full width at half maximum fixed by the energy resolution of the IN4c spectrometer. Further details on the fitting procedure are given in Appendix~\ref{app:SymAnalysisitNeutronsData}. The fits are depicted in Figs.~\ssfig{fig:2}{e}{h}, with the corresponding parameters reported in Table \ref{tab:fits}. At $T=\SI{2}{\kelvin}$, the spectra are well described using three Voigt profiles P$_{1}$, P$_{2}$ and P$_{3}$, associated with the crystal field excitations discussed above. As temperature increases up to $T=\SI{300}{\kelvin}$, all peaks shift to lower energy and broaden, while their intensities generally decrease -- with the exception of P$_2$, whose intensity strongly increases between $2$ and $\SI{30}{\kelvin}$. This behavior likely reflects transitions not only from the ground state but also from low-energy excited states thermally populated at $\SI{30}{\kelvin}$ (see section \ref{sec:LowInelastic} below). At higher temperatures ($T = 150$ and $\SI{300}{\kelvin}$), two additional excitations P$_4$ and P$_5$ are needed to reproduce the spectra, centered around $\SI{20}{\milli\electronvolt}$ and $\SI{26}{\milli\electronvolt}$, respectively. They arise from transitions between excited levels, as $\mathrm{P}_5 \simeq \mathrm{P}_3 - \mathrm{P}_1$ and $\mathrm{P}_4 \simeq \mathrm{P}_3 - \mathrm{P}_2$.

\begin{table}[t!]
\begin{ruledtabular}
{\renewcommand{\arraystretch}{1.15}
\begin{tabular}{c|c|cccc}
\multicolumn{2}{l|}{ } & $T=\SI{2}{\kelvin}$ & $T=\SI{30}{\kelvin}$ & $T=\SI{150}{\kelvin}$ & $T=\SI{300}{\kelvin}$ \\
\hline
P$_{1-5}$ & $\gamma$ & $1.46$ & $1.86$ & $2.48$ & $3.29$ \\
\hline
\multirow{2}{*}{P$_1$} & $\varepsilon_1$ & $10.0$ & $9.26$ & $8.95$ & $8.78$ \\
                       & $a_1$ & $2.25$ & $2.39$ & $1.05$ & $0.61$ \\
\hline
\multirow{2}{*}{P$_2$} & $\varepsilon_2$ & $15.9$ & $14.3$ & $14.2$ & $14.0$ \\
                       & $a_2$ & $0.44$ & $0.90$ & $0.79$ & $0.50$ \\
\hline
\multirow{2}{*}{P$_3$} & $\varepsilon_3$ & $35.6$ & $35.0$ & $34.0$ & $32.9$ \\
                       & $a_3$ & $0.63$ & $0.64$ & $0.49$ & $0.32$ \\
\hline
\multirow{2}{*}{P$_4$} & $\varepsilon_4$ & & & $20.5$ & $19.7$ \\
                       & $a_4$ & & & $0.21$ & $0.27$ \\
\hline
\multirow{2}{*}{P$_5$} & $\varepsilon_5$ & & & $26.9$ & $26.0$ \\
                       & $a_5$ & & & $0.32$ & $0.27$ \\
\end{tabular}
}
\end{ruledtabular}
\caption{Parameters of the crystal field peaks fitted from the momentum-integrated intensities  $I_{\left[1-3\right]\,\si{\angstrom^{-1}}}(E)$ obtained from IN4c measurements (see Figs.~\ssfig{fig:2}{e}{h}) : characteristic energies $\varepsilon_k$ (in $\si{\milli\electronvolt}$), intensities $a_k$, and Lorentzian half-width at half-maximum $\gamma$ (in $\si{\milli\electronvolt}$).}
\label{tab:fits}
\end{table}

\subsection{Cold inelastic neutron scattering: Low energy magnetic excitation spectra}
\label{sec:LowInelastic}

\begin{figure}
\includegraphics[width=\columnwidth]{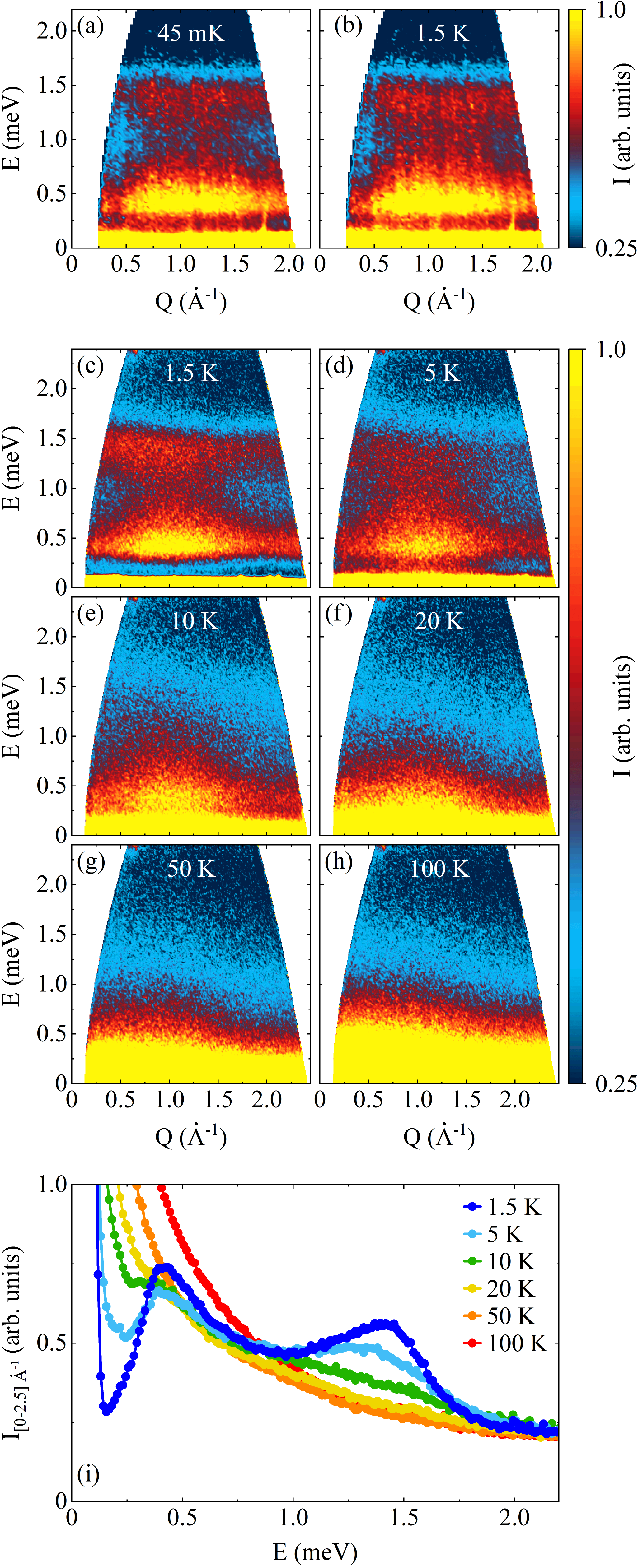}%1.008
\caption{(a,b) Scattering function intensity maps measured on IN6 at $T = 1.5$ and $\SI{45}{\milli\kelvin}$, respectively. (c-h) Scattering function intensity maps measured on IN5 at $T = 1.5$, $5$, $10$, $20$, $50$, and $\SI{100}{\kelvin}$, respectively. (i) IN5 integrated intensity over the whole accessible momentum range.}
\label{fig:3}
\end{figure}

We now turn to the low-energy magnetic excitation spectrum. The scattering function intensity $S(Q,E)$ measured between $T=\SI{45}{\milli\kelvin}$ and $T = \SI{100}{\kelvin}$ is provided in Fig.~\fig{fig:3}. At lowest temperatures, a first feature is visible around $\SI{0.4}{\milli\electronvolt}$, extending over the full $Q$ range but peaking near $Q = \SI{1.1}{\angstrom^{-1}}$, which is close to the position of the first magnetic Bragg peak of the $\Gamma_5$ order. A second less intense signal is clearly visible at $\SI{1.5}{\milli\electronvolt}$ connected to the lower one by some diffuse intensity. An intuitive interpretation is that both features originate from crystal-field levels acquiring dispersive character due to interactions (excitons), as in Tb$_2$Ti$_2$O$_7$ \cite{Roll2024}. Indeed, in Tb-based pyrochlores, the crystal-field spectrum generally consists of a ground-state doublet and a low-lying excited doublet in the energy range $E = \SIrange{1}{2}{\milli\electronvolt}$ \cite{Gingras2000,Mirebeau2007,Bertin2012,Zhang2014,Ruminy2016,Hallas2020}, though in some cases the excited doublet is found at higher energy \cite{Sibille2017, Alexanian2023b}. In \TbIr, these states are expected to be further split by the molecular field generated by the Ir sublattice. 

In line with the behavior evidenced in our diffraction measurements, no significant changes are observed between $T = \SI{45}{\milli\kelvin}$ and $T = \SI{1.5}{\kelvin}$. To have an insight on the temperature dependence of these low energy excitations at higher temperature, it is useful to look at the momentum-integrated intensity between $\SI{1.5}{\kelvin}$ and $\SI{100}{\kelvin}$ shown in Fig.~\sfig{fig:3}{i}. Indeed, above $\SI{1.5}{\kelvin}$, a quasielastic signal is observed, as in Tb$_2$ScNbO$_7$ \cite{Alexanian2023}, whose width increases with increasing temperature, filling the gap to the first level. Concomitantly, the intensity of both levels decreases while they broaden, rendering them no more visible above $T\approx\SI{20}{\kelvin}$. This temperature dependence recalls the one observed in Tb$_2$Ge$_2$O$_7$ \cite{Hallas2020} and Tb$_2$Ti$_2$O$_7$ \cite{Roll2024phd}. The features are however masked in Tb$_2$Ir$_2$O$_7$ in the maps of Fig.~\fig{fig:3} by the quasielastic signal, ascribed to a slowing down of the single-ion spin dynamics due to decreasing interactions with phonons, or to the influence of disorder. Disorder from Tb-Ir antisite or \rev{Ir$^{4+}$/Ir$^{5+}$} substitution has indeed been invoked to explain the ZFC-FC bifurcation occurring at the iridium ordering temperature \cite{Lefrancois2015}.

\section{Numerical model} 

\subsection{Hamiltonian and model parameters}

Reproducing the magnetic properties of \TbIr\ requires accounting for the crystal field, the Ir influence on the Tb ions, and the Tb–Tb interactions. Accordingly, we consider the Hamiltonian
\begin{equation}
\wh{\mathcal{H}} = \wh{\mathcal{H}}_{\mrm{cf}} + \wh{\mathcal{H}}_{\mrm{Ir}}(T) +\wh{\mathcal{H}}_{\mrm{int}}.
\label{eq:FullH}
\end{equation}

The crystal field Hamiltonian is defined as
\begin{equation} 
\begin{aligned}
\widehat{\mathcal{H}}_{\mrm{cf}} &= \sum_{i}\theta_{2}\lambda_{2}^{0}B_{0}^{2}\widehat{\mathcal{O}}_{2}^{0} + \theta_{4}\dpa*{\lambda_{4}^{0}B_{0}^{4}\widehat{\mathcal{O}}_{4}^{0} + \lambda_{4}^{3}B_{3}^{4}\widehat{\mathcal{O}}_{4}^{3}}\\
&+ \theta_{6}\dpa*{\lambda_{6}^{0}B_{0}^{6}\widehat{\mathcal{O}}_{6}^{0} + \lambda_{6}^{3}B_{3}^{6}\widehat{\mathcal{O}}_{6}^{3} + \lambda_{6}^{6}B_{6}^{6}\widehat{\mathcal{O}}_{6}^{6}},
\end{aligned}
\label{eq:Hcf}
\end{equation}
where $\widehat{\mathcal{O}}_{k}^{q} \equiv \widehat{\mathcal{O}}_{k}^{q}(\wh{\bm{J}}_{i})$ are Stevens operators, $B_{q}^{k}$ are crystal field parameters, $\theta_k$ are reduced matrix elements, and $\lambda_{k}^{q}$ are numerical coefficients. The two latter are given in Tables~\ref{Reduced_matrix_element_Tb3+} and \ref{lambda_parameters}, respectively. 

The Tb-Ir interaction Hamiltonian is written
\begin{equation}
\widehat{\mathcal{H}}_{\mrm{Ir}}(T) = -\mu_{0}H_{\mrm{m}}^{\mrm{Ir}}\dpa*{T}g_{J}\mu_{\mrm{B}}\sum_{i}\wh{J}^{z}_{i},
\label{eq:HIr}
\end{equation}
where $H_{\mathrm{m}}^{\mathrm{Ir}}(T)$ denotes the molecular field produced by the Ir and felt by the Tb$^{3+}$ ions. Since we could not determine experimentally its temperature evolution, we modelled $H_{\mrm{m}}^{\mrm{Ir}}\dpa*{T}$ as a free parameter $H_{\mrm{m}}^{\mrm{Ir}} \equiv H_{\mrm{m}}^{\mrm{Ir}}(0)$, and assumed a temperature dependence following a Brillouin function $B_{J=j_{\mathrm{eff}} = 1/2}$ (valid for fully localized Ir$^{4+}$ magnetic moments) and $T_{\mrm{N}} = \SI{130}{\kelvin}$. Note that a Stoner model (describing fully delocalized moments) yields very similar results. 

Restricting ourselves to bilinear exchange interactions between nearest neighbours, the Tb-Tb interaction Hamiltonian reads
\begin{equation}
\widehat{\mathcal{H}}_{\mrm{int}} = \sum_{\dba*{ij}}{}^{t}\wh{\bm{J}}_{i}\mathcal{J}_{ij}\wh{\bm{J}}_{j}
\label{eq:Hint},
\end{equation}
where $\mathcal{J}_{ij}$ is a $3 \times 3$ exchange matrix. A symmetry analysis of the pyrochlore lattice shows that $\mathcal{J}_{ij}$ involves only four independent parameters, denoted $\mathcal{J}_{\mathrm{zz}}$, $\mathcal{J}_{\pm}$, $\mathcal{J}_{\mathrm{z\pm}}$, and $\mathcal{J}_{\pm\pm}$ in the local basis of the Tb$^{3+}$ ions \rev{(see Appendix~\ref{app:SymAnalysis})}. Since both $\Gamma_{3}$ and $\Gamma_{5}$ orders observed experimentally are \textit{classical} magnetic configurations (i.e. are $\bm{k} = \bm{0}$ states described by tensor products of single-site states), our calculations can be restricted to the four sites of a single tetrahedron $t$ with periodic boundary conditions. Within this framework, $\mathcal{J}_{\mathrm{zz}}$ (respectively, $\mathcal{J}_{\pm}$) is the leading term that stabilizes the $\Gamma_{3}$ (respectively $\Gamma_{5}$) ordering \cite{Yan2017,Rau2019}. We thus neglect $\mathcal{J}_{\mathrm{z\pm}}$ and $\mathcal{J}_{\mathrm{\pm\pm}}$ in a first approximation, so that the Tb-Tb interaction Hamiltonian $\widehat{\mathcal{H}}_{\mrm{int}}$ reduces to
\begin{equation}
\widehat{\mathcal{H}}_{\mrm{int}} = \sum_{i,j\in t}\mathcal{J}_{\mrm{zz}}\wh{J}_{i}^{z}\wh{J}_{j}^{z} - \mathcal{J}_{\mrm{\pm}}\dpa*{\wh{J}_{i}^{+}\wh{J}_{j}^{-}+\wh{J}_{i}^{-}\wh{J}_{j}^{+}}
\label{eq:Hint_extended}
\end{equation}
which has only two free parameters.

In many pyrochlore oxides, the molecular field is absent (non-magnetic transition-metal ion) and the interactions are often treated as a perturbation on the crystal-field ground-state doublet. This approach, however, is not suitable for \TbIr\ where the Ir is magnetic and where the Tb$^{3+}$ first excited doublet can be close in energy form the ground doublet: restricting the model to a non-Kramers ground state doublet would confine the magnetic moments to the $\dba*{111}$ directions and thus preclude the onset of the $\Gamma_5$ order. Instead, we employed a mean-field method (see Appendix~\ref{app:MFmethod}) to diagonalize the full Hamiltonian of Eq.~\ref{eq:FullH} within the ${}^{7}F_{6}$ multiplet of Tb$^{3+}$. The computation of the observables later compared with experiments are described in Appendix~\ref{app:ObsCompute} for the ones calculated at the mean field level, and in Ref. \cite{Roll2024} for those within the RPA approximation. Because the full Hamiltonian involves nine parameters, we adopted a two-step strategy loosely inspired by a perturbative approach to refine them. This procedure is described in the Methods part (Section.~\ref{sec:num_details}) and in greater detail in Appendix~\ref{app:Parameters}.

\begin{table}
\begin{ruledtabular}
\begin{tabular} {ccc}
$\theta_{2}$ & $\theta_{4}$ & $\theta_{6}$ \\
\hline  \noalign{\vskip 1mm}   
$-1/99$ & $2/16335$ & $-1/891891$ \\
\end{tabular}
\end{ruledtabular}
\caption{Reduced matrix element $\theta_{k}(J)$ for Tb$^{3+}$ ($J = 6$), from Refs. \cite{Stevens1952,Hutchings1964}.}
\label{Reduced_matrix_element_Tb3+}
\end{table}

\begin{table}
\begin{ruledtabular}
\begin{tabular} {cccccc}
$\lambda_{2}^{0}$ & $\lambda_{4}^{0}$ & $\lambda_{4}^{3}$ & $\lambda_{6}^{0}$ & $\lambda_{6}^{3}$ & $\lambda_{6}^{6}$ \\
\hline \noalign{\vskip 1mm}
$1/2$ & $1/8$ & $\sqrt{35}/2$ & $1/16$ & $\sqrt{105}/8$ & $\sqrt{231}/16$ \\
\end{tabular}
\end{ruledtabular}
\caption{$\lambda_{k}^{q}$ numerical coefficients, adapted from Ref. \cite{Danielsen1972}.}
\label{lambda_parameters}
\end{table}

\subsection{Results}

We identify several parameter sets that reproduce most of the experimental observables. Interestingly, our calculations reveal that they gather into two qualitatively distinct groups. One representative parameter set for each group is listed in Table~\ref{tab:params}. The corresponding crystal-field wavefunctions (calculated with no interactions), shown in Table~\ref{tab:wavefnc}, are markedly different. In the first set, the ground and first excited doublets are dominated by $\ket{\pm 5}$ and $\ket{\pm 4}$, respectively, whereas this order is reversed in the second set. Both types of low energy level wavefunctions have been reported in Tb$_2$Ti$_2$O$_7$, Tb$_2$Sn$_2$O$_7$ and Tb$_2$Ge$_2$O$_7$ \cite{Gingras2000,Mirebeau2007,Bertin2012,Zhang2014,Ruminy2016,Hallas2020}.

\begin{table}[t]
\begin{ruledtabular}
\begin{tabular} {cccccccccc}
Set & $B_{0}^{2}$ & $B_{0}^{4}$ & $B_{3}^{4}$ & $B_{0}^{6}$ & $B_{3}^{6}$ & $B_{6}^{6}$ & $H_{\mathrm{m}}^{\mathrm{Ir}}$ & $\mathcal{J}_{\mathrm{zz}}$ & $\mathcal{J}_{\pm}$ \\
\hline\noalign{\vskip 1mm} 
1 & $\num{49}$ & $\num{251}$ & $\num{72}$ & $\num{17}$ & $\num{-99}$ & $\num{121}$ & $\num{1.8}$ & $\num{0.028}$ & $\num{0.051}$ \\
2 & $\num{46}$ & $\num{253}$ & $\num{68}$ & $\num{29}$ & $\num{-100}$ & $\num{115}$ & $\num{1.7}$ & $\num{0.005}$ & $\num{0.046}$ \\
\end{tabular}
\end{ruledtabular}
\caption{Two representative sets of crystal field $\{B_{k}^{q}\}$ (in $\si{\milli\electronvolt}$), molecular field $H_{\mathrm{m}}^{\mathrm{Ir}}$ (in $\si{\tesla}$) and interaction $\{\mathcal{J}_{\mathrm{zz}}, \mathcal{J}_{\pm}\}$ (in $\si{\kelvin}$) parameters.}
\label{tab:params}
\end{table}

\begin{table}[!ht]
\begin{ruledtabular}
{\renewcommand{\arraystretch}{1.02}
\begin{tabular} {l|cc|cc}
& \multicolumn{2}{c|}{Set $1$} & \multicolumn{2}{c}{Set $2$} \\
\hline\hline
& $\ket{\psi_{0}^{\pm}}$ ($\num{0.00}$) & $\ket{\psi_{1}^{\pm}}$ ($1.05$) & $\ket{\psi_{0}^{\pm}}$ ($\num{0.00}$) & $\ket{\psi_{1}^{\pm}}$ ($1.14$) \\
\hline
$\ket{\pm6}$ & $0$ & $0$ & $0$ & $0$ \\
$\ket{\pm5}$ & $\pm\bm{0.769}$ & $0$ & $0$ & $\pm\bm{0.853}$ \\
$\ket{\pm4}$ & $0$ & $\pm\bm{0.792}$ & $\pm\bm{0.861}$ & $0$ \\
$\ket{\pm3}$ & $0$ & $0$ & $0$ & $0$ \\
$\ket{\pm2}$ & $-0.235$ & $0$ & $0$ & $-0.148$ \\
$\ket{\pm1}$ & $0$ & $-0.118$ & $-0.058$ & $0$ \\
$\ket{\;\;\: 0}$ & $0$ & $0$ & $0$ & $0$ \\
$\ket{\mp1}$ & $\mp0.018$ & $0$ & $0$ & $\pm0.096$ \\
$\ket{\mp2}$ &$0$ & $\mp0.059$ & $\pm0.189$ & $0$ \\
$\ket{\mp3}$ & $0$ & $0$ & $0$ & $0$ \\
$\ket{\mp4}$ & $-0.594$ & $0$ & $0$ & $0.491$ \\
$\ket{\mp5}$ & $0$ & $-0.597$ & $0.469$ & $0$ \\
$\ket{\mp6}$ & $0$ & $0$ & $0$ & $0$ \\
\end{tabular}
}
\end{ruledtabular}
\caption{Crystal-field wavefunctions and corresponding energies in $\si{\milli\electronvolt}$ (in parentheses) for the four lowest levels calculated using parameter sets 1 and 2 obtained without interactions ($H_{\mathrm{m}}^{\mathrm{Ir}}$, $\mathcal{J}_{\mathrm{zz}}$ and $\mathcal{J}_{\pm}$). The coefficient of the dominant $\ket{M_{J}}$ component in each state is shown in bold.}
\label{tab:wavefnc}
\end{table}

\begin{figure}
\includegraphics[width=1\columnwidth]{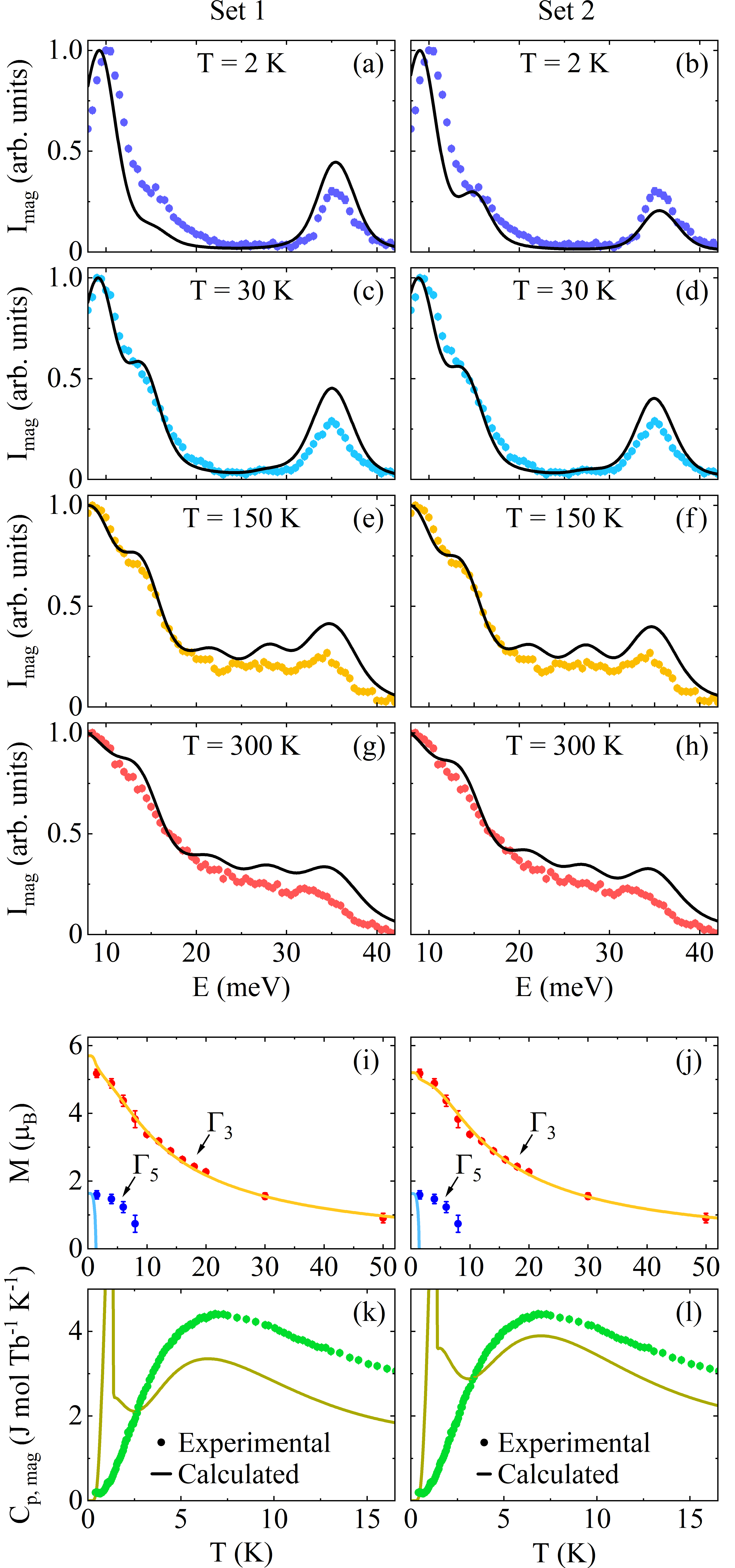}
\caption{Comparison of experimental (dots) and calculated (lines) observables at the mean-field level for parameter set 1 on the left and parameter set 2 on the right. (a-h) high-energy neutron scattering magnetic intensity at $T=2$, $30$, $150$ and $\SI{300}{\kelvin}$. (i,j) $\Gamma_{3}$ (AIAO) and $\Gamma_{5}$ components of the Tb$^{3+}$ ordered magnetic moment. (k,l) Magnetic specific heat.  }
\label{fig:4}
\end{figure}

Including the interactions, the calculated energy levels for both sets at $T=\SI{0.1}{\kelvin}$ are listed in Table~\ref{tab:Enlevels}. We note that the three lowest excited levels $E_{1}-E_{3}$ originate from the ground and first excited crystal field doublets, which are split and mixed by the molecular field and the Tb-Tb interactions. In Fig.~\ref{fig:4}, we present the comparisons between experimental and calculated observables at the mean field level. First, the high-energy neutron scattering magnetic intensity ($I_{\mrm{mag}} \equiv I_{\left[1-3\right]\,\si{\angstrom^{-1}}} - I_{\mrm{ph}}$) shown in Figs.~\ssfig{fig:4}{a}{h} at various temperatures are well reproduced. \rev{For both sets, the P$_1$ and P$_2$ features arise from singlets at $E_4\simeq\SI{9.2}{\milli\electronvolt}$ and $E_5\simeq\SI{15}{\milli\electronvolt}$, respectively. While their relative intensity is well captured by our model at $\SI{30}{\kelvin}$ and above, we note that the intensity of the latter is better reproduced at low temperature for set 2.  Meanwhile, P$_3$ originates from a cluster of four levels $E_{6-9}$ around $\num{35}-\SI{38}{\milli\electronvolt}$, with calculated intensities that agree similarly well with the data for both sets (underestimated for set 1 and overestimated for set 2 at $\SI{2}{\kelvin}$, and overestimated for both sets at $\SI{30}{\kelvin}$ and above).}

\begin{table}[!t]
\centering
\begin{ruledtabular}
\begin{tabular} {l|c|c}
& Set 1 \hspace{1.5cm} & Set 2 \hspace{1.5cm} \\
\hline \noalign{\vskip 1mm}   
$E_{12}$ & $56.1$ \hspace{1.5cm} & $55.6$ \hspace{1.5cm} \\
$E_{11}$ & $51.0$ \hspace{1.5cm} & $50.3$ \hspace{1.5cm} \\
$E_{10}$ & $50.8$ \hspace{1.5cm} & $50.0$ \hspace{1.5cm} \\
$E_{9}$ & $37.7$ \hspace{1.5cm} & $36.9$ \hspace{1.5cm} \\
$E_{8}$ & $37.3$ \hspace{1.5cm} & $36.8$ \hspace{1.5cm} \\
$E_{7}$ & $36.5$ \hspace{1.5cm} & $36.2$ \hspace{1.5cm} \\
$E_{6}$ & $34.8$ \hspace{1.5cm} & $35.0$ \hspace{1.5cm} \\
$E_{5}$ & $15.5$ \hspace{1.5cm} & $15.1$ \hspace{1.5cm} \\
$E_{4}$ &$9.39$ \hspace{1.5cm} & $9.03$ \hspace{1.5cm} \\
$E_{3}$ & $1.64$ \hspace{1.5cm} & $2.10$ \hspace{1.5cm} \\
$E_{2}$ & $1.45$ \hspace{1.5cm} & $1.50$ \hspace{1.5cm} \\
$E_{1}$ & $0.33 \hspace{1.5cm}$ & $0.37$ \hspace{1.5cm} \\
$E_{0}$ & $0$ \hspace{1.5cm} & $0$ \hspace{1.5cm} \\
\end{tabular}
\end{ruledtabular}
\caption{Calculated energy levels (in $\si{\milli\electronvolt}$) for parameter sets 1 and 2 at $T=\SI{0.1}{\kelvin}$ in presence of interactions. }
\label{tab:Enlevels}
\end{table}

\begin{figure}
\includegraphics[width=\columnwidth]{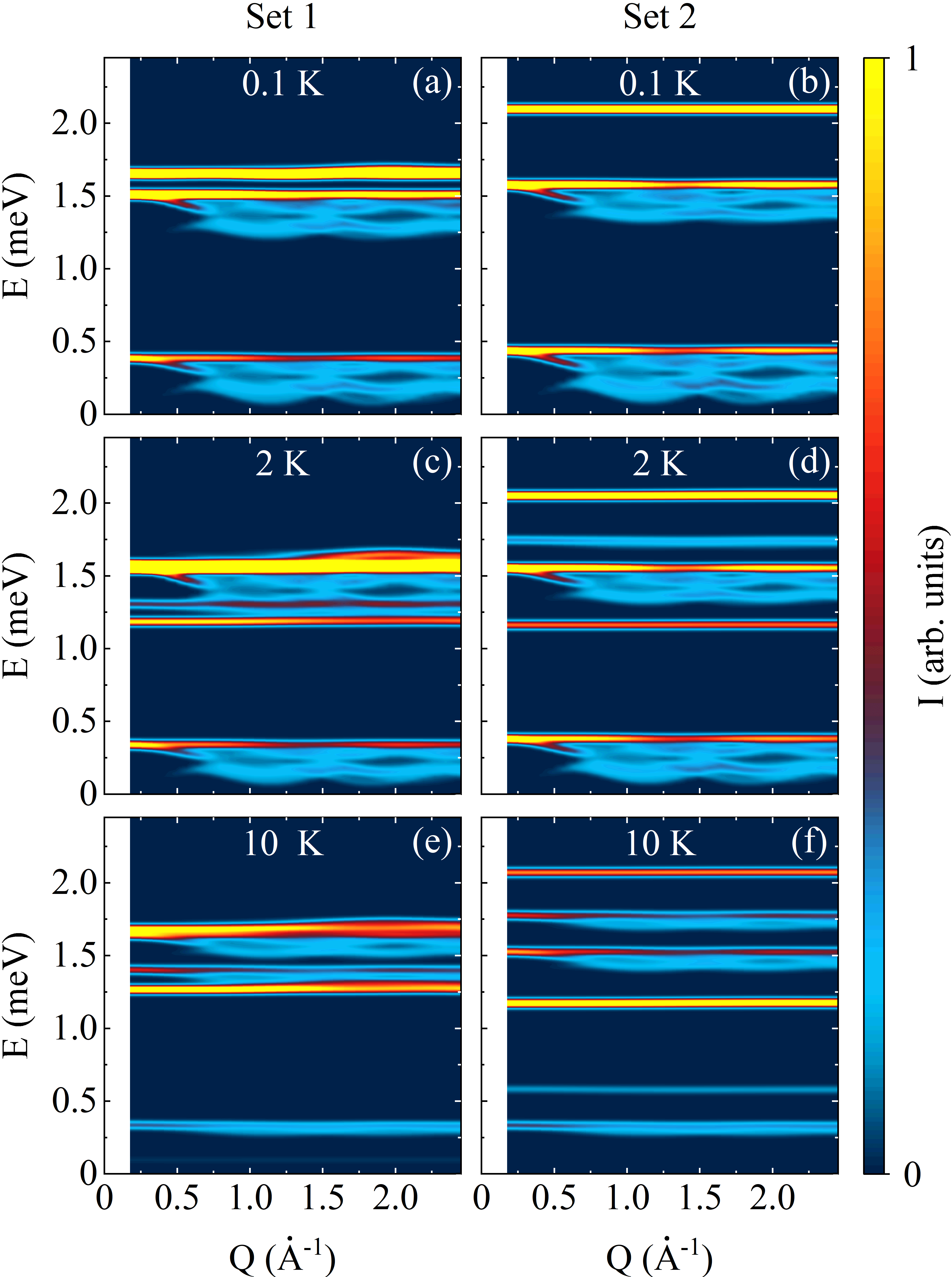}
\caption{Calculated low-energy neutron scattering spectrum using the RPA approximation for parameter set 1 on the left and parameter set 2 on the right at $T=0.1$, $2$, and $\SI{10}{\kelvin}$.}
\label{fig:5}
\end{figure}

Concerning the ordered magnetic moments (Figs.~\sfig{fig:4}{i} and \sfig{fig:4}{j}), the temperature evolution of the $\Gamma_{3}$ component is well captured in our calculations. The appearance of the $\Gamma_{5}$ moment is also calculated at low temperature, although it drops rapidly to zero as the temperature rises up to $\SI{1.5}{\kelvin}$. This contrasts with the measured $\Gamma_{5}$ moment which is still visible at $\SI{8}{\kelvin}$. Note that this is the highest ordering temperature associated with the Tb-Tb interactions reported among the members of the Tb pyrochlore family. For the magnetic specific heat (Figs.~\sfig{fig:4}{k-l}), the calculated curves capture the position of the broad bump around $\SI{7}{\kelvin}$ in the experimental data, which originates from the excited levels $E_{2}$ and $E_{3}$ in our calculations. \rev{The underestimated calculated signal (present for both sets but more pronounced for set 1) could be due to the mean-field treatment of the magnetic correlations, whose onset could produce some signal in the same temperature range.} Note that the first excited level $E_{1}$ gives rise to a calculated broad peak near $\SI{1.5}{\kelvin}$ below our experimental temperature range. Our calculations also predict a sharp peak at the temperature corresponding to the $\Gamma_{5}$ ordering temperature. Such a peak is not observed in the expected temperature range of our measured specific heat as already mentioned. This absence could be due to the disorder or to some influence of the Ir discussed later.

To inspect further the low energy behavior of our two sets of wavefunctions, neutron scattering spectra were calculated using the RPA approximation and shown in Fig.~\fig{fig:5} at $T = 10$, $2$ and $\SI{0.1}{\kelvin}$. As the temperature decreases, the spectral weight of the inter-transitions visible at $\SI{10}{\kelvin}$ diminishes until it disappears, leaving a spectrum of energy levels excited only from the ground state exhibiting a small dispersion due to the magnetic interactions. \rev{At low temperature, both sets then reveal a band near $\SI{0.4}{\milli\electronvolt}$ which corresponds to the lowest energy level visible in the neutron maps of Fig.~\fig{fig:3}. The next two excited levels are calculated close to $\SI{1.5}{\milli\electronvolt}$ for set 1, and at $\SI{1.5}{\milli\electronvolt}$ and $\SI{2}{\milli\electronvolt}$ for set 2. The calculations performed at low temperatures with set 1 resemble the experiment better than set 2 since these two levels align with the experimental signal around $\SI{1.5}{\milli\electronvolt}$, whereas no intensity is detected at $\SI{2}{\milli\electronvolt}$. However, neither sets fully reproduces all experimental features:} (i) the calculated dispersion associated with the lowest energy mode develops mainly below it where the experiment exhibits a gap; (ii) the spectral weight is not maximum near $Q=\SI{1.1}{\angstrom^{-1}}$ contrary to the experiment; (iii) the temperature evolution of the relative weight of the signals around $\SI{0.4}{\milli\electronvolt}$ and $\SI{1.5}{\milli\electronvolt}$ is not well reproduced.

\section{Discussion and conclusion}

Our model provides an overall consistent description of the magnetic specific heat, the high-energy neutron-scattering intensity reflecting the crystal field scheme of the Tb$^{3+}$ above $\SI{9}{\milli\electronvolt}$, and the temperature evolution of the $\Gamma_{3}$ ordered moment component induced by the Ir molecular field in Tb$_2$Ir$_2$O$_7$. \rev{It also predicts the presence of low-energy neutron-scattering spectral weight and the rise of a $\Gamma_{5}$ magnetic component at low temperature, both assumed to originate from Tb-Tb interactions.} \rev{In detail, both parameter sets have distinct strengths, and it is not possible to unambiguously discriminate between the two at this stage: set 2 better reproduces the intensity at $\SI{15}{\milli\electronvolt}$ at low temperature and the specific heat but introduces scattered intensity around $\SI{2}{\milli\electronvolt}$, inconsistent with the low-energy neutron scattering data more faithfully captured by set 1.}

Such a $\Gamma_{5}$ component has already been invoked in Tb$_2$Ti$_2$O$_7$, whose ground state is described at the mean field level as a superposition of order parameters from the different irreducible representations \cite{Roll2024}. According to this work, Tb$_2$Ti$_2$O$_7$ would then lie at the border between the $\Gamma_{3}\oplus\Gamma_{5}$ and the spin ice phases, explaining its spin liquid like behavior. The $\Gamma_{5}$ component is shown to be associated to the $\mathcal{J}_{\pm}$ interaction and goes along with the $\Gamma_{3}$ component. This entanglement between the $z$ and planar moment components is argued to result from virtual crystal field transitions allowed in a crystal field model considering the ground and first excited levels mixed by the magnetic interactions \cite{Roll2024}. In the present compound, the obtained $\mathcal{J}_{\pm}$ interaction is much lower than in Tb$_2$Ti$_2$O$_7$, while a significant $\Gamma_5$ component is observed in contrast to Tb$_2$Ti$_2$O$_7$. We ascribe this surprising outcome to differences in the low energy level wavefunctions.

Beyond the qualitative agreements between the calculations and the experiments, our model underestimates the $\Gamma_5$ transition temperature, predicts a marked peak in the specific heat at such temperature, and does not reproduce the details of low-energy excitations. Our results were achieved with a minimal Hamiltonian including the crystal field, the interactions of the Ir with the Tb treated as a molecular field along the Tb $z$ local axis, and the interactions $\mathcal{J}_{zz}$ and $\mathcal{J}_{\pm}$. It probably needs to be extended to capture the physics of this compound more quantitatively. Note that increasing $\mathcal{J}_{\pm}$ substantially raises the $\Gamma_{5}$ transition temperature, but at the cost of increasing the ordered $\Gamma_{5}$ component (around $\SI{4}{\mu_{\mrm{B}}}$ for a transition at $\SIrange{8}{10}{\kelvin}$), in clear disagreement with experiments. It is known that additional ingredients are present in the Tb pyrochlores that contribute to their exotic behavior: the $\mathcal{J}_{z\pm}$ and $\mathcal{J}_{\pm\pm }$ interactions, as well as quadrupolar interactions \cite{Roll2024}. Concerning the latter, our symmetry analysis identifies five out of nine independent quadrupolar exchange parameters that contribute to the $\Gamma_{3}$ and $\Gamma_{5}$ orderings. Yet, all tested parameter sets including these quadrupolar interactions still yield transition temperatures for the $\Gamma_{5}$ order that remain too low. \rev{We can also mention vibronic bound states, as they can induce a splitting and a dispersive character to the initial crystal field levels through hybridization with phonons, and are known to play an important role in Tb$_2$Ti$_2$O$_7$  \cite{Constable2017,Gaudet2018,Alexanian2023}. In their simplest form, they appear as quadrupolar mean-field terms \cite{Santini2009,Rau2019} that, as just discussed, do not improve the agreement with experiment, and are then likely not the dominant driver of the low-energy physics in Tb$_2$Ir$_2$O$_7$.} Concerning the $\mathcal{J}_{z\pm}$ and $\mathcal{J}_{\pm\pm}$ Tb-Tb interactions, while they do not affect the temperature dependence at the mean-field level as long as only the $\Gamma_{3}$ and $\Gamma_{5}$ magnetic orders are stabilized, they could become relevant beyond mean-field theory, which is however an approach beyond the scope of this paper. Note that all these interactions could also modify the dispersion of the low lying excitations \cite{Roll2024}.

Finally, the high temperature for the onset of the $\Gamma_5$ order and the absence of a concomitant sharp peak in the specific heat may point to the role of the iridium in inducing/strengthening the $\Gamma_{5}$ correlations. Such scenario may originate from more complex couplings between the Tb$^{3+}$ ions and the Ir magnetic sublattice that should be taken into account explicitly and not as a molecular field. Moreover, in our model, the Ir$^{4+}$ moments are constrained to their local $\langle 111 \rangle$ axes, which limits Tb–Ir interactions to the effective form of Eq.~\ref{eq:HIr}. Allowing deviations from this constraint and taking into account the full Ir-Ir and Ir-Tb Hamiltonian \cite{Faure2024} could introduce additional interaction channels and feedback effects \cite{Museur2026}, potentially exerting a significant influence on the $\Gamma_{5}$ temperature ordering as well.

In conclusion, our in-depth study of \TbIr\ uncovers a highly unconventional ground state, established at $T=\SI{1.5}{\kelvin}$ and stable at lower temperatures. The Tb$^{3+}$ moments deviate from their local anisotropy axes while incorporating an antiferromagnetic planar component ($\Gamma_5$) associated to Tb-Tb interactions. In addition, they exhibit a collective character revealed by a dispersive low lying excitation spectrum. Our numerical simulations, including the crystal field scheme, the molecular field produced by the Ir sublattice and Tb-Tb bilinear exchange interactions, highlight the non-trivial origin of this state. They succeed in reproducing the presence of both magnetic components at low temperature thanks to the mixing of the two lowest lying crystal field doublets. They miss other observations such as the temperature dependence of the low-energy properties. This intriguing behavior, calling for a better description of the Tb-Ir coupling, adds to the long and complex history of Tb-based pyrochlores exhibiting unexpected magnetic phenomena.

\section*{acknowledgments}

We thank L. C. Chapon for fruitful discussions. We acknowledge technical support during our experiment performed at the LLB and ILL (proposals 4-01-1380 and TEST-2380), and P. Lachkar for the technical assistance for the specific heat measurements. This work was supported by the Fédération Française de Diffusion Neutronique (2FDN). We would also like to pay tribute to Bjorn F{\aa}k, who passed away in 2024.

\section*{Data availability}

Neutron scattering data that support the findings of this article are openly available \cite{ILLdata1}. Specific heat data are available from the authors upon reasonable request.

%%%%%%%%%%%%%%%%%%%%%%%%%%%%%%%%%%%%%%%%%%%%%%%%%%%%%%%%%%%%%%%%%%%%%%%%%%%%%%%%%%%%%%%%%%%%%%%%%%%%%%%%%%%%%%%%

\begin{appendix}

\section{ADDITIONNAL SPECIFIC HEAT DATA}
\label{app:AddCpData}
\setcounter{figure}{0}
\setcounter{table}{0}
\renewcommand{\thefigure}{A\arabic{figure}}
\renewcommand{\thetable}{A\arabic{table}}

\begin{figure}[t]
\includegraphics[width=1\columnwidth]{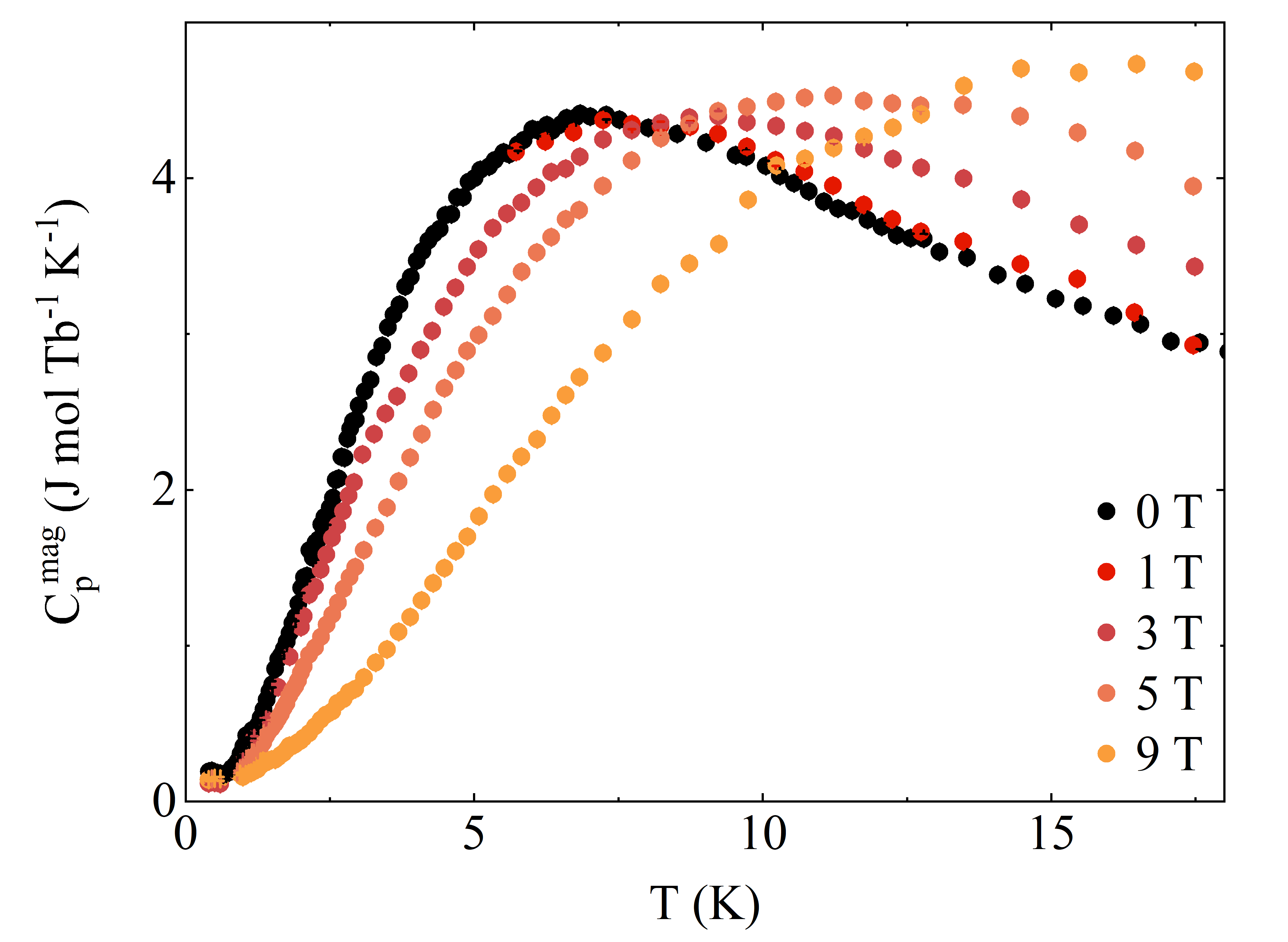}
\caption{Temperature dependence of the magnetic specific heat at different applied magnetic fields.}
\label{fig:SM0}
\end{figure}

\rev{Fig.~\fig{fig:SM0} shows the magnetic specific heat of Tb$_2$Ir$_2$O$_7$ at various applied magnetic fields. It was obtained by subtracting the rescaled zero-field heat capacity of the isostructural non-magnetic compound Eu$_2$Ir$_2$O$_7$ from the total specific heat of Tb$_2$Ir$_2$O$_7$ measured at each applied field. The broad anomaly observed around $T=\SI{7}{\kelvin}$ at zero field shifts to higher temperature and is progressively suppressed under applied magnetic field, confirming its magnetic origin.}

\section{ADDITIONNAL INELASTIC NEUTRON SCATTERING DATA}
\label{app:AddNeutronsData}

The negative energy transfer part of the momentum-integrated neutron scattering intensity measured on the IN5 spectrometer ($\lambda = \SI{4.8}{\angstrom}$) is shown in Fig.~\fig{fig:SM1}. Clear inelastic scattering is observed around $E = \SIrange{-7}{-10}{\milli\electronvolt}$ and $E = \SI{-14}{\milli\electronvolt}$ in the $T = \SI{50}{\kelvin}$ and $\SI{100}{\kelvin}$ datasets, supporting the presence of crystal field transitions at these energies. The intensity of the former consists of two peaks, one at $E = \SI{-9.5}{\milli\electronvolt}$ and another one at $E = \SI{-8}{\milli\electronvolt}$ for $T = \SI{50}{\kelvin}$, which shift closer at $T = \SI{100}{\kelvin}$. These features are interpreted as signatures of multiple transitions to the crystal field level near $E \simeq \SI{9.5}{\milli\electronvolt}$, originating from the ground state and low-lying excited levels. These are more strongly split at lower temperatures, probably reflecting the stronger molecular field from the Ir sublattice.

The $S(Q,E)$ scattering function intensity map and the corresponding energy-averaged intensity in the range $E = \SIrange{32}{38}{\milli\electronvolt}$, measured at $T = \SI{2}{\kelvin}$ on the IN4c spectrometer with $\lambda = \SI{0.8}{\angstrom}$, are presented in Fig.~\fig{fig:SM2}. No scattering is observed above $\SI{50}{\milli\electronvolt}$. The energy-integrated spectrum confirms the presence of a mixed magnetic and phononic signal centered at $E = \SI{36}{\milli\electronvolt}$. The magnetic component accounts for most of the intensity below $Q \simeq \SI{5}{\angstrom^{-1}}$, while the phonon contribution becomes dominant at higher momentum transfers.

\section{COMPLEMENTS ON THE FITS OF IN4c MOMENTUM-AVERAGED INTENSITY}
\label{app:SymAnalysisitNeutronsData}

\setcounter{figure}{0}
\setcounter{table}{0}
\renewcommand{\thefigure}{B\arabic{figure}}
\renewcommand{\thetable}{B\arabic{table}}

We detail in this appendix the phononic and crystal field contributions used to fit the momentum-integrated intensities $I_{\left[1-3\right]\si{\angstrom^{-1}}}(E,T)$ shown in Fig.~\fig{fig:2}. Note, this analysis assumes a complete separation of the powder-averaged dynamical structure factor into a phononic $S^{\mathrm{d}}_{\mathrm{ph}}$ and a crystal field $S^{\mathrm{d}}_{\mathrm{cf}}$ contribution.

\begin{figure}
\includegraphics[width=1\columnwidth]{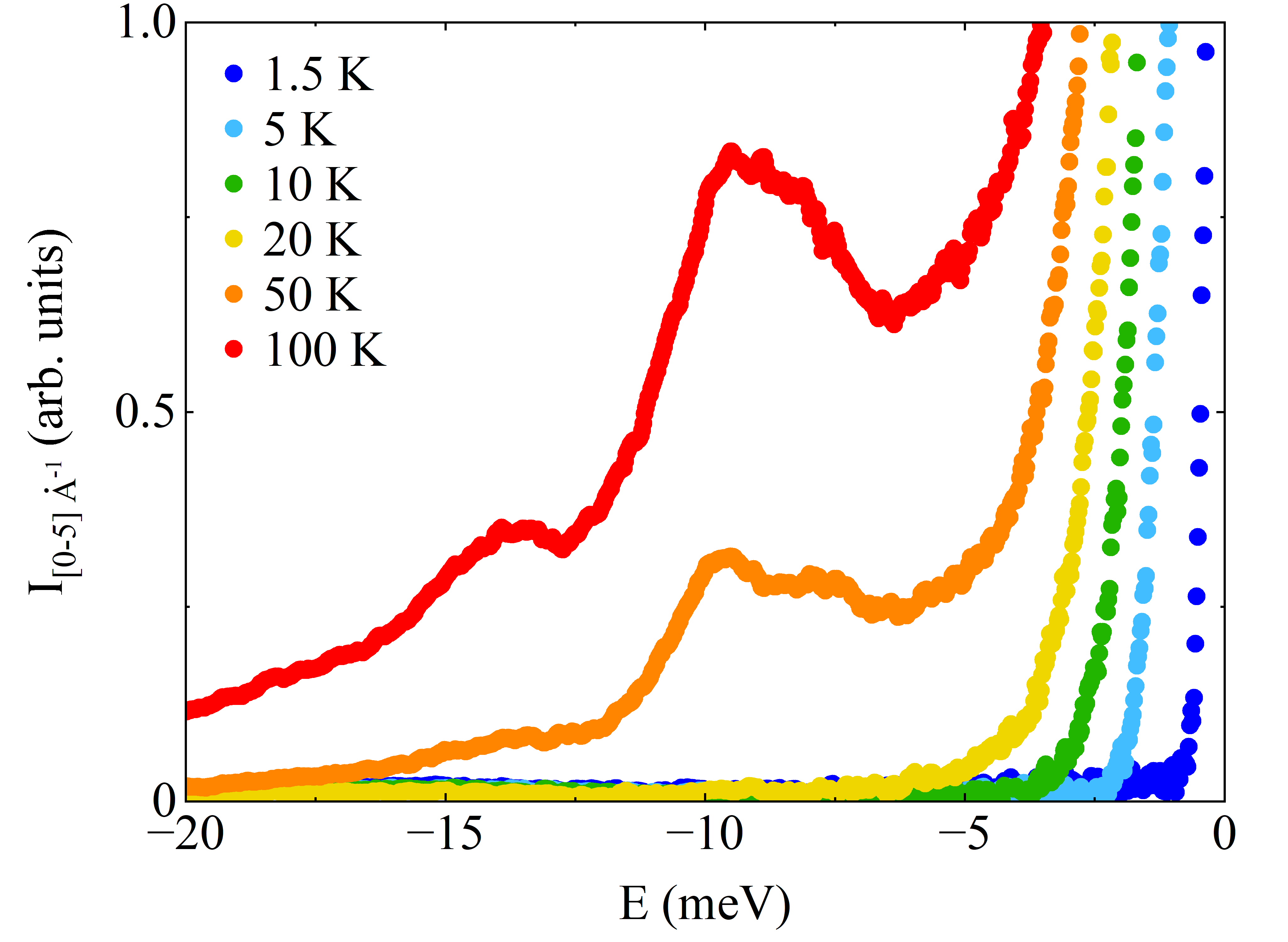}
\caption{Negative energy transfer part of the momentum-integrated intensities over the range $Q = \SIrange{0}{5}{\angstrom^{-1}}$, measured at various temperatures on the IN5 spectrometer ($\lambda_{\mathrm{i}} = \SI{4.8}{\angstrom}$).}
\label{fig:SM1}
\end{figure}

\subsection{Phononic contribution}

At sufficiently low temperature, when the phonon density of state $Z_{\mrm{ph}}$ is nearly temperature-independent, the incoherent phonon scattering dynamical structure factor can be approximated as
\begin{equation}
S^{\mrm{d}}_{\mrm{ph, inc}}(Q,E,T) \simeq \frac{a+bQ^{2}\mrm{e}^{-2W(Q)}}{1-\mrm{e}^{-E/k_{\mrm{B}}T}}\frac{Z_{\mrm{ph}}(E)}{E}.
\label{eq:phononIncScattering}
\end{equation}
In Eq. \ref{eq:phononIncScattering}, $a$ and $b$ are numerical factors accounting for the multiphonon scattering and the average polarization dependence in the incoherent one-phonon cross section, respectively, and $\mrm{e}^{-2W(Q)}$ is the Debye-Waller factor. This expression no longer holds for coherent phonon scattering, but averaging the coherent scattering at sufficiently large momentum and over a sufficiently large range $2\Delta Q = Q_{2} - Q_{1}$ also results in a signal proportional to $Z_{\mrm{ph}}(E)/E$ , this is the \textit{incoherent approximation}. Neglecting the Debye-Waller factor, it comes,
\begin{equation}
\begin{gathered}
\int_{Q_{1}}^{Q_{2}} \frac{S^{\mrm{d}}_{\mrm{ph, coh}}(Q,E,T)}{2\Delta Q}\mathrm{d}Q \propto  \frac{a+b\dba{Q^{2}}_{Q_{1}}^{Q_{2}}}{1-\mrm{e}^{-E/k_{\mrm{B}}T}} \frac{Z_{\mrm{ph}}(E)}{E},\\
\dba{Q^{2}}_{Q_{1}}^{Q_{2}} \equiv \frac{1}{2\Delta Q}\int_{Q_{1}}^{Q_{2}}Q^{2}\mrm{d}Q.
\end{gathered}
\label{eq:phononCohScatteringInteg}
\end{equation}

\begin{figure}
\includegraphics[width=1\columnwidth]{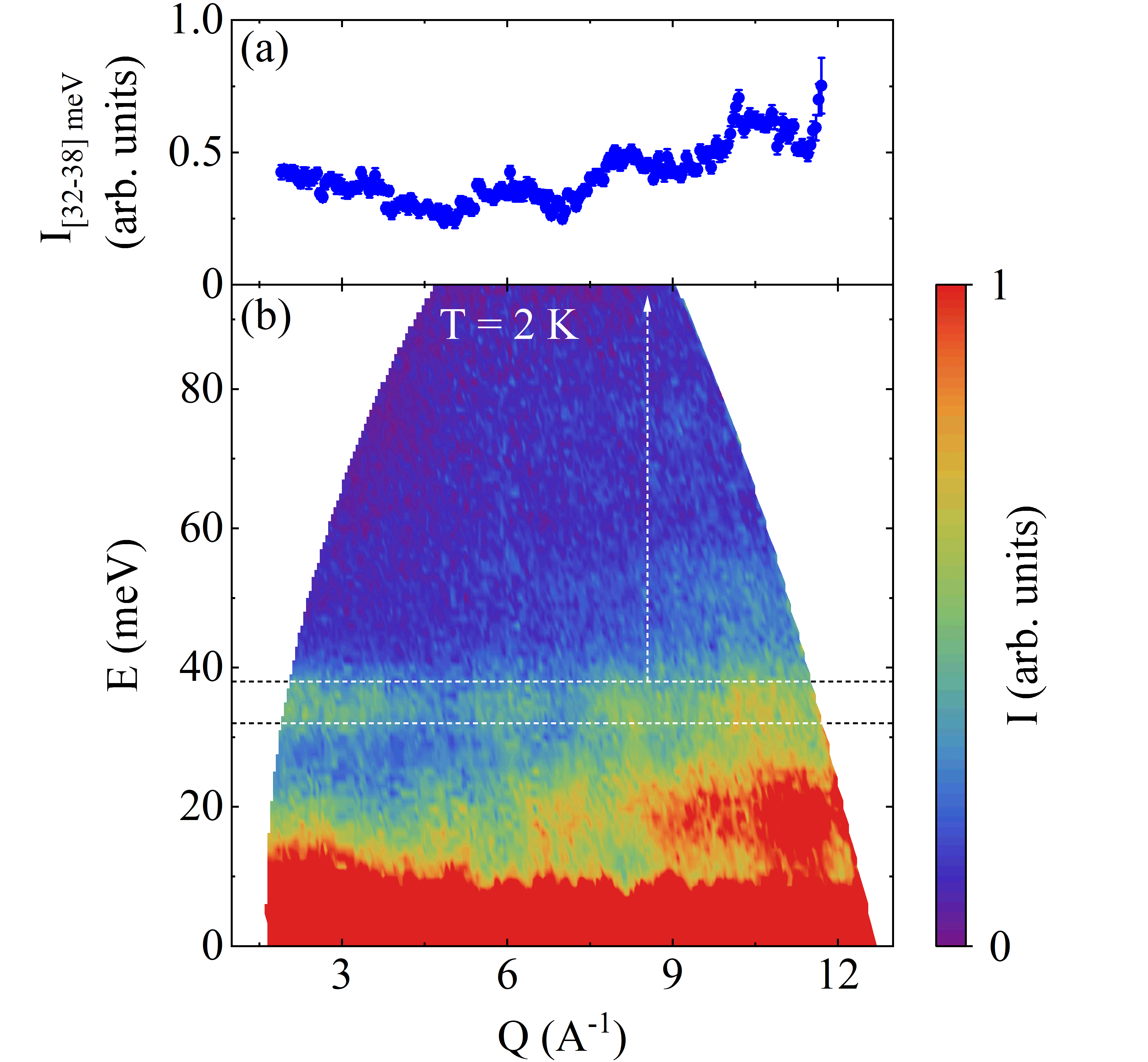}
\caption{(a) Scattering function intensity map $S(Q,E)$ measured on the IN4c spectrometer at $T = \SI{2}{\kelvin}$ with an incident wavelength $\lambda_{\mathrm{i}} = \SI{0.8}{\angstrom}$. (b) Energy-integrated intensity over the range $E = \SIrange{32}{38}{\milli\electronvolt}$.}
\label{fig:SM2}
\end{figure}

\setcounter{figure}{0}
\setcounter{table}{0}
\renewcommand{\thefigure}{C\arabic{figure}}
\renewcommand{\thetable}{C\arabic{table}}

\begin{figure}
\includegraphics[width=1\columnwidth]{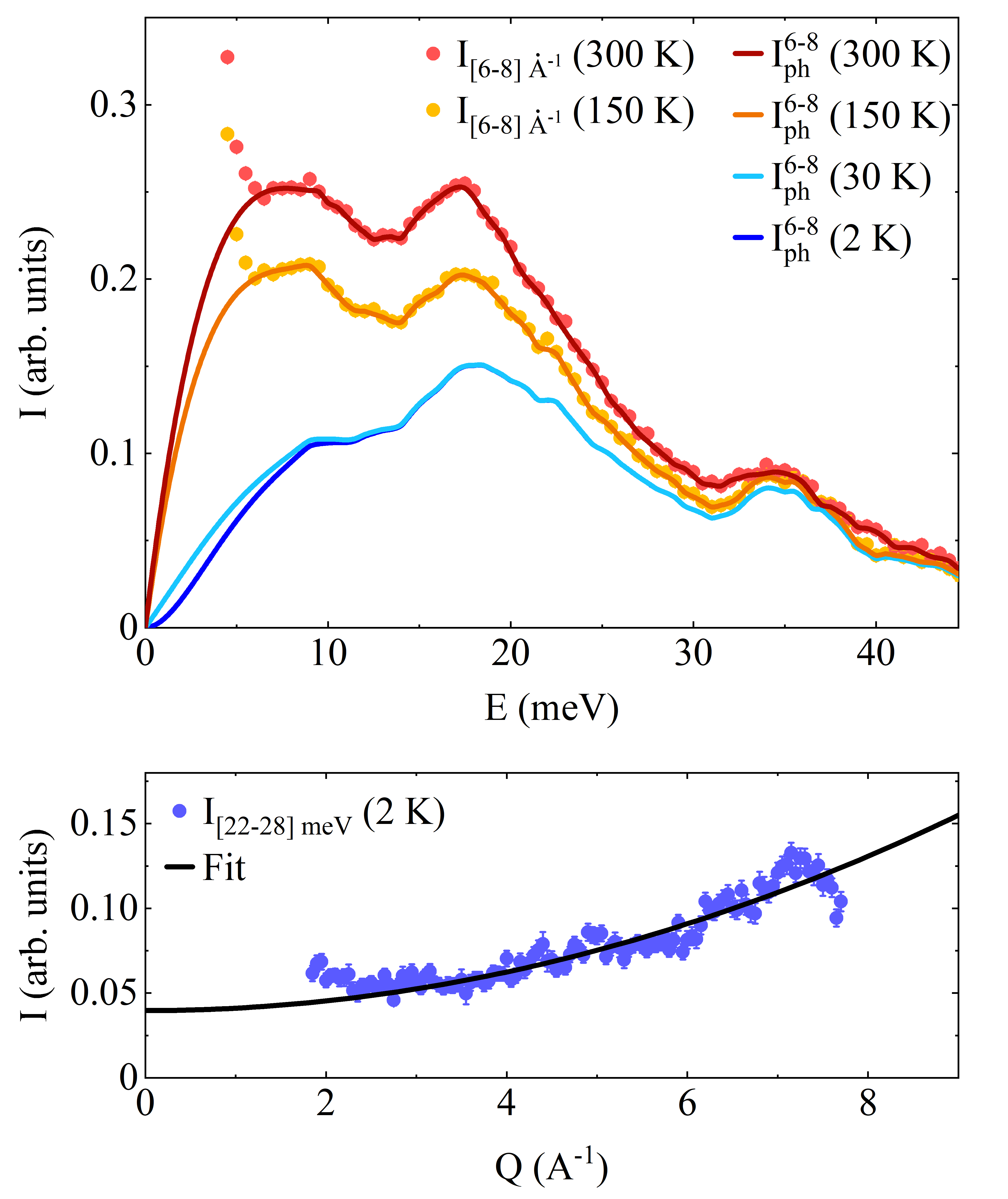}
\caption{(a) Estimated phononic signal between $6$ and $\SI{8}{\angstrom^{-1}}$ (lines) extracted from the momentum-averaged intensities $I_{\left[6-8\right]\si{\angstrom^{-1}}}$ measured on the IN4c spectrometer ($\lambda_{\mathrm{i}} = \SI{1.2}{\angstrom}$) at $T= 150$ and $\SI{300}{\kelvin}$ (dots). (b) Energy-integrated intensity over the range $E = \SIrange{22}{28}{\milli\electronvolt}$, along with its fit between $Q = \SI{3}{\angstrom^{-1}}$ and $Q = \SI{7.7}{\angstrom^{-1}}$ using the function $a + b Q^{2}$. The fit yields $a = 0.0397$ and $b = \SI{0.0014}{\angstrom^{2}}$.}
\label{fig:SM3}
\end{figure}

Thus, we could estimate the phonon density of states beyond the elastic peak $E \ge \SI{8}{\milli\electronvolt}$ by $I_{[6-8]\si{\angstrom^{-1}}}$, which covers a sufficiently large momentum range while maximizing the ratio $S^{\mathrm{d}}_{\mathrm{ph}} / S^{\mathrm{d}}_{\mathrm{cf}}$ thanks to the large $Q$ value. We found that such estimation of $Z_{\mathrm{ph}} / E$ by $I_{[6-8]\si{\angstrom^{-1}}}$ is nearly temperature-independent for $T \le \SI{150}{\kelvin}$, apart from residual magnetic contributions. Since these contributions diminish at higher temperatures, we defined our phononic signal from the high-temperatures spectra, i.e.  
\begin{equation}
I_{\mrm{ph}}^{6-8}(E,T) = \frac{1-\mrm{e}^{-E/k_{\mrm{B}}T'}}{1-\mrm{e}^{-E/k_{\mrm{B}}T}}I_{\left[6-8\right]\si{\angstrom^{-1}}}(E,T')
\label{Eq:phononSignal}
\end{equation}
with $T'=\SI{150}{\kelvin}$ for $T \le \SI{150}{\kelvin}$ and $T'=T$ for $T = \SI{300}{\kelvin}$. Below $E = \SI{8}{\milli\electronvolt}$, where the phonons are dominated by acoustic modes, we approximated the phonon density of states as $Z_{\mathrm{ph}}/E \propto E$. This yielded the phononic signals shown in Fig.~\sfig{fig:SM3}{a}.

We finally had to account for the reduction of phonon intensity when averaging over the lower momentum range $\SIrange{1}{3}{\angstrom^{-1}}$ compared to $\SIrange{6}{8}{\angstrom^{-1}}$, i.e.
\begin{equation}
I_{\mrm{ph}}(E,T) \equiv I_{\mrm{ph}}^{1-3}(E,T) = \frac{1 + c \langle Q^2 \rangle_{1}^{3}}{1 + c \langle Q^2 \rangle_{6}^{8}} I_{\mrm{ph}}^{6-8}(E,T)
\end{equation}
with $c = b/a$. Note that $\langle Q^2 \rangle_{1}^{3}$ and $\langle Q^2 \rangle_{6}^{8}$ are energy-dependent due to kinematic constraints limiting part of the integration windows. We determined $c$ by fitting the energy-integrated intensity in a region with the least expected magnetic contribution ($T = \SI{2}{\kelvin}$, $E = \SIrange{22}{28}{\milli\electronvolt}$, $Q\ge\SI{3}{\angstrom^{-1}}$) using the form $a + b Q^2$, see Fig.~\sfig{fig:SM3}{b}. This yielded $c \simeq \SI{0.04}{\angstrom^{2}}$, which we adopted for our fits at all temperatures.

\subsection{Crystal field contribution}

\begin{figure}[t!]
\includegraphics[width=1\columnwidth]{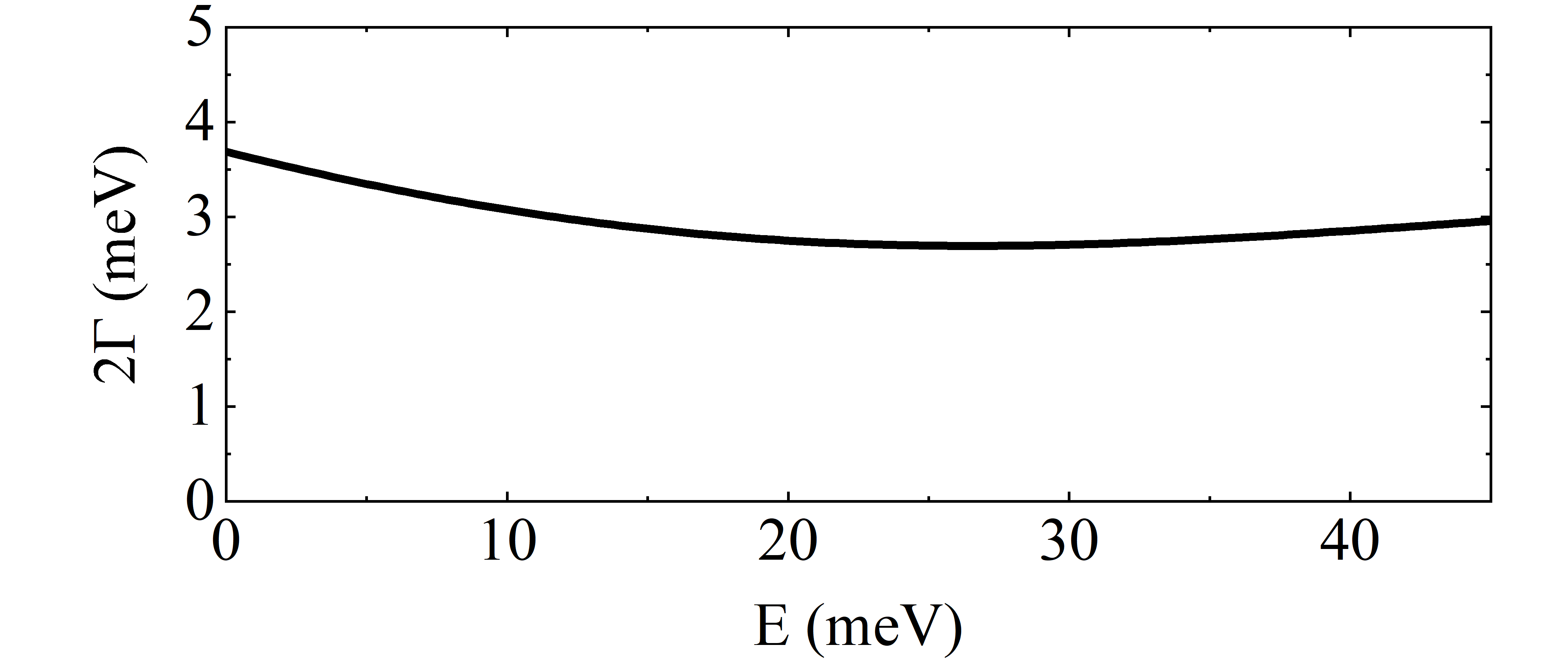}
\caption{Experimental resolution of the IN4c spectrometer for $\lambda_{\mrm{i}}=\SI{1.2}{\angstrom}$.}
\label{fig:SMX}
\end{figure}

The crystal field part $I_{\mrm{cf}}(E,T)$ used to fit the experimental data $I_{\left[1-3\right]\si{\angstrom^{-1}}}$ was defined as
\begin{equation}
\begin{gathered}
I_{\mrm{cf}}(E,T) = \sum_{k} \frac{a_{k}\left[V_{k}(E; \varepsilon_{k})-V_{k}(E; -\varepsilon_{k})\right]}{1-\mrm{e}^{-\beta E}}
\end{gathered}
\label{eq:CEF_contrib}
\end{equation}
where $k$ labels the different crystal field transition P$_{k}$. In Eq. \ref{eq:CEF_contrib}, $a_{k}$ is the intensity of the transition integrated over the range $Q = \SIrange{1}{3}{\angstrom^{-1}}$, $1/(1-\mrm{e}^{-\beta E})$ ensures the detailed balance principle, and $V_{k}(E; \varepsilon_{k}) \equiv V(E; \varepsilon_{k},\gamma,\Gamma(E))$ is a normalized Voigt profile centered at $\varepsilon_{k}$. It represents the convolution of a Lorentzian of full width at half maximum (FWHM) $2\gamma$ $-$ the intrinsic width of the crystal-field levels $-$ and a Gaussian of FWHM $2\Gamma(E)$ $-$ the energy-dependent experimental resolution of IN4c spectrometer (see Fig.~\fig{fig:SMX}).

\section{PYROCHLORES SYMMETRY ANALYSIS}
\label{app:SymAnalysis}
\setcounter{table}{0}
\renewcommand{\thetable}{D\arabic{table}}

The general interaction Hamiltonian between rare-earth ions $\wh{\mathcal{H}}_{\mrm{int}}^{\mrm{gen}}$ can be expressed as \cite{Santini2009,Rau2019}
\begin{equation}
\wh{\mathcal{H}}_{\mrm{int}}^{\mrm{gen.}} = \sum_{\dba*{i,j}}\sum_{k,q}\sum_{k',q'}\wh{\mathcal{T}}_{k}^{\,q}\dpa*{\wh{\bm{J}}_{i}}\mathcal{M}_{ij}^{k,q;k',q'}\wh{\mathcal{T}}_{k'}^{\,q'}\dpa*{\wh{\bm{J}}_{j}}
\label{eq:Hint_multipole}
\end{equation}
where $\wh{\mathcal{T}}_{k}^{\,q}$ denotes a multipolar operator of rank $k$ and order $q$, and $\mathcal{M}_{ij}^{k,q;k',q'}$ the interaction matrix. Considering only dipolar operators ($k = 1$), Eq. \ref{eq:Hint_multipole} becomes
\begin{equation}
\wh{\mathcal{H}}_{\mrm{int}} = \sum_{\dba*{i,j}}\wh{\bm{J}}_{i}\mathcal{J}_{ij}\wh{\bm{J}}_{j}
\end{equation}
with $\widetilde{\mathcal{J}}_{ij}$ a $3\times 3$ interacting matrix. This matrix is strongly constrained by the symmetry of the pyrochlore lattice, resulting in only four independent parameters $\mathcal{J}_{\mathrm{zz}}$, $\mathcal{J}_{\pm}$, $\mathcal{J}_{\pm\pm}$ and $\mathcal{J}_{\mathrm{z}\pm}$ \cite{Yan2017} such that
\begin{equation}
\begin{split}
\wh{\mathcal{H}}_{\mrm{int}} &= \sum_{ij\in t} \mathcal{J}_{\mathrm{zz}}\wh{J}_{i}^{z}\wh{J}_{j}^{z} - \mathcal{J}_{\pm}\left(\wh{J}_{i}^{+}\wh{J}_{j}^{-}+\wh{J}_{i}^{-}\wh{J}_{j}^{+}\right) \\
&+ \mathcal{J}_{\pm\pm}\left(\gamma_{ij}\wh{J}_{i}^{+}\wh{J}_{j}^{+}+\gamma_{ij}^{*}\wh{J}_{i}^{-}\wh{J}_{j}^{-}\right)\\
&+ \mathcal{J}_{\mathrm{z}\pm}\left(\wh{J}_{i}^{z}\left[\zeta_{ij}\wh{J}_{j}^{+}+\zeta_{ij}^{*}\wh{J}_{j}^{-}\right] + i\leftrightarrow j\right)
\label{eq:Hloc}
\end{split}
\end{equation}
where the matrices $\zeta = -\gamma^{*}$ account for the bond-dependent part of the Hamiltonian and are given by
\begin{equation}
\zeta = 
\begin{pmatrix}
0 & -1 & \textrm{e}^{\mathrm{i}\pi/3} & \textrm{e}^{-\mathrm{i}\pi/3} \\
-1 & 0 & \textrm{e}^{-\mathrm{i}\pi/3} & \textrm{e}^{\mathrm{i}\pi/3} \\
\textrm{e}^{\mathrm{i}\pi/3} & \textrm{e}^{-\mathrm{i}\pi/3} & 0 & -1 \\
\textrm{e}^{-\mathrm{i}\pi/3} & \textrm{e}^{\mathrm{i}\pi/3} & -1 & 0
\end{pmatrix} = -\gamma^{*}\textrm{.}
\label{eq:met_zeta_gamma}
\end{equation}
Note that Eqs.~\ref{eq:Hloc} and \ref{eq:met_zeta_gamma} have the same form as those derived for a pyrochlore pseudospin model based on a Kramers $\Gamma_4$ crystal-field doublet. Indeed, in that case, all components of the pseudospin transform as magnetic dipoles under the symmetry operations of the local point group $\mathcal{D}_{3d}$, similarly to $\wh{\bm{J}}$. However, while the $3\times3$ interaction matrix in the pseudospin model encodes all multipolar interactions, it includes only dipolar ($k=1$) terms in Eq.~\ref{eq:Hloc}. This is because $\wh{\mathcal{H}}_{\mrm{int}}$ acts here on the same level as the crystal field hamiltonian $\wh{\mathcal{H}}_{\mrm{cf}}$ (the ${}^7F_{6}$ Tb$^{3+}$ multiplet), allowing interactions to mix the ground-state with excited crystal-field levels.

\section{MEAN-FIELD APPROXIMATION}
\label{app:MFmethod}
\setcounter{figure}{0}
\setcounter{table}{0}
\renewcommand{\thefigure}{E\arabic{figure}}
\renewcommand{\thetable}{E\arabic{table}}

When Tb–Tb interactions are included through $\widehat{\mathcal{H}}_{\mathrm{int}}$, one should consider operator products involving different Tb sites, $\widehat{J}_i^\alpha \widehat{J}_j^\beta$ with $\alpha, \beta = z, \pm$ (see Eq.~\ref{eq:Hint}). In our study, we employed the mean-field approximation, in which two-site operator products are replaced by sums of single-site terms. Specifically, each operator is decomposed as $\widehat{J}_i^\alpha = \langle \widehat{J}_i^\alpha \rangle + \delta \widehat{J}_i^\alpha$, where $\delta \widehat{J}_i^\alpha$ represents the fluctuation around the quantum statistical average. Thus,
\begin{equation}
\wh{J}_{i}^{\alpha}\wh{J}_{j}^{\beta} = \dba{\wh{J}_{i}^{\alpha}}\dba{\wh{J}_{j}^{\beta}} + \dba{\wh{J}_{i}^{\alpha}}\delta\wh{J}_{j}^{\beta} + \delta\wh{J}_{i}^{\alpha}\dba{\wh{J}_{j}^{\beta}} + \delta\wh{J}_{i}^{\alpha}\delta\wh{J}_{j}^{\beta}.
\end{equation}
Neglecting the second-order fluctuation term $\delta \widehat{J}_i^\alpha \delta \widehat{J}_j^\beta$, we obtain a sum of effective single-site Hamiltonians
\begin{equation}
\begin{gathered}
\wh{\mathcal{H}}_{\mrm{int}}^{\mrm{MF}} = \sum_{i}\sum_{\alpha=z,\pm}h_{i}^{\alpha}\wh{J}_{i}^{\alpha},\\
h_{i}^{\alpha} = \sum_{j}\sum_{\beta=z,\pm}\mathcal{J}_{ij}^{\alpha\beta}\dba{J_{j}^{\beta}}
\end{gathered}
\label{eq:MF}
\end{equation}
where $h_{i}^{\alpha}$ is the mean field acting on site $i$ ion due to all other ions. The system defined by Eq.~\ref{eq:MF} is solved iteratively. Starting from an initial configuration $\{\dba{J_{i}^{\alpha}}\}_{i,\alpha}$, all single-site Hamiltonians are diagonalized, and the resulting expectation values are used as input for the next iteration. This procedure is repeated until convergence is reached, defined by
\begin{equation}
\begin{gathered}
\abs{F_{\mrm{tot}}^{(p)}(T)-F_{\mrm{tot}}^{(p-1)}(T)}<n_{\mrm{c}},\\
F_{\mrm{tot}}(T) = -\frac{1}{\beta}\ln\mathcal{Z} = -\frac{1}{\beta}\sum_{i=1}^{4}\ln \sum_{n=1}^{13}\mrm{e}^{-\beta E_{i,n}}
\end{gathered}
\end{equation}
with $\beta = 1/k_{\mrm{B}}T$. Here, $F_{\mathrm{tot}}^{(p)}(T)$ denotes the total free energy at iteration $p$ (computed using the partition function $\mathcal{Z}$ obtained from the $2J+1 = 13$ energy levels $E_{i,n}$ of the four ions $i$ of the tetrahedron), and $n_{\mathrm{c}}$ is the numerical convergence threshold (typically set to $\SI{e-5}{\milli\electronvolt}$). To minimize the risk of convergence to a local minimum, the diagonalization was performed multiple times with different initial values of $\{\dba{J_{i}^{\alpha}}\}_{i,\alpha}$, and the solution with the lowest total free energy was retained.

Note that this approach prevents access to any entangled states: mean-field eigenstates are tensor products of single-site states since the original interaction Hamiltonian is replaced by a sum of single-site Hamiltonians.

\section{MEAN-FIELD OBSERVABLE COMPUTATION} 
\label{app:ObsCompute}
\setcounter{figure}{0}
\setcounter{table}{0}
\renewcommand{\thefigure}{F\arabic{figure}}
\renewcommand{\thetable}{F\arabic{table}}

We describe here the method used to compute the specific heat, ordered magnetic moment component, and neutron scattering magnetic intensity shown in Fig.~\fig{fig:4}, based on the single-site eigenstates $\ket{\psi_{n}}$ and eigenvalues $E_{n}$ of each Tb$^{3+}$ ion in a tetrahedron calculated at the mean-field level. For clarity, the site index $i\in\left[1,4\right]$ and the corresponding sum over $i$ are omitted. Note also that $\ket{\psi_{n}}$ and $E_{n}$ explicitly depend on temperature $T$, due to both the temperature dependence of the Ir molecular field Hamiltonian $\wh{\mathcal{H}}_{\mrm{Ir}}$ and to the interaction Hamiltonian, which modifies the magnetic moment orientation at low temperatures.

\subsection{Macroscopic quantities}

The magnetic specific heat is given by
\begin{equation}
C_{\mrm{p, mag}} = \frac{\mrm{d}U}{\mrm{d}T}  = \frac{\mrm{d}}{\mrm{d}T} \sum_{n}p_{n}E_{n}, \quad p_{n} = \frac{e^{-\beta E_n}}{\mathcal{Z}}.
\end{equation}

The ordered magnetic moment component along the $\alpha = x, y, z$ direction is calculated from the expression
\begin{equation}
M^{\alpha} = \dba{\wh{M}^{\alpha}}_{T}  = g_{J}\mu_{\mrm{B}}\sum_{n}p_{n}\mel{\psi_{n}}{\wh{J}^{\alpha}}{\psi_{n}},
\end{equation}
where $\dba{\cdot}_{T}$ denote the quantum statistical expectation value. $\Gamma_{3}$ (AIAO) and $\Gamma_{5}$ magnetic moments are then calculated from the definition of the order parameters provided in Appendix~\ref{app:SymAnalysis}. If there is no other non-null order parameter, they correspond to $\alpha = z$ and $\alpha = x/y$, respectively.

\subsection{Neutron scattering intensity}
\label{app:ObsComputeNeutrons}

Up to an experimental scaling factor, the neutron-scattered intensity outside the elastic peak corresponds to the dynamical structure factor $S^{\mrm{d}}(Q,E)$. For a single magnetic species (Tb$^{3+}$ ions) in a polycrystalline sample, its magnetic contribution can be expressed as
\begin{equation}
\begin{gathered}
S^{\mrm{d}}_{\mrm{mag}}(Q,E) \propto f^{2}(Q)\mrm{e}^{-2W(Q)} \frac{2}{3}\sum_{\alpha} \frac{\mrm{Im}\left[\chi^{\alpha\alpha}(E)\right]}{1-\mrm{e}^{-\beta E}}, \\
\mrm{Im}\left[\chi^{\alpha\alpha}\right] = \sum_{n,m}^{En\ne E_{m}}(p_{n}-p_{m})\abs{\mel{\psi_{m}}{\wh{J}^{\alpha}}{\psi_{n}}}^{2} \kappa\dpa*{E,E_{mn}},
\end{gathered}
\label{eq:dynamical_scattering_fnc}
\end{equation}
where the $2/3$ factor arises from powder averaging the polarisation factor.

The energy dependence of $S^{\mrm{d}}(Q, E)$ is straightforward to calculate, apart from the energy-conservation factor $\kappa(E, E_{mn})$. In principle, this is a Dirac delta distribution centered at $E_{mn} = E_{m}-E_{n}$. However, in real materials, excitations have a finite lifetime $\tau$, leading to a Lorentzian broadening with half-width at half-maximum (HWHM) $\gamma = 1/\tau$, which must be further convolved with a Gaussian function of HWHM $\Gamma$ representing the instrumental resolution. The resulting lineshape is a Voigt profile, $V(E; E_{mn}, \gamma, \Gamma)$. In our calculations, we set $\gamma = 1.1$, $1.2$, $2.2$, and $\SI{3.0}{\milli\electronvolt}$ at $T = 2$, $30$, $150$, and $\SI{300}{\kelvin}$, respectively.

The momentum dependence of $S^{\mathrm{d}}(Q,E)$ is given by the product of the magnetic form factor squared $f^2(Q)$ and the Debye–Waller factor $\mathrm{e}^{-2W(Q)}$. The effect of the latter is negligible at small $Q$ and was therefore omitted in our calculations. For the magnetic form factor (the Fourier transform of the single-ion magnetization density), we assumed LS coupling and used the dipolar approximation (valid at small $Q$) leading to
\begin{equation}
f(Q) = \frac{g_{S}}{2}\dba*{\,j_{0}(Q)} + \frac{g_{L}}{2}\dpa*{\dba*{\,j_{0}(Q)}+\dba*{\,j_{2}(Q)}}
\end{equation}
where $g_{S} = 1$ and $g_{L} = 1/2$ are the Landé factors of the Tb$^{3+}$ ions, and $\dba*{\,j_{k}(Q)}$ are the radial integral of the spherical Bessel functions multiplied by the normalized radial wavefunctions of the 4$f$ electrons. These can be approximated by
\begin{equation}
\begin{gathered}
\dba*{\,j_{k}(Q)} \approx Q^{k} \\
\times\left(A_{k}\mathrm{e}^{-a_{k}Q^{2}} + B_{k}\mathrm{e}^{-b_{k}Q^{2}} + C_{k}\mathrm{e}^{-c_{k}Q^{2}} + D_{k}\right),
\end{gathered}
\end{equation}
where $a_{k}$, $A_{k}$, $b_{k}$, $B_{k}$, $c_{k}$, $C_{k}$ and $D_{k}$ ($k=0,2$) are numercial coefficients (see Table \ref{tab:form_factor}).

\begin{table}[t!]
\begin{ruledtabular}
\begin{tabular} {cccccccc}
$k$ & $A_{k}$ & $a_{k}$ & $B_{k}$ & $b_{k}$ & $C_{k}$ & $c_{k}$ & $D_{k}$ \\
\hline \noalign{\vskip 1mm}    
$0$ & $\num{0.0177}$ & $\num{25.510}$ & $\num{0.2921}$ & $\num{10.577}$ & $\num{0.7133}$ & $\num{3.512}$ & $\num{-0.0231}$ \\
$2$ & $\num{0.2892}$ & $\num{18.497}$ & $\num{1.1678}$ & $\num{6.797}$ & $\num{0.9437}$ & $\num{2.257}$ & $\num{0.0232}$ \\
\end{tabular}
\end{ruledtabular}
\caption{Numerical parameters used in the magnetic form factor calculation of the Tb$^{3+}$ ions \cite{Brown2006}.}
\label{tab:form_factor}
\end{table}

Finally, once $S^{\mrm{d}}_{\mrm{mag}}(Q,E)$ is computed, its average over an arbitrary $Q$-range $\left[Q_{1},Q_{2}\right]$ can be evaluated using
\begin{equation}
I_{\mrm{mag}}(E) = \int_{\max(Q_{1};Q_{\mrm{l}}^{E})}^{\min(Q_{2};Q_{\mrm{h}}^{E})}\frac{S_{\mrm{mag}}^{\mrm{d}}(Q,E)\mathrm{d}Q}{\max(Q_{1};Q_{\mrm{l}}^{E})-\min(Q_{2};Q_{\mrm{h}}^{E})},
\label{eq:ICalc}
\end{equation}
where $Q_{\mrm{l}}$ is the lowest and $Q_{\mrm{h}}$ the highest accessible momentum transfers for neutrons of outgoing energy $E$. They are set by kinematic constraints and instrument characteristics,
\begin{equation}
Q_{\mrm{l,h}}^{E}=\sqrt{\frac{2m_{\mrm{n}}E_{\mrm{i}}}{\hbar^{2}}\dpa*{2-2\sqrt{1-\frac{E}{E_{\mrm{i}}}}\cos 2\theta_{\mrm{l,h}} - \frac{E}{E_{\mrm{i}}}}},
\label{eq:kinematics}
\end{equation}
with $E_{\mrm{i}}$ the neutron’s incoming energy, and $\theta_{\mrm{l}}$, $\theta_{\mrm{h}}$ the scattering angles of the first and last detectors, respectively. For the spectra depicted in Fig.~\fig{fig:4}, $Q_{1} = \SI{1}{\angstrom^{-1}}$ and $Q_{2} = \SI{3}{\angstrom^{-1}}$. The detector angles are $\theta_{\mrm{l}}=\SI{12}{\degree}$ and $\theta_{\mrm{h}}=\SI{115}{\degree}$, the settings of the IN4c spectrometer.

\section{PARAMETER REFINEMENTS}
\label{app:Parameters}
\setcounter{figure}{0}
\setcounter{table}{0}
\renewcommand{\thefigure}{G\arabic{figure}}
\renewcommand{\thetable}{G\arabic{table}}

We describe in this appendix our procedure to refine the crystal-field parameters $\{B_{k}^{q}\}$ and the interaction parameters $\{H_{\mrm{m}}^{\mrm{Ir}},\mathcal{J}_{zz},\mathcal{J}_{\pm}\}$ of our Hamiltonian (see Eq. \ref{eq:FullH}). We followed a two-step strategy. In the first step, the interaction parameters were set to zero, and we only refined the crystal-field parameters by a reverse Monte Carlo method. In the second step, the crystal-field parameters were fixed while we systematically searched for the optimal interaction parameters.

\subsection{Reverse Monte-Carlo method for crystal field parameters}

To identify sets of crystal-field parameters $\{B_{k}^{q}\}$ that best reproduce $I_{\left[1-3\right]\,\si{\angstrom^{-1}}}(T,E)$, we defined a functional $\phi$ that quantifies the discrepancy between the experimental spectra and those calculated using a given set of $\{B_{k}^{q}\}$. This functional was minimized using a Metropolis algorithm with simulated annealing. At each iteration $n$, a new set of parameters was randomly generated and used to compute $\phi^{(n)}$. The new set was accepted if $\Delta\phi = \phi^{(n)} - \phi^{(n-1)} \le 0$, or with probability $\mathrm{e}^{-\beta\Delta\phi}$ if $\Delta\phi > 0$. The inverse fictitious temperature $\beta$ was gradually increased with the iteration number ($\beta \propto n$) to reduce the risk of getting trapped in local minima. This process was typically repeated up to $n=10000$ to ensure convergence to a satisfactory solution.

Specifically, the functional was defined as
\begin{equation}
\begin{aligned}
\phi\dca*{\{B_{k}^{q}\}} = \sum_{T,E}\frac{1}{\sigma(T)}\dca*{I_{\mrm{mag}}^{\mrm{calc}}\dpa*{T,E}-I_{\mrm{mag}}^{\mrm{exp}}\dpa*{T,E}}^{2},
\end{aligned}
\label{eq:fonctionnelle}
\end{equation}
where $I_{\mathrm{mag}}^{\mathrm{exp}} = I_{\left[1-3\right]\,\si{\angstrom^{-1}}} - I_{\mrm{ph}}$ is the magnetic contribution to the experimental signal, and $I_{\mathrm{mag}}^{\mathrm{calc}}(T,E)$ 
denotes its calculated counterpart (see Appendix~\ref{app:ObsCompute}). Finally, $\sigma(T)$ are weighting factors used to adjust the relative importance of datasets at different temperatures. We typically set $\sigma(\SI{2}{\kelvin},\SI{30}{\kelvin})=1$ and $\sigma(\SI{150}{\kelvin},\SI{300}{\kelvin})=4$.

\subsection{Selection of interaction parameters}

\begin{figure}[!t]
\includegraphics[width=1\columnwidth]{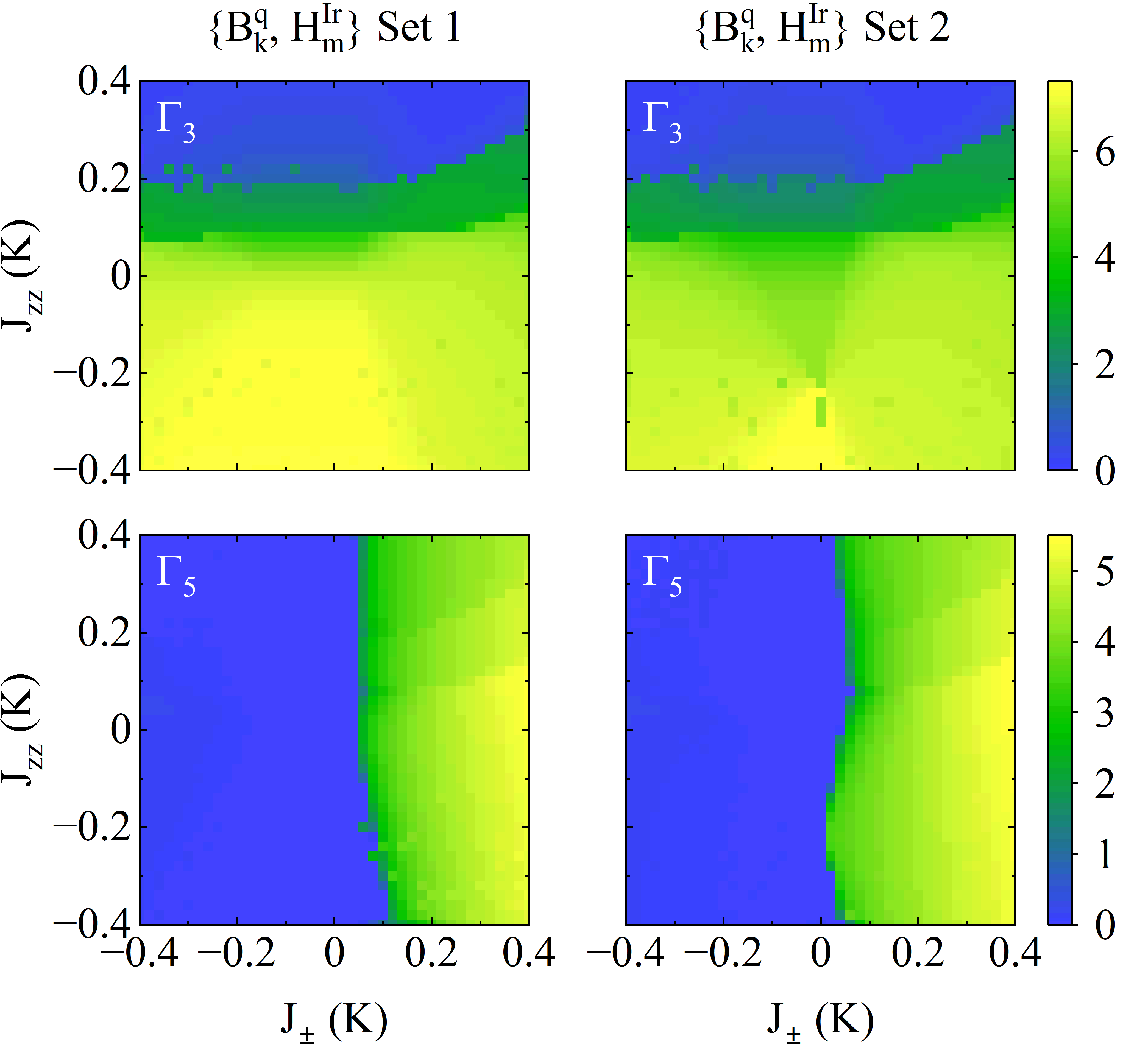}
\caption{$\Gamma_{3}$ (top) and $\Gamma_{5}$ (bottom) order parameters in $\upmu_{\mrm{B}}$ as a function of the interaction parameters $\mathcal{J}_{zz}$ and $\mathcal{J}_{\pm}$, for the crystal-field and iridium molecular field values corresponding to set 1 (left) and set 2 (right), listed in Table~\ref{tab:params}.}
\label{fig:SM4}
\end{figure}

For each set of crystal-field parameters $\{B_{k}^{q}\}$ identified, we first selected all possible triplets of interaction parameters $\{{H_{\mathrm{m}}^{\mathrm{Ir}}, \mathcal{J}_{zz}, \mathcal{J}_{\pm}}\}$ that reproduce the experimental low-temperature values of the magnetic moments. Examples of phase diagrams showing  $\Gamma_{3}$ and $\Gamma_{5}$ order parameters as a function of $\mathcal{J}_{zz}$ and $\mathcal{J}_{\pm}$ are shown in Fig.~\fig{fig:SM4}. Interestingly, almost any molecular field value $H_{\mathrm{m}}^{\mathrm{Ir}}$ in the range $\mu_{0}H = \SIrange{1}{2}{\tesla}$ could provide a reasonable match, provided the two other interaction parameters were chosen accordingly.

\mbox{}

We then examined the calculated temperature dependence of the $\Gamma_{3}$ and $\Gamma_{5}$ magnetic moments and compared it with the experimental data to select a unique set of interaction parameters for each crystal-field parameter set. It is worth noting that the temperature dependence of the $\Gamma_{5}$ magnetic moment was not discriminating, as none of the solutions yielded a behavior consistent with the experimental one. Ultimately, we kept only the nine-parameter sets that best reproduced the overall experimental temperature dependence of the specific heat.

The retained nine-parameter sets cluster into two distinct groups: in one, the ground and first excited doublets are dominated by $\ket{\pm5}$ and $\ket{\pm4}$, respectively, while in the other this order is reversed. Set 1 and Set 2 in the main text are representative parameter sets from each group.

\end{appendix}

\clearpage

%%%%%%%%%%%%%%%%%%%%%%%%%%%%%%%%%%%%%%%%%%%%%%%%%%%%%%%%%%%%%%%%%%%%%%%%%%%%%%%%%%%%%%%%%%%%%%%%%%%%%%%%%%%%%%



\begin{thebibliography}{65}%
\makeatletter
\providecommand \@ifxundefined [1]{%
 \@ifx{#1\undefined}
}%
\providecommand \@ifnum [1]{%
 \ifnum #1\expandafter \@firstoftwo
 \else \expandafter \@secondoftwo
 \fi
}%
\providecommand \@ifx [1]{%
 \ifx #1\expandafter \@firstoftwo
 \else \expandafter \@secondoftwo
 \fi
}%
\providecommand \natexlab [1]{#1}%
\providecommand \enquote  [1]{``#1''}%
\providecommand \bibnamefont  [1]{#1}%
\providecommand \bibfnamefont [1]{#1}%
\providecommand \citenamefont [1]{#1}%
\providecommand \href@noop [0]{\@secondoftwo}%
\providecommand \href [0]{\begingroup \@sanitize@url \@href}%
\providecommand \@href[1]{\@@startlink{#1}\@@href}%
\providecommand \@@href[1]{\endgroup#1\@@endlink}%
\providecommand \@sanitize@url [0]{\catcode `\\12\catcode `\$12\catcode `\&12\catcode `\#12\catcode `\^12\catcode `\_12\catcode `\%12\relax}%
\providecommand \@@startlink[1]{}%
\providecommand \@@endlink[0]{}%
\providecommand \url  [0]{\begingroup\@sanitize@url \@url }%
\providecommand \@url [1]{\endgroup\@href {#1}{\urlprefix }}%
\providecommand \urlprefix  [0]{URL }%
\providecommand \Eprint [0]{\href }%
\providecommand \doibase [0]{https://doi.org/}%
\providecommand \selectlanguage [0]{\@gobble}%
\providecommand \bibinfo  [0]{\@secondoftwo}%
\providecommand \bibfield  [0]{\@secondoftwo}%
\providecommand \translation [1]{[#1]}%
\providecommand \BibitemOpen [0]{}%
\providecommand \bibitemStop [0]{}%
\providecommand \bibitemNoStop [0]{.\EOS\space}%
\providecommand \EOS [0]{\spacefactor3000\relax}%
\providecommand \BibitemShut  [1]{\csname bibitem#1\endcsname}%
\let\auto@bib@innerbib\@empty
%</preamble>
\bibitem [{\citenamefont {Pesin}\ and\ \citenamefont {Balents}(2010)}]{Pesin2010}%
  \BibitemOpen
  \bibfield  {author} {\bibinfo {author} {\bibfnamefont {D.}~\bibnamefont {Pesin}}\ and\ \bibinfo {author} {\bibfnamefont {L.}~\bibnamefont {Balents}},\ }\bibfield  {title} {\bibinfo {title} {Mott physics and band topology in materials with strong spin–orbit interaction},\ }\href {https://doi.org/10.1038/nphys1606} {\bibfield  {journal} {\bibinfo  {journal} {Nature Physics}\ }\textbf {\bibinfo {volume} {6}},\ \bibinfo {pages} {376} (\bibinfo {year} {2010})}\BibitemShut {NoStop}%
\bibitem [{\citenamefont {Witczak-Krempa}\ \emph {et~al.}(2014)\citenamefont {Witczak-Krempa}, \citenamefont {Chen}, \citenamefont {Kim},\ and\ \citenamefont {Balents}}]{WitczakKrempa2014}%
  \BibitemOpen
  \bibfield  {author} {\bibinfo {author} {\bibfnamefont {W.}~\bibnamefont {Witczak-Krempa}}, \bibinfo {author} {\bibfnamefont {G.}~\bibnamefont {Chen}}, \bibinfo {author} {\bibfnamefont {Y.~B.}\ \bibnamefont {Kim}},\ and\ \bibinfo {author} {\bibfnamefont {L.}~\bibnamefont {Balents}},\ }\bibfield  {title} {\bibinfo {title} {Correlated quantum phenomena in the strong spin-orbit regime},\ }\href {https://doi.org/10.1146/annurev-conmatphys-020911-125138} {\bibfield  {journal} {\bibinfo  {journal} {Annual Review of Condensed Matter Physics}\ }\textbf {\bibinfo {volume} {5}},\ \bibinfo {pages} {57} (\bibinfo {year} {2014})}\BibitemShut {NoStop}%
\bibitem [{\citenamefont {Kim}\ \emph {et~al.}(2008)\citenamefont {Kim}, \citenamefont {Jin}, \citenamefont {Moon}, \citenamefont {Kim}, \citenamefont {Park}, \citenamefont {Leem}, \citenamefont {Yu}, \citenamefont {Noh}, \citenamefont {Kim}, \citenamefont {Oh}, \citenamefont {Park}, \citenamefont {Durairaj}, \citenamefont {Cao},\ and\ \citenamefont {Rotenberg}}]{Kim2008}%
  \BibitemOpen
  \bibfield  {author} {\bibinfo {author} {\bibfnamefont {B.~J.}\ \bibnamefont {Kim}}, \bibinfo {author} {\bibfnamefont {H.}~\bibnamefont {Jin}}, \bibinfo {author} {\bibfnamefont {S.~J.}\ \bibnamefont {Moon}}, \bibinfo {author} {\bibfnamefont {J.-Y.}\ \bibnamefont {Kim}}, \bibinfo {author} {\bibfnamefont {B.-G.}\ \bibnamefont {Park}}, \bibinfo {author} {\bibfnamefont {C.~S.}\ \bibnamefont {Leem}}, \bibinfo {author} {\bibfnamefont {J.}~\bibnamefont {Yu}}, \bibinfo {author} {\bibfnamefont {T.~W.}\ \bibnamefont {Noh}}, \bibinfo {author} {\bibfnamefont {C.}~\bibnamefont {Kim}}, \bibinfo {author} {\bibfnamefont {S.-J.}\ \bibnamefont {Oh}}, \bibinfo {author} {\bibfnamefont {J.-H.}\ \bibnamefont {Park}}, \bibinfo {author} {\bibfnamefont {V.}~\bibnamefont {Durairaj}}, \bibinfo {author} {\bibfnamefont {G.}~\bibnamefont {Cao}},\ and\ \bibinfo {author} {\bibfnamefont {E.}~\bibnamefont {Rotenberg}},\ }\bibfield  {title} {\bibinfo {title} {Novel ${J_{\mathrm{eff}}} =1/2$ mott state induced by relativistic spin-orbit coupling in {Sr$_{2}$IrO$_{4}$}},\ }\href {https://doi.org/10.1103/physrevlett.101.076402} {\bibfield  {journal} {\bibinfo  {journal} {Physical Review Letters}\ }\textbf {\bibinfo {volume} {101}},\ \bibinfo {pages} {076402} (\bibinfo {year} {2008})}\BibitemShut {NoStop}%
\bibitem [{\citenamefont {Kim}\ \emph {et~al.}(2009)\citenamefont {Kim}, \citenamefont {Ohsumi}, \citenamefont {Komesu}, \citenamefont {Sakai}, \citenamefont {Morita}, \citenamefont {Takagi},\ and\ \citenamefont {Arima}}]{Kim2009}%
  \BibitemOpen
  \bibfield  {author} {\bibinfo {author} {\bibfnamefont {B.~J.}\ \bibnamefont {Kim}}, \bibinfo {author} {\bibfnamefont {H.}~\bibnamefont {Ohsumi}}, \bibinfo {author} {\bibfnamefont {T.}~\bibnamefont {Komesu}}, \bibinfo {author} {\bibfnamefont {S.}~\bibnamefont {Sakai}}, \bibinfo {author} {\bibfnamefont {T.}~\bibnamefont {Morita}}, \bibinfo {author} {\bibfnamefont {H.}~\bibnamefont {Takagi}},\ and\ \bibinfo {author} {\bibfnamefont {T.}~\bibnamefont {Arima}},\ }\bibfield  {title} {\bibinfo {title} {Phase-sensitive observation of a spin-orbital mott state in {Sr$_{2}$IrO$_{4}$}},\ }\href {https://doi.org/10.1126/science.1167106} {\bibfield  {journal} {\bibinfo  {journal} {Science}\ }\textbf {\bibinfo {volume} {323}},\ \bibinfo {pages} {1329} (\bibinfo {year} {2009})}\BibitemShut {NoStop}%
\bibitem [{\citenamefont {Rau}\ \emph {et~al.}(2016)\citenamefont {Rau}, \citenamefont {Lee},\ and\ \citenamefont {Kee}}]{Rau2016}%
  \BibitemOpen
  \bibfield  {author} {\bibinfo {author} {\bibfnamefont {J.~G.}\ \bibnamefont {Rau}}, \bibinfo {author} {\bibfnamefont {E.~K.-H.}\ \bibnamefont {Lee}},\ and\ \bibinfo {author} {\bibfnamefont {H.-Y.}\ \bibnamefont {Kee}},\ }\bibfield  {title} {\bibinfo {title} {Spin-orbit physics giving rise to novel phases in correlated systems: Iridates and related materials},\ }\href {https://doi.org/10.1146/annurev-conmatphys-031115-011319} {\bibfield  {journal} {\bibinfo  {journal} {Annual Review of Condensed Matter Physics}\ }\textbf {\bibinfo {volume} {7}},\ \bibinfo {pages} {195} (\bibinfo {year} {2016})}\BibitemShut {NoStop}%
\bibitem [{\citenamefont {Schaffer}\ \emph {et~al.}(2016)\citenamefont {Schaffer}, \citenamefont {Kin-Ho~Lee}, \citenamefont {Yang},\ and\ \citenamefont {Kim}}]{Schaffer2016}%
  \BibitemOpen
  \bibfield  {author} {\bibinfo {author} {\bibfnamefont {R.}~\bibnamefont {Schaffer}}, \bibinfo {author} {\bibfnamefont {E.}~\bibnamefont {Kin-Ho~Lee}}, \bibinfo {author} {\bibfnamefont {B.-J.}\ \bibnamefont {Yang}},\ and\ \bibinfo {author} {\bibfnamefont {Y.~B.}\ \bibnamefont {Kim}},\ }\bibfield  {title} {\bibinfo {title} {Recent progress on correlated electron systems with strong spin–orbit coupling},\ }\href {https://doi.org/10.1088/0034-4885/79/9/094504} {\bibfield  {journal} {\bibinfo  {journal} {Reports on Progress in Physics}\ }\textbf {\bibinfo {volume} {79}},\ \bibinfo {pages} {094504} (\bibinfo {year} {2016})}\BibitemShut {NoStop}%
\bibitem [{\citenamefont {Subramanian}\ \emph {et~al.}(1983)\citenamefont {Subramanian}, \citenamefont {Aravamudan},\ and\ \citenamefont {Subba~Rao}}]{Subramanian1983}%
  \BibitemOpen
  \bibfield  {author} {\bibinfo {author} {\bibfnamefont {M.}~\bibnamefont {Subramanian}}, \bibinfo {author} {\bibfnamefont {G.}~\bibnamefont {Aravamudan}},\ and\ \bibinfo {author} {\bibfnamefont {G.}~\bibnamefont {Subba~Rao}},\ }\bibfield  {title} {\bibinfo {title} {Oxide pyrochlores — a review},\ }\href {https://doi.org/10.1016/0079-6786(83)90001-8} {\bibfield  {journal} {\bibinfo  {journal} {Progress in Solid State Chemistry}\ }\textbf {\bibinfo {volume} {15}},\ \bibinfo {pages} {55} (\bibinfo {year} {1983})}\BibitemShut {NoStop}%
\bibitem [{\citenamefont {Yanagishima}\ and\ \citenamefont {Maeno}(2001)}]{Yanagishima2001}%
  \BibitemOpen
  \bibfield  {author} {\bibinfo {author} {\bibfnamefont {D.}~\bibnamefont {Yanagishima}}\ and\ \bibinfo {author} {\bibfnamefont {Y.}~\bibnamefont {Maeno}},\ }\bibfield  {title} {\bibinfo {title} {Metal-nonmetal changeover in pyrochlore iridates},\ }\href {https://doi.org/10.1143/jpsj.70.2880} {\bibfield  {journal} {\bibinfo  {journal} {Journal of the Physical Society of Japan}\ }\textbf {\bibinfo {volume} {70}},\ \bibinfo {pages} {2880–2883} (\bibinfo {year} {2001})}\BibitemShut {NoStop}%
\bibitem [{\citenamefont {Matsuhira}\ \emph {et~al.}(2011)\citenamefont {Matsuhira}, \citenamefont {Wakeshima}, \citenamefont {Hinatsu},\ and\ \citenamefont {Takagi}}]{Matsuhira2011}%
  \BibitemOpen
  \bibfield  {author} {\bibinfo {author} {\bibfnamefont {K.}~\bibnamefont {Matsuhira}}, \bibinfo {author} {\bibfnamefont {M.}~\bibnamefont {Wakeshima}}, \bibinfo {author} {\bibfnamefont {Y.}~\bibnamefont {Hinatsu}},\ and\ \bibinfo {author} {\bibfnamefont {S.}~\bibnamefont {Takagi}},\ }\bibfield  {title} {\bibinfo {title} {Metal–insulator transitions in pyrochlore oxides {$Ln_{2}$Ir$_{2}$O$_{7}$}},\ }\href {https://doi.org/10.1143/jpsj.80.094701} {\bibfield  {journal} {\bibinfo  {journal} {Journal of the Physical Society of Japan}\ }\textbf {\bibinfo {volume} {80}},\ \bibinfo {pages} {094701} (\bibinfo {year} {2011})}\BibitemShut {NoStop}%
\bibitem [{\citenamefont {Disseler}\ \emph {et~al.}(2012)\citenamefont {Disseler}, \citenamefont {Dhital}, \citenamefont {Amato}, \citenamefont {Giblin}, \citenamefont {de~la Cruz}, \citenamefont {Wilson},\ and\ \citenamefont {Graf}}]{Disseler2012}%
  \BibitemOpen
  \bibfield  {author} {\bibinfo {author} {\bibfnamefont {S.~M.}\ \bibnamefont {Disseler}}, \bibinfo {author} {\bibfnamefont {C.}~\bibnamefont {Dhital}}, \bibinfo {author} {\bibfnamefont {A.}~\bibnamefont {Amato}}, \bibinfo {author} {\bibfnamefont {S.~R.}\ \bibnamefont {Giblin}}, \bibinfo {author} {\bibfnamefont {C.}~\bibnamefont {de~la Cruz}}, \bibinfo {author} {\bibfnamefont {S.~D.}\ \bibnamefont {Wilson}},\ and\ \bibinfo {author} {\bibfnamefont {M.~J.}\ \bibnamefont {Graf}},\ }\bibfield  {title} {\bibinfo {title} {Magnetic order in the pyrochlore iridates {$A_{2}$Ir$_{2}$O$_{7}$ ($A=$ Y, Yb)}},\ }\href {https://doi.org/10.1103/physrevb.86.014428} {\bibfield  {journal} {\bibinfo  {journal} {Physical Review B}\ }\textbf {\bibinfo {volume} {86}},\ \bibinfo {pages} {014428} (\bibinfo {year} {2012})}\BibitemShut {NoStop}%
\bibitem [{\citenamefont {Sagayama}\ \emph {et~al.}(2013)\citenamefont {Sagayama}, \citenamefont {Uematsu}, \citenamefont {Arima}, \citenamefont {Sugimoto}, \citenamefont {Ishikawa}, \citenamefont {O’Farrell},\ and\ \citenamefont {Nakatsuji}}]{Sagayama2013}%
  \BibitemOpen
  \bibfield  {author} {\bibinfo {author} {\bibfnamefont {H.}~\bibnamefont {Sagayama}}, \bibinfo {author} {\bibfnamefont {D.}~\bibnamefont {Uematsu}}, \bibinfo {author} {\bibfnamefont {T.}~\bibnamefont {Arima}}, \bibinfo {author} {\bibfnamefont {K.}~\bibnamefont {Sugimoto}}, \bibinfo {author} {\bibfnamefont {J.~J.}\ \bibnamefont {Ishikawa}}, \bibinfo {author} {\bibfnamefont {E.}~\bibnamefont {O’Farrell}},\ and\ \bibinfo {author} {\bibfnamefont {S.}~\bibnamefont {Nakatsuji}},\ }\bibfield  {title} {\bibinfo {title} {Determination of long-range all-in-all-out ordering of {Ir$^{4+}$} moment in a pyrochlore iridate {Eu$_{2}$Ir$_{2}$O$_{7}$} by resonant x-ray diffraction},\ }\href {https://doi.org/10.1103/physrevb.87.100403} {\bibfield  {journal} {\bibinfo  {journal} {Physical Review B}\ }\textbf {\bibinfo {volume} {87}},\ \bibinfo {pages} {100403(R)} (\bibinfo {year} {2013})}\BibitemShut {NoStop}%
\bibitem [{\citenamefont {Disseler}(2014)}]{Disseler2014}%
  \BibitemOpen
  \bibfield  {author} {\bibinfo {author} {\bibfnamefont {S.~M.}\ \bibnamefont {Disseler}},\ }\bibfield  {title} {\bibinfo {title} {Direct evidence for the all-in/all-out magnetic structure in the pyrochlore iridates from muon spin relaxation},\ }\href {https://doi.org/10.1103/physrevb.89.140413} {\bibfield  {journal} {\bibinfo  {journal} {Physical Review B}\ }\textbf {\bibinfo {volume} {89}},\ \bibinfo {pages} {140413(R)} (\bibinfo {year} {2014})}\BibitemShut {NoStop}%
\bibitem [{\citenamefont {Lefran\c{c}ois}\ \emph {et~al.}(2015)\citenamefont {Lefran\c{c}ois}, \citenamefont {Simonet}, \citenamefont {Ballou}, \citenamefont {Lhotel}, \citenamefont {Hadj-Azzem}, \citenamefont {Kodjikian}, \citenamefont {Lejay}, \citenamefont {Manuel}, \citenamefont {Khalyavin},\ and\ \citenamefont {Chapon}}]{Lefrancois2015}%
  \BibitemOpen
  \bibfield  {author} {\bibinfo {author} {\bibfnamefont {E.}~\bibnamefont {Lefran\c{c}ois}}, \bibinfo {author} {\bibfnamefont {V.}~\bibnamefont {Simonet}}, \bibinfo {author} {\bibfnamefont {R.}~\bibnamefont {Ballou}}, \bibinfo {author} {\bibfnamefont {E.}~\bibnamefont {Lhotel}}, \bibinfo {author} {\bibfnamefont {A.}~\bibnamefont {Hadj-Azzem}}, \bibinfo {author} {\bibfnamefont {S.}~\bibnamefont {Kodjikian}}, \bibinfo {author} {\bibfnamefont {P.}~\bibnamefont {Lejay}}, \bibinfo {author} {\bibfnamefont {P.}~\bibnamefont {Manuel}}, \bibinfo {author} {\bibfnamefont {D.}~\bibnamefont {Khalyavin}},\ and\ \bibinfo {author} {\bibfnamefont {L.~C.}\ \bibnamefont {Chapon}},\ }\bibfield  {title} {\bibinfo {title} {Anisotropy-tuned magnetic order in pyrochlore iridates},\ }\href {https://doi.org/10.1103/physrevlett.114.247202} {\bibfield  {journal} {\bibinfo  {journal} {Physical Review Letters}\ }\textbf {\bibinfo {volume} {114}},\ \bibinfo {pages} {247202} (\bibinfo {year} {2015})}\BibitemShut {NoStop}%
\bibitem [{\citenamefont {Guo}\ \emph {et~al.}(2016)\citenamefont {Guo}, \citenamefont {Ritter},\ and\ \citenamefont {Komarek}}]{Guo2016}%
  \BibitemOpen
  \bibfield  {author} {\bibinfo {author} {\bibfnamefont {H.}~\bibnamefont {Guo}}, \bibinfo {author} {\bibfnamefont {C.}~\bibnamefont {Ritter}},\ and\ \bibinfo {author} {\bibfnamefont {A.~C.}\ \bibnamefont {Komarek}},\ }\bibfield  {title} {\bibinfo {title} {Direct determination of the spin structure of {Nd$_{2}$Ir$_{2}$O$_{7}$} by means of neutron diffraction},\ }\href {https://doi.org/10.1103/physrevb.94.161102} {\bibfield  {journal} {\bibinfo  {journal} {Physical Review B}\ }\textbf {\bibinfo {volume} {94}},\ \bibinfo {pages} {161102(R)} (\bibinfo {year} {2016})}\BibitemShut {NoStop}%
\bibitem [{\citenamefont {Yang}\ and\ \citenamefont {Kim}(2010)}]{Yang2010}%
  \BibitemOpen
  \bibfield  {author} {\bibinfo {author} {\bibfnamefont {B.-J.}\ \bibnamefont {Yang}}\ and\ \bibinfo {author} {\bibfnamefont {Y.~B.}\ \bibnamefont {Kim}},\ }\bibfield  {title} {\bibinfo {title} {Topological insulators and metal-insulator transition in the pyrochlore iridates},\ }\href {https://doi.org/10.1103/physrevb.82.085111} {\bibfield  {journal} {\bibinfo  {journal} {Physical Review B}\ }\textbf {\bibinfo {volume} {82}},\ \bibinfo {pages} {085111} (\bibinfo {year} {2010})}\BibitemShut {NoStop}%
\bibitem [{\citenamefont {Wan}\ \emph {et~al.}(2011)\citenamefont {Wan}, \citenamefont {Turner}, \citenamefont {Vishwanath},\ and\ \citenamefont {Savrasov}}]{Wan2011}%
  \BibitemOpen
  \bibfield  {author} {\bibinfo {author} {\bibfnamefont {X.}~\bibnamefont {Wan}}, \bibinfo {author} {\bibfnamefont {A.~M.}\ \bibnamefont {Turner}}, \bibinfo {author} {\bibfnamefont {A.}~\bibnamefont {Vishwanath}},\ and\ \bibinfo {author} {\bibfnamefont {S.~Y.}\ \bibnamefont {Savrasov}},\ }\bibfield  {title} {\bibinfo {title} {Topological semimetal and fermi-arc surface states in the electronic structure of pyrochlore iridates},\ }\href {https://doi.org/10.1103/physrevb.83.205101} {\bibfield  {journal} {\bibinfo  {journal} {Physical Review B}\ }\textbf {\bibinfo {volume} {83}},\ \bibinfo {pages} {205101} (\bibinfo {year} {2011})}\BibitemShut {NoStop}%
\bibitem [{\citenamefont {Go}\ \emph {et~al.}(2012)\citenamefont {Go}, \citenamefont {Witczak-Krempa}, \citenamefont {Jeon}, \citenamefont {Park},\ and\ \citenamefont {Kim}}]{Go2012}%
  \BibitemOpen
  \bibfield  {author} {\bibinfo {author} {\bibfnamefont {A.}~\bibnamefont {Go}}, \bibinfo {author} {\bibfnamefont {W.}~\bibnamefont {Witczak-Krempa}}, \bibinfo {author} {\bibfnamefont {G.~S.}\ \bibnamefont {Jeon}}, \bibinfo {author} {\bibfnamefont {K.}~\bibnamefont {Park}},\ and\ \bibinfo {author} {\bibfnamefont {Y.~B.}\ \bibnamefont {Kim}},\ }\bibfield  {title} {\bibinfo {title} {Correlation effects on {3D} topological phases: From bulk to boundary},\ }\href {https://doi.org/10.1103/physrevlett.109.066401} {\bibfield  {journal} {\bibinfo  {journal} {Physical Review Letters}\ }\textbf {\bibinfo {volume} {109}},\ \bibinfo {pages} {066401} (\bibinfo {year} {2012})}\BibitemShut {NoStop}%
\bibitem [{\citenamefont {Varnava}\ and\ \citenamefont {Vanderbilt}(2018)}]{Varnava2018}%
  \BibitemOpen
  \bibfield  {author} {\bibinfo {author} {\bibfnamefont {N.}~\bibnamefont {Varnava}}\ and\ \bibinfo {author} {\bibfnamefont {D.}~\bibnamefont {Vanderbilt}},\ }\bibfield  {title} {\bibinfo {title} {Surfaces of axion insulators},\ }\href {https://doi.org/10.1103/physrevb.98.245117} {\bibfield  {journal} {\bibinfo  {journal} {Physical Review B}\ }\textbf {\bibinfo {volume} {98}},\ \bibinfo {pages} {245117} (\bibinfo {year} {2018})}\BibitemShut {NoStop}%
\bibitem [{\citenamefont {Machida}\ \emph {et~al.}(2007)\citenamefont {Machida}, \citenamefont {Nakatsuji}, \citenamefont {Maeno}, \citenamefont {Tayama}, \citenamefont {Sakakibara},\ and\ \citenamefont {Onoda}}]{Machida2007}%
  \BibitemOpen
  \bibfield  {author} {\bibinfo {author} {\bibfnamefont {Y.}~\bibnamefont {Machida}}, \bibinfo {author} {\bibfnamefont {S.}~\bibnamefont {Nakatsuji}}, \bibinfo {author} {\bibfnamefont {Y.}~\bibnamefont {Maeno}}, \bibinfo {author} {\bibfnamefont {T.}~\bibnamefont {Tayama}}, \bibinfo {author} {\bibfnamefont {T.}~\bibnamefont {Sakakibara}},\ and\ \bibinfo {author} {\bibfnamefont {S.}~\bibnamefont {Onoda}},\ }\bibfield  {title} {\bibinfo {title} {Unconventional anomalous hall effect enhanced by a noncoplanar spin texture in the frustrated kondo lattice {Pr$_{2}$Ir$_{2}$O$_{7}$}},\ }\href {https://doi.org/10.1103/physrevlett.98.057203} {\bibfield  {journal} {\bibinfo  {journal} {Physical Review Letters}\ }\textbf {\bibinfo {volume} {98}},\ \bibinfo {pages} {057203} (\bibinfo {year} {2007})}\BibitemShut {NoStop}%
\bibitem [{\citenamefont {Machida}\ \emph {et~al.}(2009)\citenamefont {Machida}, \citenamefont {Nakatsuji}, \citenamefont {Onoda}, \citenamefont {Tayama},\ and\ \citenamefont {Sakakibara}}]{Machida2009}%
  \BibitemOpen
  \bibfield  {author} {\bibinfo {author} {\bibfnamefont {Y.}~\bibnamefont {Machida}}, \bibinfo {author} {\bibfnamefont {S.}~\bibnamefont {Nakatsuji}}, \bibinfo {author} {\bibfnamefont {S.}~\bibnamefont {Onoda}}, \bibinfo {author} {\bibfnamefont {T.}~\bibnamefont {Tayama}},\ and\ \bibinfo {author} {\bibfnamefont {T.}~\bibnamefont {Sakakibara}},\ }\bibfield  {title} {\bibinfo {title} {Time-reversal symmetry breaking and spontaneous hall effect without magnetic dipole order},\ }\href {https://doi.org/10.1038/nature08680} {\bibfield  {journal} {\bibinfo  {journal} {Nature}\ }\textbf {\bibinfo {volume} {463}},\ \bibinfo {pages} {210} (\bibinfo {year} {2009})}\BibitemShut {NoStop}%
\bibitem [{\citenamefont {Sushkov}\ \emph {et~al.}(2015)\citenamefont {Sushkov}, \citenamefont {Hofmann}, \citenamefont {Jenkins}, \citenamefont {Ishikawa}, \citenamefont {Nakatsuji}, \citenamefont {Das~Sarma},\ and\ \citenamefont {Drew}}]{Sushkov2015}%
  \BibitemOpen
  \bibfield  {author} {\bibinfo {author} {\bibfnamefont {A.~B.}\ \bibnamefont {Sushkov}}, \bibinfo {author} {\bibfnamefont {J.~B.}\ \bibnamefont {Hofmann}}, \bibinfo {author} {\bibfnamefont {G.~S.}\ \bibnamefont {Jenkins}}, \bibinfo {author} {\bibfnamefont {J.}~\bibnamefont {Ishikawa}}, \bibinfo {author} {\bibfnamefont {S.}~\bibnamefont {Nakatsuji}}, \bibinfo {author} {\bibfnamefont {S.}~\bibnamefont {Das~Sarma}},\ and\ \bibinfo {author} {\bibfnamefont {H.~D.}\ \bibnamefont {Drew}},\ }\bibfield  {title} {\bibinfo {title} {Optical evidence for a weyl semimetal state in pyrochlore {Eu$_{2}$Ir$_{2}$O$_{7}$}},\ }\href {https://doi.org/10.1103/physrevb.92.241108} {\bibfield  {journal} {\bibinfo  {journal} {Physical Review B}\ }\textbf {\bibinfo {volume} {92}},\ \bibinfo {pages} {241108(R)} (\bibinfo {year} {2015})}\BibitemShut {NoStop}%
\bibitem [{\citenamefont {Ueda}\ \emph {et~al.}(2018)\citenamefont {Ueda}, \citenamefont {Kaneko}, \citenamefont {Ishizuka}, \citenamefont {Fujioka}, \citenamefont {Nagaosa},\ and\ \citenamefont {Tokura}}]{Ueda2018}%
  \BibitemOpen
  \bibfield  {author} {\bibinfo {author} {\bibfnamefont {K.}~\bibnamefont {Ueda}}, \bibinfo {author} {\bibfnamefont {R.}~\bibnamefont {Kaneko}}, \bibinfo {author} {\bibfnamefont {H.}~\bibnamefont {Ishizuka}}, \bibinfo {author} {\bibfnamefont {J.}~\bibnamefont {Fujioka}}, \bibinfo {author} {\bibfnamefont {N.}~\bibnamefont {Nagaosa}},\ and\ \bibinfo {author} {\bibfnamefont {Y.}~\bibnamefont {Tokura}},\ }\bibfield  {title} {\bibinfo {title} {Spontaneous hall effect in the weyl semimetal candidate of all-in all-out pyrochlore iridate},\ }\href {https://doi.org/10.1038/s41467-018-05530-9} {\bibfield  {journal} {\bibinfo  {journal} {Nature Communications}\ }\textbf {\bibinfo {volume} {9}},\ \bibinfo {pages} {3032} (\bibinfo {year} {2018})}\BibitemShut {NoStop}%
\bibitem [{\citenamefont {Tomiyasu}\ \emph {et~al.}(2012)\citenamefont {Tomiyasu}, \citenamefont {Matsuhira}, \citenamefont {Iwasa}, \citenamefont {Watahiki}, \citenamefont {Takagi}, \citenamefont {Wakeshima}, \citenamefont {Hinatsu}, \citenamefont {Yokoyama}, \citenamefont {Ohoyama},\ and\ \citenamefont {Yamada}}]{Tomiyasu2012}%
  \BibitemOpen
  \bibfield  {author} {\bibinfo {author} {\bibfnamefont {K.}~\bibnamefont {Tomiyasu}}, \bibinfo {author} {\bibfnamefont {K.}~\bibnamefont {Matsuhira}}, \bibinfo {author} {\bibfnamefont {K.}~\bibnamefont {Iwasa}}, \bibinfo {author} {\bibfnamefont {M.}~\bibnamefont {Watahiki}}, \bibinfo {author} {\bibfnamefont {S.}~\bibnamefont {Takagi}}, \bibinfo {author} {\bibfnamefont {M.}~\bibnamefont {Wakeshima}}, \bibinfo {author} {\bibfnamefont {Y.}~\bibnamefont {Hinatsu}}, \bibinfo {author} {\bibfnamefont {M.}~\bibnamefont {Yokoyama}}, \bibinfo {author} {\bibfnamefont {K.}~\bibnamefont {Ohoyama}},\ and\ \bibinfo {author} {\bibfnamefont {K.}~\bibnamefont {Yamada}},\ }\bibfield  {title} {\bibinfo {title} {Emergence of magnetic long-range order in frustrated pyrochlore {Nd$_{2}$Ir$_{2}$O$_{7}$} with metal–insulator transition},\ }\href {https://doi.org/10.1143/jpsj.81.034709} {\bibfield  {journal} {\bibinfo  {journal} {Journal of the Physical Society of Japan}\ }\textbf {\bibinfo {volume} {81}},\ \bibinfo {pages} {034709} (\bibinfo {year} {2012})}\BibitemShut {NoStop}%
\bibitem [{\citenamefont {Lefran\c{c}ois}\ \emph {et~al.}(2017)\citenamefont {Lefran\c{c}ois}, \citenamefont {Cathelin}, \citenamefont {Lhotel}, \citenamefont {Robert}, \citenamefont {Lejay}, \citenamefont {Colin}, \citenamefont {Canals}, \citenamefont {Damay}, \citenamefont {Ollivier}, \citenamefont {Fåk}, \citenamefont {Chapon}, \citenamefont {Ballou},\ and\ \citenamefont {Simonet}}]{Lefrancois2017}%
  \BibitemOpen
  \bibfield  {author} {\bibinfo {author} {\bibfnamefont {E.}~\bibnamefont {Lefran\c{c}ois}}, \bibinfo {author} {\bibfnamefont {V.}~\bibnamefont {Cathelin}}, \bibinfo {author} {\bibfnamefont {E.}~\bibnamefont {Lhotel}}, \bibinfo {author} {\bibfnamefont {J.}~\bibnamefont {Robert}}, \bibinfo {author} {\bibfnamefont {P.}~\bibnamefont {Lejay}}, \bibinfo {author} {\bibfnamefont {C.~V.}\ \bibnamefont {Colin}}, \bibinfo {author} {\bibfnamefont {B.}~\bibnamefont {Canals}}, \bibinfo {author} {\bibfnamefont {F.}~\bibnamefont {Damay}}, \bibinfo {author} {\bibfnamefont {J.}~\bibnamefont {Ollivier}}, \bibinfo {author} {\bibfnamefont {B.}~\bibnamefont {Fåk}}, \bibinfo {author} {\bibfnamefont {L.~C.}\ \bibnamefont {Chapon}}, \bibinfo {author} {\bibfnamefont {R.}~\bibnamefont {Ballou}},\ and\ \bibinfo {author} {\bibfnamefont {V.}~\bibnamefont {Simonet}},\ }\bibfield  {title} {\bibinfo {title} {Fragmentation in spin ice from magnetic charge injection},\ }\href {https://doi.org/10.1038/s41467-017-00277-1} {\bibfield  {journal} {\bibinfo  {journal} {Nature Communications}\ }\textbf {\bibinfo {volume} {8}},\ \bibinfo {pages} {209} (\bibinfo {year} {2017})}\BibitemShut {NoStop}%
\bibitem [{\citenamefont {Lefran\c{c}ois}\ \emph {et~al.}(2019)\citenamefont {Lefran\c{c}ois}, \citenamefont {Mangin-Thro}, \citenamefont {Lhotel}, \citenamefont {Robert}, \citenamefont {Petit}, \citenamefont {Cathelin}, \citenamefont {Fischer}, \citenamefont {Colin}, \citenamefont {Damay}, \citenamefont {Ollivier}, \citenamefont {Lejay}, \citenamefont {Chapon}, \citenamefont {Simonet},\ and\ \citenamefont {Ballou}}]{Lefrancois2019}%
  \BibitemOpen
  \bibfield  {author} {\bibinfo {author} {\bibfnamefont {E.}~\bibnamefont {Lefran\c{c}ois}}, \bibinfo {author} {\bibfnamefont {L.}~\bibnamefont {Mangin-Thro}}, \bibinfo {author} {\bibfnamefont {E.}~\bibnamefont {Lhotel}}, \bibinfo {author} {\bibfnamefont {J.}~\bibnamefont {Robert}}, \bibinfo {author} {\bibfnamefont {S.}~\bibnamefont {Petit}}, \bibinfo {author} {\bibfnamefont {V.}~\bibnamefont {Cathelin}}, \bibinfo {author} {\bibfnamefont {H.~E.}\ \bibnamefont {Fischer}}, \bibinfo {author} {\bibfnamefont {C.~V.}\ \bibnamefont {Colin}}, \bibinfo {author} {\bibfnamefont {F.}~\bibnamefont {Damay}}, \bibinfo {author} {\bibfnamefont {J.}~\bibnamefont {Ollivier}}, \bibinfo {author} {\bibfnamefont {P.}~\bibnamefont {Lejay}}, \bibinfo {author} {\bibfnamefont {L.~C.}\ \bibnamefont {Chapon}}, \bibinfo {author} {\bibfnamefont {V.}~\bibnamefont {Simonet}},\ and\ \bibinfo {author} {\bibfnamefont {R.}~\bibnamefont {Ballou}},\ }\bibfield  {title} {\bibinfo {title} {Spin decoupling under a staggered field in the {Gd$_{2}$Ir$_{2}$O$_{7}$} pyrochlore},\ }\href {https://doi.org/10.1103/physrevb.99.060401} {\bibfield  {journal} {\bibinfo  {journal} {Physical Review B}\ }\textbf {\bibinfo {volume} {99}},\ \bibinfo {pages} {060401} (\bibinfo {year} {2019})}\BibitemShut {NoStop}%
\bibitem [{\citenamefont {Cathelin}\ \emph {et~al.}(2020)\citenamefont {Cathelin}, \citenamefont {Lefran\c{c}ois}, \citenamefont {Robert}, \citenamefont {Guruciaga}, \citenamefont {Paulsen}, \citenamefont {Prabhakaran}, \citenamefont {Lejay}, \citenamefont {Damay}, \citenamefont {Ollivier}, \citenamefont {Fåk}, \citenamefont {Chapon}, \citenamefont {Ballou}, \citenamefont {Simonet}, \citenamefont {Holdsworth},\ and\ \citenamefont {Lhotel}}]{Cathelin2020}%
  \BibitemOpen
  \bibfield  {author} {\bibinfo {author} {\bibfnamefont {V.}~\bibnamefont {Cathelin}}, \bibinfo {author} {\bibfnamefont {E.}~\bibnamefont {Lefran\c{c}ois}}, \bibinfo {author} {\bibfnamefont {J.}~\bibnamefont {Robert}}, \bibinfo {author} {\bibfnamefont {P.~C.}\ \bibnamefont {Guruciaga}}, \bibinfo {author} {\bibfnamefont {C.}~\bibnamefont {Paulsen}}, \bibinfo {author} {\bibfnamefont {D.}~\bibnamefont {Prabhakaran}}, \bibinfo {author} {\bibfnamefont {P.}~\bibnamefont {Lejay}}, \bibinfo {author} {\bibfnamefont {F.}~\bibnamefont {Damay}}, \bibinfo {author} {\bibfnamefont {J.}~\bibnamefont {Ollivier}}, \bibinfo {author} {\bibfnamefont {B.}~\bibnamefont {Fåk}}, \bibinfo {author} {\bibfnamefont {L.~C.}\ \bibnamefont {Chapon}}, \bibinfo {author} {\bibfnamefont {R.}~\bibnamefont {Ballou}}, \bibinfo {author} {\bibfnamefont {V.}~\bibnamefont {Simonet}}, \bibinfo {author} {\bibfnamefont {P.~C.~W.}\ \bibnamefont {Holdsworth}},\ and\ \bibinfo {author} {\bibfnamefont {E.}~\bibnamefont {Lhotel}},\ }\bibfield  {title} {\bibinfo {title} {Fragmented monopole crystal, dimer entropy, and coulomb interactions in {Dy$_{2}$Ir$_{2}$O$_{7}$}},\ }\href {https://doi.org/10.1103/physrevresearch.2.032073} {\bibfield  {journal} {\bibinfo  {journal} {Physical Review Research}\ }\textbf {\bibinfo {volume} {2}},\ \bibinfo {pages} {032073(R)} (\bibinfo {year} {2020})}\BibitemShut {NoStop}%
\bibitem [{\citenamefont {Brooks-Bartlett}\ \emph {et~al.}(2014)\citenamefont {Brooks-Bartlett}, \citenamefont {Banks}, \citenamefont {Jaubert}, \citenamefont {Harman-Clarke},\ and\ \citenamefont {Holdsworth}}]{BrooksBartlett2014}%
  \BibitemOpen
  \bibfield  {author} {\bibinfo {author} {\bibfnamefont {M.~E.}\ \bibnamefont {Brooks-Bartlett}}, \bibinfo {author} {\bibfnamefont {S.~T.}\ \bibnamefont {Banks}}, \bibinfo {author} {\bibfnamefont {L.~D.~C.}\ \bibnamefont {Jaubert}}, \bibinfo {author} {\bibfnamefont {A.}~\bibnamefont {Harman-Clarke}},\ and\ \bibinfo {author} {\bibfnamefont {P.~C.~W.}\ \bibnamefont {Holdsworth}},\ }\bibfield  {title} {\bibinfo {title} {Magnetic-moment fragmentation and monopole crystallization},\ }\href {https://doi.org/10.1103/physrevx.4.011007} {\bibfield  {journal} {\bibinfo  {journal} {Physical Review X}\ }\textbf {\bibinfo {volume} {4}},\ \bibinfo {pages} {011007} (\bibinfo {year} {2014})}\BibitemShut {NoStop}%
\bibitem [{\citenamefont {Jacobsen}\ \emph {et~al.}(2020)\citenamefont {Jacobsen}, \citenamefont {Dashwood}, \citenamefont {Lhotel}, \citenamefont {Khalyavin}, \citenamefont {Manuel}, \citenamefont {Stewart}, \citenamefont {Prabhakaran}, \citenamefont {McMorrow},\ and\ \citenamefont {Boothroyd}}]{Jacobsen2020}%
  \BibitemOpen
  \bibfield  {author} {\bibinfo {author} {\bibfnamefont {H.}~\bibnamefont {Jacobsen}}, \bibinfo {author} {\bibfnamefont {C.~D.}\ \bibnamefont {Dashwood}}, \bibinfo {author} {\bibfnamefont {E.}~\bibnamefont {Lhotel}}, \bibinfo {author} {\bibfnamefont {D.}~\bibnamefont {Khalyavin}}, \bibinfo {author} {\bibfnamefont {P.}~\bibnamefont {Manuel}}, \bibinfo {author} {\bibfnamefont {R.}~\bibnamefont {Stewart}}, \bibinfo {author} {\bibfnamefont {D.}~\bibnamefont {Prabhakaran}}, \bibinfo {author} {\bibfnamefont {D.~F.}\ \bibnamefont {McMorrow}},\ and\ \bibinfo {author} {\bibfnamefont {A.~T.}\ \bibnamefont {Boothroyd}},\ }\bibfield  {title} {\bibinfo {title} {Strong quantum fluctuations from competition between magnetic phases in a pyrochlore iridate},\ }\href {https://doi.org/10.1103/physrevb.101.104404} {\bibfield  {journal} {\bibinfo  {journal} {Physical Review B}\ }\textbf {\bibinfo {volume} {101}},\ \bibinfo {pages} {104404} (\bibinfo {year} {2020})}\BibitemShut {NoStop}%
\bibitem [{\citenamefont {Guo}\ \emph {et~al.}(2017)\citenamefont {Guo}, \citenamefont {Ritter},\ and\ \citenamefont {Komarek}}]{Guo2017}%
  \BibitemOpen
  \bibfield  {author} {\bibinfo {author} {\bibfnamefont {H.}~\bibnamefont {Guo}}, \bibinfo {author} {\bibfnamefont {C.}~\bibnamefont {Ritter}},\ and\ \bibinfo {author} {\bibfnamefont {A.~C.}\ \bibnamefont {Komarek}},\ }\bibfield  {title} {\bibinfo {title} {Magnetic structure of {Tb$_{2}$Ir$_{2}$O$_{7}$} determined by powder neutron diffraction},\ }\href {https://doi.org/10.1103/physrevb.96.144415} {\bibfield  {journal} {\bibinfo  {journal} {Physical Review B}\ }\textbf {\bibinfo {volume} {96}},\ \bibinfo {pages} {144415} (\bibinfo {year} {2017})}\BibitemShut {NoStop}%
\bibitem [{\citenamefont {Rau}\ and\ \citenamefont {Gingras}(2019)}]{Rau2019}%
  \BibitemOpen
  \bibfield  {author} {\bibinfo {author} {\bibfnamefont {J.~G.}\ \bibnamefont {Rau}}\ and\ \bibinfo {author} {\bibfnamefont {M.~J.}\ \bibnamefont {Gingras}},\ }\bibfield  {title} {\bibinfo {title} {Frustrated quantum rare-earth pyrochlores},\ }\href {https://doi.org/10.1146/annurev-conmatphys-022317-110520} {\bibfield  {journal} {\bibinfo  {journal} {Annual Review of Condensed Matter Physics}\ }\textbf {\bibinfo {volume} {10}},\ \bibinfo {pages} {357} (\bibinfo {year} {2019})}\BibitemShut {NoStop}%
\bibitem [{\citenamefont {Mirebeau}\ \emph {et~al.}(2005)\citenamefont {Mirebeau}, \citenamefont {Apetrei}, \citenamefont {Rodríguez-Carvajal}, \citenamefont {Bonville}, \citenamefont {Forget}, \citenamefont {Colson}, \citenamefont {Glazkov}, \citenamefont {Sanchez}, \citenamefont {Isnard},\ and\ \citenamefont {Suard}}]{Mirebeau2005}%
  \BibitemOpen
  \bibfield  {author} {\bibinfo {author} {\bibfnamefont {I.}~\bibnamefont {Mirebeau}}, \bibinfo {author} {\bibfnamefont {A.}~\bibnamefont {Apetrei}}, \bibinfo {author} {\bibfnamefont {J.}~\bibnamefont {Rodríguez-Carvajal}}, \bibinfo {author} {\bibfnamefont {P.}~\bibnamefont {Bonville}}, \bibinfo {author} {\bibfnamefont {A.}~\bibnamefont {Forget}}, \bibinfo {author} {\bibfnamefont {D.}~\bibnamefont {Colson}}, \bibinfo {author} {\bibfnamefont {V.}~\bibnamefont {Glazkov}}, \bibinfo {author} {\bibfnamefont {J.~P.}\ \bibnamefont {Sanchez}}, \bibinfo {author} {\bibfnamefont {O.}~\bibnamefont {Isnard}},\ and\ \bibinfo {author} {\bibfnamefont {E.}~\bibnamefont {Suard}},\ }\bibfield  {title} {\bibinfo {title} {Ordered spin ice state and magnetic fluctuations in {Tb$_{2}$Sn$_{2}$O$_{7}$}},\ }\href {https://doi.org/10.1103/physrevlett.94.246402} {\bibfield  {journal} {\bibinfo  {journal} {Physical Review Letters}\ }\textbf {\bibinfo {volume} {94}},\ \bibinfo {pages} {246402} (\bibinfo {year} {2005})}\BibitemShut {NoStop}%
\bibitem [{\citenamefont {Dalmas~de Réotier}\ \emph {et~al.}(2006)\citenamefont {Dalmas~de Réotier}, \citenamefont {Yaouanc}, \citenamefont {Keller}, \citenamefont {Cervellino}, \citenamefont {Roessli}, \citenamefont {Baines}, \citenamefont {Forget}, \citenamefont {Vaju}, \citenamefont {Gubbens}, \citenamefont {Amato},\ and\ \citenamefont {King}}]{DalmasdeReotier2006}%
  \BibitemOpen
  \bibfield  {author} {\bibinfo {author} {\bibfnamefont {P.}~\bibnamefont {Dalmas~de Réotier}}, \bibinfo {author} {\bibfnamefont {A.}~\bibnamefont {Yaouanc}}, \bibinfo {author} {\bibfnamefont {L.}~\bibnamefont {Keller}}, \bibinfo {author} {\bibfnamefont {A.}~\bibnamefont {Cervellino}}, \bibinfo {author} {\bibfnamefont {B.}~\bibnamefont {Roessli}}, \bibinfo {author} {\bibfnamefont {C.}~\bibnamefont {Baines}}, \bibinfo {author} {\bibfnamefont {A.}~\bibnamefont {Forget}}, \bibinfo {author} {\bibfnamefont {C.}~\bibnamefont {Vaju}}, \bibinfo {author} {\bibfnamefont {P.~C.~M.}\ \bibnamefont {Gubbens}}, \bibinfo {author} {\bibfnamefont {A.}~\bibnamefont {Amato}},\ and\ \bibinfo {author} {\bibfnamefont {P.~J.~C.}\ \bibnamefont {King}},\ }\bibfield  {title} {\bibinfo {title} {Spin dynamics and magnetic order in magnetically frustrated {Tb$_{2}$Sn$_{2}$O$_{7}$}},\ }\href {https://doi.org/10.1103/physrevlett.96.127202} {\bibfield  {journal} {\bibinfo  {journal} {Physical Review Letters}\ }\textbf {\bibinfo {volume} {96}},\ \bibinfo {pages} {127202} (\bibinfo {year} {2006})}\BibitemShut {NoStop}%
\bibitem [{\citenamefont {Hallas}\ \emph {et~al.}(2020)\citenamefont {Hallas}, \citenamefont {Jin}, \citenamefont {Gaudet}, \citenamefont {Tonita}, \citenamefont {Pomaranski}, \citenamefont {Buhariwalla}, \citenamefont {Tachibana}, \citenamefont {Butch}, \citenamefont {Calder}, \citenamefont {Stone}, \citenamefont {Luke}, \citenamefont {Wiebe}, \citenamefont {Kycia}, \citenamefont {Gingras},\ and\ \citenamefont {Gaulin}}]{Hallas2020}%
  \BibitemOpen
  \bibfield  {author} {\bibinfo {author} {\bibfnamefont {A.~M.}\ \bibnamefont {Hallas}}, \bibinfo {author} {\bibfnamefont {W.}~\bibnamefont {Jin}}, \bibinfo {author} {\bibfnamefont {J.}~\bibnamefont {Gaudet}}, \bibinfo {author} {\bibfnamefont {E.~M.}\ \bibnamefont {Tonita}}, \bibinfo {author} {\bibfnamefont {D.}~\bibnamefont {Pomaranski}}, \bibinfo {author} {\bibfnamefont {C.~R.~C.}\ \bibnamefont {Buhariwalla}}, \bibinfo {author} {\bibfnamefont {M.}~\bibnamefont {Tachibana}}, \bibinfo {author} {\bibfnamefont {N.~P.}\ \bibnamefont {Butch}}, \bibinfo {author} {\bibfnamefont {S.}~\bibnamefont {Calder}}, \bibinfo {author} {\bibfnamefont {M.~B.}\ \bibnamefont {Stone}}, \bibinfo {author} {\bibfnamefont {G.~M.}\ \bibnamefont {Luke}}, \bibinfo {author} {\bibfnamefont {C.~R.}\ \bibnamefont {Wiebe}}, \bibinfo {author} {\bibfnamefont {J.~B.}\ \bibnamefont {Kycia}}, \bibinfo {author} {\bibfnamefont {M.~J.~P.}\ \bibnamefont {Gingras}},\ and\ \bibinfo {author} {\bibfnamefont {B.~D.}\ \bibnamefont {Gaulin}},\ }\bibfield  {title} {\bibinfo {title} {Intertwined magnetic dipolar and electric quadrupolar correlations in the pyrochlore {Tb$_{2}$Ge$_{2}$O$_{7}$}},\ }\Eprint {https://arxiv.org/abs/2009.05036v2} {arXiv:2009.05036v2}  (\bibinfo {year} {2020})\BibitemShut {NoStop}%
\bibitem [{\citenamefont {Sibille}\ \emph {et~al.}(2017)\citenamefont {Sibille}, \citenamefont {Lhotel}, \citenamefont {Ciomaga~Hatnean}, \citenamefont {Nilsen}, \citenamefont {Ehlers}, \citenamefont {Cervellino}, \citenamefont {Ressouche}, \citenamefont {Frontzek}, \citenamefont {Zaharko}, \citenamefont {Pomjakushin}, \citenamefont {Stuhr}, \citenamefont {Walker}, \citenamefont {Adroja}, \citenamefont {Luetkens}, \citenamefont {Baines}, \citenamefont {Amato}, \citenamefont {Balakrishnan}, \citenamefont {Fennell},\ and\ \citenamefont {Kenzelmann}}]{Sibille2017}%
  \BibitemOpen
  \bibfield  {author} {\bibinfo {author} {\bibfnamefont {R.}~\bibnamefont {Sibille}}, \bibinfo {author} {\bibfnamefont {E.}~\bibnamefont {Lhotel}}, \bibinfo {author} {\bibfnamefont {M.}~\bibnamefont {Ciomaga~Hatnean}}, \bibinfo {author} {\bibfnamefont {G.~J.}\ \bibnamefont {Nilsen}}, \bibinfo {author} {\bibfnamefont {G.}~\bibnamefont {Ehlers}}, \bibinfo {author} {\bibfnamefont {A.}~\bibnamefont {Cervellino}}, \bibinfo {author} {\bibfnamefont {E.}~\bibnamefont {Ressouche}}, \bibinfo {author} {\bibfnamefont {M.}~\bibnamefont {Frontzek}}, \bibinfo {author} {\bibfnamefont {O.}~\bibnamefont {Zaharko}}, \bibinfo {author} {\bibfnamefont {V.}~\bibnamefont {Pomjakushin}}, \bibinfo {author} {\bibfnamefont {U.}~\bibnamefont {Stuhr}}, \bibinfo {author} {\bibfnamefont {H.~C.}\ \bibnamefont {Walker}}, \bibinfo {author} {\bibfnamefont {D.~T.}\ \bibnamefont {Adroja}}, \bibinfo {author} {\bibfnamefont {H.}~\bibnamefont {Luetkens}}, \bibinfo {author} {\bibfnamefont {C.}~\bibnamefont {Baines}}, \bibinfo {author} {\bibfnamefont {A.}~\bibnamefont {Amato}}, \bibinfo {author} {\bibfnamefont {G.}~\bibnamefont {Balakrishnan}}, \bibinfo {author} {\bibfnamefont {T.}~\bibnamefont {Fennell}},\ and\ \bibinfo {author} {\bibfnamefont {M.}~\bibnamefont {Kenzelmann}},\ }\bibfield  {title} {\bibinfo {title} {Coulomb spin liquid in anion-disordered pyrochlore {Tb$_{2}$Hf$_{2}$O$_{7}$}},\ }\href {https://doi.org/10.1038/s41467-017-00905-w} {\bibfield  {journal} {\bibinfo  {journal} {Nature Communications}\ }\textbf {\bibinfo {volume} {8}},\ \bibinfo {pages} {892} (\bibinfo {year} {2017})}\BibitemShut {NoStop}%
\bibitem [{\citenamefont {Anand}\ \emph {et~al.}(2018)\citenamefont {Anand}, \citenamefont {Opherden}, \citenamefont {Xu}, \citenamefont {Adroja}, \citenamefont {Hillier}, \citenamefont {Biswas}, \citenamefont {Herrmannsd\"{o}rfer}, \citenamefont {Uhlarz}, \citenamefont {Hornung}, \citenamefont {Wosnitza}, \citenamefont {Canévet},\ and\ \citenamefont {Lake}}]{Anand2018}%
  \BibitemOpen
  \bibfield  {author} {\bibinfo {author} {\bibfnamefont {V.~K.}\ \bibnamefont {Anand}}, \bibinfo {author} {\bibfnamefont {L.}~\bibnamefont {Opherden}}, \bibinfo {author} {\bibfnamefont {J.}~\bibnamefont {Xu}}, \bibinfo {author} {\bibfnamefont {D.~T.}\ \bibnamefont {Adroja}}, \bibinfo {author} {\bibfnamefont {A.~D.}\ \bibnamefont {Hillier}}, \bibinfo {author} {\bibfnamefont {P.~K.}\ \bibnamefont {Biswas}}, \bibinfo {author} {\bibfnamefont {T.}~\bibnamefont {Herrmannsd\"{o}rfer}}, \bibinfo {author} {\bibfnamefont {M.}~\bibnamefont {Uhlarz}}, \bibinfo {author} {\bibfnamefont {J.}~\bibnamefont {Hornung}}, \bibinfo {author} {\bibfnamefont {J.}~\bibnamefont {Wosnitza}}, \bibinfo {author} {\bibfnamefont {E.}~\bibnamefont {Canévet}},\ and\ \bibinfo {author} {\bibfnamefont {B.}~\bibnamefont {Lake}},\ }\bibfield  {title} {\bibinfo {title} {Evidence for a dynamical ground state in the frustrated pyrohafnate {Tb$_{2}$Hf$_{2}$O$_{7}$}},\ }\href {https://doi.org/10.1103/physrevb.97.094402} {\bibfield  {journal} {\bibinfo  {journal} {Physical Review B}\ }\textbf {\bibinfo {volume} {97}},\ \bibinfo {pages} {094402} (\bibinfo {year} {2018})}\BibitemShut {NoStop}%
\bibitem [{\citenamefont {Alexanian}\ \emph {et~al.}(2023{\natexlab{a}})\citenamefont {Alexanian}, \citenamefont {Lhotel}, \citenamefont {Ballou}, \citenamefont {Colin}, \citenamefont {Klein}, \citenamefont {Le~Priol}, \citenamefont {Museur}, \citenamefont {Robert}, \citenamefont {Pachoud}, \citenamefont {Lejay}, \citenamefont {Hadj-Azzem}, \citenamefont {Fåk}, \citenamefont {Berrod}, \citenamefont {Zanotti}, \citenamefont {Suard}, \citenamefont {Dejoie}, \citenamefont {de~Brion},\ and\ \citenamefont {Simonet}}]{Alexanian2023b}%
  \BibitemOpen
  \bibfield  {author} {\bibinfo {author} {\bibfnamefont {Y.}~\bibnamefont {Alexanian}}, \bibinfo {author} {\bibfnamefont {E.}~\bibnamefont {Lhotel}}, \bibinfo {author} {\bibfnamefont {R.}~\bibnamefont {Ballou}}, \bibinfo {author} {\bibfnamefont {C.~V.}\ \bibnamefont {Colin}}, \bibinfo {author} {\bibfnamefont {H.}~\bibnamefont {Klein}}, \bibinfo {author} {\bibfnamefont {A.}~\bibnamefont {Le~Priol}}, \bibinfo {author} {\bibfnamefont {F.}~\bibnamefont {Museur}}, \bibinfo {author} {\bibfnamefont {J.}~\bibnamefont {Robert}}, \bibinfo {author} {\bibfnamefont {E.}~\bibnamefont {Pachoud}}, \bibinfo {author} {\bibfnamefont {P.}~\bibnamefont {Lejay}}, \bibinfo {author} {\bibfnamefont {A.}~\bibnamefont {Hadj-Azzem}}, \bibinfo {author} {\bibfnamefont {B.}~\bibnamefont {Fåk}}, \bibinfo {author} {\bibfnamefont {Q.}~\bibnamefont {Berrod}}, \bibinfo {author} {\bibfnamefont {J.-M.}\ \bibnamefont {Zanotti}}, \bibinfo {author} {\bibfnamefont {E.}~\bibnamefont {Suard}}, \bibinfo {author} {\bibfnamefont {C.}~\bibnamefont {Dejoie}}, \bibinfo {author} {\bibfnamefont {S.}~\bibnamefont {de~Brion}},\ and\ \bibinfo {author} {\bibfnamefont {V.}~\bibnamefont {Simonet}},\ }\bibfield  {title} {\bibinfo {title} {Collective magnetic state induced by charge disorder in the non-kramers rare-earth pyrochlore {Tb$_{2}$ScNbO$_{7}$}},\ }\href {https://doi.org/10.1103/physrevmaterials.7.094403} {\bibfield  {journal} {\bibinfo  {journal} {Physical Review Materials}\ }\textbf {\bibinfo {volume} {7}},\ \bibinfo {pages} {094403} (\bibinfo {year} {2023}{\natexlab{a}})}\BibitemShut {NoStop}%
\bibitem [{\citenamefont {Gardner}\ \emph {et~al.}(1999)\citenamefont {Gardner}, \citenamefont {Dunsiger}, \citenamefont {Gaulin}, \citenamefont {Gingras}, \citenamefont {Greedan}, \citenamefont {Kiefl}, \citenamefont {Lumsden}, \citenamefont {MacFarlane}, \citenamefont {Raju}, \citenamefont {Sonier}, \citenamefont {Swainson},\ and\ \citenamefont {Tun}}]{Gardner1999}%
  \BibitemOpen
  \bibfield  {author} {\bibinfo {author} {\bibfnamefont {J.~S.}\ \bibnamefont {Gardner}}, \bibinfo {author} {\bibfnamefont {S.~R.}\ \bibnamefont {Dunsiger}}, \bibinfo {author} {\bibfnamefont {B.~D.}\ \bibnamefont {Gaulin}}, \bibinfo {author} {\bibfnamefont {M.~J.~P.}\ \bibnamefont {Gingras}}, \bibinfo {author} {\bibfnamefont {J.~E.}\ \bibnamefont {Greedan}}, \bibinfo {author} {\bibfnamefont {R.~F.}\ \bibnamefont {Kiefl}}, \bibinfo {author} {\bibfnamefont {M.~D.}\ \bibnamefont {Lumsden}}, \bibinfo {author} {\bibfnamefont {W.~A.}\ \bibnamefont {MacFarlane}}, \bibinfo {author} {\bibfnamefont {N.~P.}\ \bibnamefont {Raju}}, \bibinfo {author} {\bibfnamefont {J.~E.}\ \bibnamefont {Sonier}}, \bibinfo {author} {\bibfnamefont {I.}~\bibnamefont {Swainson}},\ and\ \bibinfo {author} {\bibfnamefont {Z.}~\bibnamefont {Tun}},\ }\bibfield  {title} {\bibinfo {title} {Cooperative paramagnetism in the geometrically frustrated pyrochlore antiferromagnet {${\mathrm{Tb}}_{2}{\mathrm{Ti}}_{2}{\mathrm{O}}_{7}$}},\ }\href {https://doi.org/10.1103/PhysRevLett.82.1012} {\bibfield  {journal} {\bibinfo  {journal} {Phys. Rev. Lett.}\ }\textbf {\bibinfo {volume} {82}},\ \bibinfo {pages} {1012} (\bibinfo {year} {1999})}\BibitemShut {NoStop}%
\bibitem [{\citenamefont {Fennell}\ \emph {et~al.}(2012)\citenamefont {Fennell}, \citenamefont {Kenzelmann}, \citenamefont {Roessli}, \citenamefont {Haas},\ and\ \citenamefont {Cava}}]{Fennell2012}%
  \BibitemOpen
  \bibfield  {author} {\bibinfo {author} {\bibfnamefont {T.}~\bibnamefont {Fennell}}, \bibinfo {author} {\bibfnamefont {M.}~\bibnamefont {Kenzelmann}}, \bibinfo {author} {\bibfnamefont {B.}~\bibnamefont {Roessli}}, \bibinfo {author} {\bibfnamefont {M.~K.}\ \bibnamefont {Haas}},\ and\ \bibinfo {author} {\bibfnamefont {R.~J.}\ \bibnamefont {Cava}},\ }\bibfield  {title} {\bibinfo {title} {Power-law spin correlations in the pyrochlore antiferromagnet {Tb$_{2}$Ti$_{2}$O$_{7}$}},\ }\href {https://doi.org/10.1103/physrevlett.109.017201} {\bibfield  {journal} {\bibinfo  {journal} {Physical Review Letters}\ }\textbf {\bibinfo {volume} {109}},\ \bibinfo {pages} {017201} (\bibinfo {year} {2012})}\BibitemShut {NoStop}%
\bibitem [{\citenamefont {Taniguchi}\ \emph {et~al.}(2013)\citenamefont {Taniguchi}, \citenamefont {Kadowaki}, \citenamefont {Takatsu}, \citenamefont {F\aa{}k}, \citenamefont {Ollivier}, \citenamefont {Yamazaki}, \citenamefont {Sato}, \citenamefont {Yoshizawa}, \citenamefont {Shimura}, \citenamefont {Sakakibara}, \citenamefont {Hong}, \citenamefont {Goto}, \citenamefont {Yaraskavitch},\ and\ \citenamefont {Kycia}}]{Taniguchi2013}%
  \BibitemOpen
  \bibfield  {author} {\bibinfo {author} {\bibfnamefont {T.}~\bibnamefont {Taniguchi}}, \bibinfo {author} {\bibfnamefont {H.}~\bibnamefont {Kadowaki}}, \bibinfo {author} {\bibfnamefont {H.}~\bibnamefont {Takatsu}}, \bibinfo {author} {\bibfnamefont {B.}~\bibnamefont {F\aa{}k}}, \bibinfo {author} {\bibfnamefont {J.}~\bibnamefont {Ollivier}}, \bibinfo {author} {\bibfnamefont {T.}~\bibnamefont {Yamazaki}}, \bibinfo {author} {\bibfnamefont {T.~J.}\ \bibnamefont {Sato}}, \bibinfo {author} {\bibfnamefont {H.}~\bibnamefont {Yoshizawa}}, \bibinfo {author} {\bibfnamefont {Y.}~\bibnamefont {Shimura}}, \bibinfo {author} {\bibfnamefont {T.}~\bibnamefont {Sakakibara}}, \bibinfo {author} {\bibfnamefont {T.}~\bibnamefont {Hong}}, \bibinfo {author} {\bibfnamefont {K.}~\bibnamefont {Goto}}, \bibinfo {author} {\bibfnamefont {L.~R.}\ \bibnamefont {Yaraskavitch}},\ and\ \bibinfo {author} {\bibfnamefont {J.~B.}\ \bibnamefont {Kycia}},\ }\bibfield  {title} {\bibinfo {title} {Long-range order and spin-liquid states of polycrystalline {${\mathrm{Tb}}_{2+x}{\mathrm{Ti}}_{2-x}{\mathrm{O}}_{7+y}$}},\ }\href {https://doi.org/10.1103/PhysRevB.87.060408} {\bibfield  {journal} {\bibinfo  {journal} {Phys. Rev. B}\ }\textbf {\bibinfo {volume} {87}},\ \bibinfo {pages} {060408(R)} (\bibinfo {year} {2013})}\BibitemShut {NoStop}%
\bibitem [{\citenamefont {Guitteny}\ \emph {et~al.}(2013)\citenamefont {Guitteny}, \citenamefont {Robert}, \citenamefont {Bonville}, \citenamefont {Ollivier}, \citenamefont {Decorse}, \citenamefont {Steffens}, \citenamefont {Boehm}, \citenamefont {Mutka}, \citenamefont {Mirebeau},\ and\ \citenamefont {Petit}}]{Guitteny2013}%
  \BibitemOpen
  \bibfield  {author} {\bibinfo {author} {\bibfnamefont {S.}~\bibnamefont {Guitteny}}, \bibinfo {author} {\bibfnamefont {J.}~\bibnamefont {Robert}}, \bibinfo {author} {\bibfnamefont {P.}~\bibnamefont {Bonville}}, \bibinfo {author} {\bibfnamefont {J.}~\bibnamefont {Ollivier}}, \bibinfo {author} {\bibfnamefont {C.}~\bibnamefont {Decorse}}, \bibinfo {author} {\bibfnamefont {P.}~\bibnamefont {Steffens}}, \bibinfo {author} {\bibfnamefont {M.}~\bibnamefont {Boehm}}, \bibinfo {author} {\bibfnamefont {H.}~\bibnamefont {Mutka}}, \bibinfo {author} {\bibfnamefont {I.}~\bibnamefont {Mirebeau}},\ and\ \bibinfo {author} {\bibfnamefont {S.}~\bibnamefont {Petit}},\ }\bibfield  {title} {\bibinfo {title} {Anisotropic propagating excitations and quadrupolar effects in {Tb$_{2}$Ti$_{2}$O$_{7}$}},\ }\href {https://doi.org/10.1103/physrevlett.111.087201} {\bibfield  {journal} {\bibinfo  {journal} {Physical Review Letters}\ }\textbf {\bibinfo {volume} {111}},\ \bibinfo {pages} {087201} (\bibinfo {year} {2013})}\BibitemShut {NoStop}%
\bibitem [{\citenamefont {Fritsch}\ \emph {et~al.}(2013)\citenamefont {Fritsch}, \citenamefont {Ross}, \citenamefont {Qiu}, \citenamefont {Copley}, \citenamefont {Guidi}, \citenamefont {Bewley}, \citenamefont {Dabkowska},\ and\ \citenamefont {Gaulin}}]{Fritsch2013}%
  \BibitemOpen
  \bibfield  {author} {\bibinfo {author} {\bibfnamefont {K.}~\bibnamefont {Fritsch}}, \bibinfo {author} {\bibfnamefont {K.~A.}\ \bibnamefont {Ross}}, \bibinfo {author} {\bibfnamefont {Y.}~\bibnamefont {Qiu}}, \bibinfo {author} {\bibfnamefont {J.~R.~D.}\ \bibnamefont {Copley}}, \bibinfo {author} {\bibfnamefont {T.}~\bibnamefont {Guidi}}, \bibinfo {author} {\bibfnamefont {R.~I.}\ \bibnamefont {Bewley}}, \bibinfo {author} {\bibfnamefont {H.~A.}\ \bibnamefont {Dabkowska}},\ and\ \bibinfo {author} {\bibfnamefont {B.~D.}\ \bibnamefont {Gaulin}},\ }\bibfield  {title} {\bibinfo {title} {Antiferromagnetic spin ice correlations at $(\frac{1}{2},\frac{1}{2},\frac{1}{2})$ in the ground state of the pyrochlore magnet {${\mathrm{Tb}}_{2}{\mathrm{Ti}}_{2}{\mathrm{O}}_{7}$}},\ }\href {https://doi.org/10.1103/physrevb.87.094410} {\bibfield  {journal} {\bibinfo  {journal} {Physical Review B}\ }\textbf {\bibinfo {volume} {87}},\ \bibinfo {pages} {094410} (\bibinfo {year} {2013})}\BibitemShut {NoStop}%
\bibitem [{\citenamefont {Fennell}\ \emph {et~al.}(2014)\citenamefont {Fennell}, \citenamefont {Kenzelmann}, \citenamefont {Roessli}, \citenamefont {Mutka}, \citenamefont {Ollivier}, \citenamefont {Ruminy}, \citenamefont {Stuhr}, \citenamefont {Zaharko}, \citenamefont {Bovo}, \citenamefont {Cervellino}, \citenamefont {Haas},\ and\ \citenamefont {Cava}}]{Fennell2014}%
  \BibitemOpen
  \bibfield  {author} {\bibinfo {author} {\bibfnamefont {T.}~\bibnamefont {Fennell}}, \bibinfo {author} {\bibfnamefont {M.}~\bibnamefont {Kenzelmann}}, \bibinfo {author} {\bibfnamefont {B.}~\bibnamefont {Roessli}}, \bibinfo {author} {\bibfnamefont {H.}~\bibnamefont {Mutka}}, \bibinfo {author} {\bibfnamefont {J.}~\bibnamefont {Ollivier}}, \bibinfo {author} {\bibfnamefont {M.}~\bibnamefont {Ruminy}}, \bibinfo {author} {\bibfnamefont {U.}~\bibnamefont {Stuhr}}, \bibinfo {author} {\bibfnamefont {O.}~\bibnamefont {Zaharko}}, \bibinfo {author} {\bibfnamefont {L.}~\bibnamefont {Bovo}}, \bibinfo {author} {\bibfnamefont {A.}~\bibnamefont {Cervellino}}, \bibinfo {author} {\bibfnamefont {M.~K.}\ \bibnamefont {Haas}},\ and\ \bibinfo {author} {\bibfnamefont {R.~J.}\ \bibnamefont {Cava}},\ }\bibfield  {title} {\bibinfo {title} {Magnetoelastic excitations in the pyrochlore spin liquid {${\mathrm{Tb}}_{2}{\mathrm{Ti}}_{2}{\mathrm{O}}_{7}$}},\ }\href {https://doi.org/10.1103/physrevlett.112.017203} {\bibfield  {journal} {\bibinfo  {journal} {Physical Review Letters}\ }\textbf {\bibinfo {volume} {112}},\ \bibinfo {pages} {017203} (\bibinfo {year} {2014})}\BibitemShut {NoStop}%
\bibitem [{\citenamefont {Kermarrec}\ \emph {et~al.}(2015)\citenamefont {Kermarrec}, \citenamefont {Maharaj}, \citenamefont {Gaudet}, \citenamefont {Fritsch}, \citenamefont {Pomaranski}, \citenamefont {Kycia}, \citenamefont {Qiu}, \citenamefont {Copley}, \citenamefont {Couchman}, \citenamefont {Morningstar}, \citenamefont {Dabkowska},\ and\ \citenamefont {Gaulin}}]{Kermarrec2015}%
  \BibitemOpen
  \bibfield  {author} {\bibinfo {author} {\bibfnamefont {E.}~\bibnamefont {Kermarrec}}, \bibinfo {author} {\bibfnamefont {D.~D.}\ \bibnamefont {Maharaj}}, \bibinfo {author} {\bibfnamefont {J.}~\bibnamefont {Gaudet}}, \bibinfo {author} {\bibfnamefont {K.}~\bibnamefont {Fritsch}}, \bibinfo {author} {\bibfnamefont {D.}~\bibnamefont {Pomaranski}}, \bibinfo {author} {\bibfnamefont {J.~B.}\ \bibnamefont {Kycia}}, \bibinfo {author} {\bibfnamefont {Y.}~\bibnamefont {Qiu}}, \bibinfo {author} {\bibfnamefont {J.~R.~D.}\ \bibnamefont {Copley}}, \bibinfo {author} {\bibfnamefont {M.~M.~P.}\ \bibnamefont {Couchman}}, \bibinfo {author} {\bibfnamefont {A.~O.~R.}\ \bibnamefont {Morningstar}}, \bibinfo {author} {\bibfnamefont {H.~A.}\ \bibnamefont {Dabkowska}},\ and\ \bibinfo {author} {\bibfnamefont {B.~D.}\ \bibnamefont {Gaulin}},\ }\bibfield  {title} {\bibinfo {title} {Gapped and gapless short-range-ordered magnetic states with $(\frac{1}{2},\frac{1}{2},\frac{1}{2})$ wave vectors in the pyrochlore magnet ${\mathrm{tb}}_{2+x}{\mathrm{ti}}_{2\ensuremath{-}x}{\mathrm{o}}_{7+\ensuremath{\delta}}$},\ }\href {https://doi.org/10.1103/PhysRevB.92.245114} {\bibfield  {journal} {\bibinfo  {journal} {Phys. Rev. B}\ }\textbf {\bibinfo {volume} {92}},\ \bibinfo {pages} {245114} (\bibinfo {year} {2015})}\BibitemShut {NoStop}%
\bibitem [{\citenamefont {Takatsu}\ \emph {et~al.}(2016)\citenamefont {Takatsu}, \citenamefont {Onoda}, \citenamefont {Kittaka}, \citenamefont {Kasahara}, \citenamefont {Kono}, \citenamefont {Sakakibara}, \citenamefont {Kato}, \citenamefont {Fåk}, \citenamefont {Ollivier}, \citenamefont {Lynn}, \citenamefont {Taniguchi}, \citenamefont {Wakita},\ and\ \citenamefont {Kadowaki}}]{Takatsu2016}%
  \BibitemOpen
  \bibfield  {author} {\bibinfo {author} {\bibfnamefont {H.}~\bibnamefont {Takatsu}}, \bibinfo {author} {\bibfnamefont {S.}~\bibnamefont {Onoda}}, \bibinfo {author} {\bibfnamefont {S.}~\bibnamefont {Kittaka}}, \bibinfo {author} {\bibfnamefont {A.}~\bibnamefont {Kasahara}}, \bibinfo {author} {\bibfnamefont {Y.}~\bibnamefont {Kono}}, \bibinfo {author} {\bibfnamefont {T.}~\bibnamefont {Sakakibara}}, \bibinfo {author} {\bibfnamefont {Y.}~\bibnamefont {Kato}}, \bibinfo {author} {\bibfnamefont {B.}~\bibnamefont {Fåk}}, \bibinfo {author} {\bibfnamefont {J.}~\bibnamefont {Ollivier}}, \bibinfo {author} {\bibfnamefont {J.}~\bibnamefont {Lynn}}, \bibinfo {author} {\bibfnamefont {T.}~\bibnamefont {Taniguchi}}, \bibinfo {author} {\bibfnamefont {M.}~\bibnamefont {Wakita}},\ and\ \bibinfo {author} {\bibfnamefont {H.}~\bibnamefont {Kadowaki}},\ }\bibfield  {title} {\bibinfo {title} {Quadrupole order in the frustrated pyrochlore {${\mathrm{Tb}}_{2+x}{\mathrm{Ti}}_{2-x}{\mathrm{O}}_{7+y}$}},\ }\href {https://doi.org/10.1103/physrevlett.116.217201} {\bibfield  {journal} {\bibinfo  {journal} {Physical Review Letters}\ }\textbf {\bibinfo {volume} {116}},\ \bibinfo {pages} {217201} (\bibinfo {year} {2016})}\BibitemShut {NoStop}%
\bibitem [{\citenamefont {Kadowaki}\ \emph {et~al.}(2022)\citenamefont {Kadowaki}, \citenamefont {Wakita}, \citenamefont {F\aa{}k}, \citenamefont {Ollivier},\ and\ \citenamefont {Ohira-Kawamura}}]{Kadowaki2022}%
  \BibitemOpen
  \bibfield  {author} {\bibinfo {author} {\bibfnamefont {H.}~\bibnamefont {Kadowaki}}, \bibinfo {author} {\bibfnamefont {M.}~\bibnamefont {Wakita}}, \bibinfo {author} {\bibfnamefont {B.}~\bibnamefont {F\aa{}k}}, \bibinfo {author} {\bibfnamefont {J.}~\bibnamefont {Ollivier}},\ and\ \bibinfo {author} {\bibfnamefont {S.}~\bibnamefont {Ohira-Kawamura}},\ }\bibfield  {title} {\bibinfo {title} {Spin and quadrupole correlations by three-spin interaction in the frustrated pyrochlore magnet {${\mathrm{Tb}}_{2+x}{\mathrm{Ti}}_{2\ensuremath{-}x}{\mathrm{O}}_{7+y}$}},\ }\href {https://doi.org/10.1103/physrevb.105.014439} {\bibfield  {journal} {\bibinfo  {journal} {Physical Review B}\ }\textbf {\bibinfo {volume} {105}},\ \bibinfo {pages} {014439} (\bibinfo {year} {2022})}\BibitemShut {NoStop}%
\bibitem [{\citenamefont {Alexanian}\ \emph {et~al.}(2023{\natexlab{b}})\citenamefont {Alexanian}, \citenamefont {Robert}, \citenamefont {Simonet}, \citenamefont {Langér\^ome}, \citenamefont {Brubach}, \citenamefont {Roy}, \citenamefont {Decorse}, \citenamefont {Lhotel}, \citenamefont {Constable}, \citenamefont {Ballou},\ and\ \citenamefont {De~Brion}}]{Alexanian2023}%
  \BibitemOpen
  \bibfield  {author} {\bibinfo {author} {\bibfnamefont {Y.}~\bibnamefont {Alexanian}}, \bibinfo {author} {\bibfnamefont {J.}~\bibnamefont {Robert}}, \bibinfo {author} {\bibfnamefont {V.}~\bibnamefont {Simonet}}, \bibinfo {author} {\bibfnamefont {B.}~\bibnamefont {Langér\^ome}}, \bibinfo {author} {\bibfnamefont {J.-B.}\ \bibnamefont {Brubach}}, \bibinfo {author} {\bibfnamefont {P.}~\bibnamefont {Roy}}, \bibinfo {author} {\bibfnamefont {C.}~\bibnamefont {Decorse}}, \bibinfo {author} {\bibfnamefont {E.}~\bibnamefont {Lhotel}}, \bibinfo {author} {\bibfnamefont {E.}~\bibnamefont {Constable}}, \bibinfo {author} {\bibfnamefont {R.}~\bibnamefont {Ballou}},\ and\ \bibinfo {author} {\bibfnamefont {S.}~\bibnamefont {De~Brion}},\ }\bibfield  {title} {\bibinfo {title} {Vibronic collapse of ordered quadrupolar ice in the pyrochlore magnet {Tb$_{2}$Ti$_{2}$O$_{7}$}},\ }\href {https://doi.org/10.1103/physrevb.107.224404} {\bibfield  {journal} {\bibinfo  {journal} {Physical Review B}\ }\textbf {\bibinfo {volume} {107}},\ \bibinfo {pages} {224404} (\bibinfo {year} {2023}{\natexlab{b}})}\BibitemShut {NoStop}%
\bibitem [{\citenamefont {Roll}\ \emph {et~al.}(2024)\citenamefont {Roll}, \citenamefont {Balédent}, \citenamefont {Robert}, \citenamefont {Ollivier}, \citenamefont {Decorse}, \citenamefont {Guitteny}, \citenamefont {Mirebeau},\ and\ \citenamefont {Petit}}]{Roll2024}%
  \BibitemOpen
  \bibfield  {author} {\bibinfo {author} {\bibfnamefont {A.}~\bibnamefont {Roll}}, \bibinfo {author} {\bibfnamefont {V.}~\bibnamefont {Balédent}}, \bibinfo {author} {\bibfnamefont {J.}~\bibnamefont {Robert}}, \bibinfo {author} {\bibfnamefont {J.}~\bibnamefont {Ollivier}}, \bibinfo {author} {\bibfnamefont {C.}~\bibnamefont {Decorse}}, \bibinfo {author} {\bibfnamefont {S.}~\bibnamefont {Guitteny}}, \bibinfo {author} {\bibfnamefont {I.}~\bibnamefont {Mirebeau}},\ and\ \bibinfo {author} {\bibfnamefont {S.}~\bibnamefont {Petit}},\ }\bibfield  {title} {\bibinfo {title} {Magnetic interactions in the cooperative paramagnet {Tb$_{2}$Ti$_{2}$O$_{7}$}},\ }\href {https://doi.org/10.1103/physrevresearch.6.043011} {\bibfield  {journal} {\bibinfo  {journal} {Physical Review Research}\ }\textbf {\bibinfo {volume} {6}},\ \bibinfo {pages} {043011} (\bibinfo {year} {2024})}\BibitemShut {NoStop}%
\bibitem [{\citenamefont {Yan}\ \emph {et~al.}(2017)\citenamefont {Yan}, \citenamefont {Benton}, \citenamefont {Jaubert},\ and\ \citenamefont {Shannon}}]{Yan2017}%
  \BibitemOpen
  \bibfield  {author} {\bibinfo {author} {\bibfnamefont {H.}~\bibnamefont {Yan}}, \bibinfo {author} {\bibfnamefont {O.}~\bibnamefont {Benton}}, \bibinfo {author} {\bibfnamefont {L.}~\bibnamefont {Jaubert}},\ and\ \bibinfo {author} {\bibfnamefont {N.}~\bibnamefont {Shannon}},\ }\bibfield  {title} {\bibinfo {title} {Theory of multiple-phase competition in pyrochlore magnets with anisotropic exchange with application to {Yb$_{2}$Ti$_{2}$O$_{7}$}, {Er$_{2}$Ti$_{2}$O$_{7}$}, and {Er$_{2}$Sn$_{2}$O$_{7}$}},\ }\href {https://doi.org/10.1103/physrevb.95.094422} {\bibfield  {journal} {\bibinfo  {journal} {Physical Review B}\ }\textbf {\bibinfo {volume} {95}},\ \bibinfo {pages} {094422} (\bibinfo {year} {2017})}\BibitemShut {NoStop}%
\bibitem [{\citenamefont {Hardy}\ \emph {et~al.}(2003)\citenamefont {Hardy}, \citenamefont {Lambert}, \citenamefont {Lees},\ and\ \citenamefont {McK.~Paul}}]{Hardy2003}%
  \BibitemOpen
  \bibfield  {author} {\bibinfo {author} {\bibfnamefont {V.}~\bibnamefont {Hardy}}, \bibinfo {author} {\bibfnamefont {S.}~\bibnamefont {Lambert}}, \bibinfo {author} {\bibfnamefont {M.~R.}\ \bibnamefont {Lees}},\ and\ \bibinfo {author} {\bibfnamefont {D.}~\bibnamefont {McK.~Paul}},\ }\bibfield  {title} {\bibinfo {title} {Specific heat and magnetization study on single crystals of the frustrated quasi-one-dimensional oxide {Ca}$_{3}${Co}$_{2}${O}$_{6}$},\ }\href {http://dx.doi.org/10.1103/PhysRevB.68.014424} {\bibfield  {journal} {\bibinfo  {journal} {Physical Review B}\ }\textbf {\bibinfo {volume} {68}},\ \bibinfo {pages} {014424} (\bibinfo {year} {2003})}\BibitemShut {NoStop}%
\bibitem [{\citenamefont {Gingras}\ \emph {et~al.}(2000)\citenamefont {Gingras}, \citenamefont {den Hertog}, \citenamefont {Faucher}, \citenamefont {Gardner}, \citenamefont {Dunsiger}, \citenamefont {Chang}, \citenamefont {Gaulin}, \citenamefont {Raju},\ and\ \citenamefont {Greedan}}]{Gingras2000}%
  \BibitemOpen
  \bibfield  {author} {\bibinfo {author} {\bibfnamefont {M.~J.~P.}\ \bibnamefont {Gingras}}, \bibinfo {author} {\bibfnamefont {B.~C.}\ \bibnamefont {den Hertog}}, \bibinfo {author} {\bibfnamefont {M.}~\bibnamefont {Faucher}}, \bibinfo {author} {\bibfnamefont {J.~S.}\ \bibnamefont {Gardner}}, \bibinfo {author} {\bibfnamefont {S.~R.}\ \bibnamefont {Dunsiger}}, \bibinfo {author} {\bibfnamefont {L.~J.}\ \bibnamefont {Chang}}, \bibinfo {author} {\bibfnamefont {B.~D.}\ \bibnamefont {Gaulin}}, \bibinfo {author} {\bibfnamefont {N.~P.}\ \bibnamefont {Raju}},\ and\ \bibinfo {author} {\bibfnamefont {J.~E.}\ \bibnamefont {Greedan}},\ }\bibfield  {title} {\bibinfo {title} {Thermodynamic and single-ion properties of {Tb$^{3+}$} within the collective paramagnetic-spin liquid state of the frustrated pyrochlore antiferromagnet {Tb$_{2}$Ti$_{2}$O$_{7}$}},\ }\href {https://doi.org/10.1103/PhysRevB.62.6496} {\bibfield  {journal} {\bibinfo  {journal} {Phys. Rev. B}\ }\textbf {\bibinfo {volume} {62}},\ \bibinfo {pages} {6496} (\bibinfo {year} {2000})}\BibitemShut {NoStop}%
\bibitem [{\citenamefont {Mirebeau}\ \emph {et~al.}(2007)\citenamefont {Mirebeau}, \citenamefont {Bonville},\ and\ \citenamefont {Hennion}}]{Mirebeau2007}%
  \BibitemOpen
  \bibfield  {author} {\bibinfo {author} {\bibfnamefont {I.}~\bibnamefont {Mirebeau}}, \bibinfo {author} {\bibfnamefont {P.}~\bibnamefont {Bonville}},\ and\ \bibinfo {author} {\bibfnamefont {M.}~\bibnamefont {Hennion}},\ }\bibfield  {title} {\bibinfo {title} {Magnetic excitations in {Tb$_{2}$Sn$_{2}$O$_{7}$} and {Tb$_{2}$Ti$_{2}$O$_{7}$} as measured by inelastic neutron scattering},\ }\href {https://doi.org/10.1103/physrevb.76.184436} {\bibfield  {journal} {\bibinfo  {journal} {Physical Review B}\ }\textbf {\bibinfo {volume} {76}},\ \bibinfo {pages} {184436} (\bibinfo {year} {2007})}\BibitemShut {NoStop}%
\bibitem [{\citenamefont {Bertin}\ \emph {et~al.}(2012)\citenamefont {Bertin}, \citenamefont {Chapuis}, \citenamefont {Dalmas~de Réotier},\ and\ \citenamefont {Yaouanc}}]{Bertin2012}%
  \BibitemOpen
  \bibfield  {author} {\bibinfo {author} {\bibfnamefont {A.}~\bibnamefont {Bertin}}, \bibinfo {author} {\bibfnamefont {Y.}~\bibnamefont {Chapuis}}, \bibinfo {author} {\bibfnamefont {P.}~\bibnamefont {Dalmas~de Réotier}},\ and\ \bibinfo {author} {\bibfnamefont {A.}~\bibnamefont {Yaouanc}},\ }\bibfield  {title} {\bibinfo {title} {Crystal electric field in the {$R_{2}$Ti$_{2}$O$_{7}$} pyrochlore compounds},\ }\href {https://doi.org/10.1088/0953-8984/24/25/256003} {\bibfield  {journal} {\bibinfo  {journal} {Journal of Physics: Condensed Matter}\ }\textbf {\bibinfo {volume} {24}},\ \bibinfo {pages} {256003} (\bibinfo {year} {2012})}\BibitemShut {NoStop}%
\bibitem [{\citenamefont {Zhang}\ \emph {et~al.}(2014)\citenamefont {Zhang}, \citenamefont {Fritsch}, \citenamefont {Hao}, \citenamefont {Bagheri}, \citenamefont {Gingras}, \citenamefont {Granroth}, \citenamefont {Jiramongkolchai}, \citenamefont {Cava},\ and\ \citenamefont {Gaulin}}]{Zhang2014}%
  \BibitemOpen
  \bibfield  {author} {\bibinfo {author} {\bibfnamefont {J.}~\bibnamefont {Zhang}}, \bibinfo {author} {\bibfnamefont {K.}~\bibnamefont {Fritsch}}, \bibinfo {author} {\bibfnamefont {Z.}~\bibnamefont {Hao}}, \bibinfo {author} {\bibfnamefont {B.~V.}\ \bibnamefont {Bagheri}}, \bibinfo {author} {\bibfnamefont {M.~J.~P.}\ \bibnamefont {Gingras}}, \bibinfo {author} {\bibfnamefont {G.~E.}\ \bibnamefont {Granroth}}, \bibinfo {author} {\bibfnamefont {P.}~\bibnamefont {Jiramongkolchai}}, \bibinfo {author} {\bibfnamefont {R.~J.}\ \bibnamefont {Cava}},\ and\ \bibinfo {author} {\bibfnamefont {B.~D.}\ \bibnamefont {Gaulin}},\ }\bibfield  {title} {\bibinfo {title} {Neutron spectroscopic study of crystal field excitations in {Tb$_{2}$Ti$_{2}$O$_{7}$} and {Tb$_{2}$Sn$_{2}$O$_{7}$}},\ }\href {https://doi.org/10.1103/PhysRevB.89.134410} {\bibfield  {journal} {\bibinfo  {journal} {Phys. Rev. B}\ }\textbf {\bibinfo {volume} {89}},\ \bibinfo {pages} {134410} (\bibinfo {year} {2014})}\BibitemShut {NoStop}%
\bibitem [{\citenamefont {Ruminy}\ \emph {et~al.}(2016)\citenamefont {Ruminy}, \citenamefont {Pomjakushina}, \citenamefont {Iida}, \citenamefont {Kamazawa}, \citenamefont {Adroja}, \citenamefont {Stuhr},\ and\ \citenamefont {Fennell}}]{Ruminy2016}%
  \BibitemOpen
  \bibfield  {author} {\bibinfo {author} {\bibfnamefont {M.}~\bibnamefont {Ruminy}}, \bibinfo {author} {\bibfnamefont {E.}~\bibnamefont {Pomjakushina}}, \bibinfo {author} {\bibfnamefont {K.}~\bibnamefont {Iida}}, \bibinfo {author} {\bibfnamefont {K.}~\bibnamefont {Kamazawa}}, \bibinfo {author} {\bibfnamefont {D.~T.}\ \bibnamefont {Adroja}}, \bibinfo {author} {\bibfnamefont {U.}~\bibnamefont {Stuhr}},\ and\ \bibinfo {author} {\bibfnamefont {T.}~\bibnamefont {Fennell}},\ }\bibfield  {title} {\bibinfo {title} {Crystal-field parameters of the rare-earth pyrochlores {$R_{2}$Ti$_{2}$O$_{7}$ ($R=\text{Tb}$, Dy, and Ho)}},\ }\href {https://doi.org/10.1103/PhysRevB.94.024430} {\bibfield  {journal} {\bibinfo  {journal} {Phys. Rev. B}\ }\textbf {\bibinfo {volume} {94}},\ \bibinfo {pages} {024430} (\bibinfo {year} {2016})}\BibitemShut {NoStop}%
\bibitem [{\citenamefont {Roll}(2024)}]{Roll2024phd}%
  \BibitemOpen
  \bibfield  {author} {\bibinfo {author} {\bibfnamefont {A.}~\bibnamefont {Roll}},\ }\emph {\bibinfo {title} {Spectroscopie et dynamique de spin des systèmes frustrés : des pnictures aux pyrochlores}},\ \href {https://theses.fr/2024UPASP141} {Ph.D. thesis},\ \bibinfo  {school} {Université Paris-Saclay} (\bibinfo {year} {2024})\BibitemShut {NoStop}%
\bibitem [{\citenamefont {Stevens}(1952)}]{Stevens1952}%
  \BibitemOpen
  \bibfield  {author} {\bibinfo {author} {\bibfnamefont {K.~W.~H.}\ \bibnamefont {Stevens}},\ }\bibfield  {title} {\bibinfo {title} {Matrix elements and operator equivalents connected with the magnetic properties of rare earth ions},\ }\href {https://doi.org/10.1088/0370-1298/65/3/308} {\bibfield  {journal} {\bibinfo  {journal} {Proc. Phys. Soc. Sect. A}\ }\textbf {\bibinfo {volume} {65}},\ \bibinfo {pages} {209} (\bibinfo {year} {1952})}\BibitemShut {NoStop}%
\bibitem [{\citenamefont {Hutchings}(1964)}]{Hutchings1964}%
  \BibitemOpen
  \bibfield  {author} {\bibinfo {author} {\bibfnamefont {M.~T.}\ \bibnamefont {Hutchings}},\ }\bibfield  {title} {\bibinfo {title} {Point-charge calculations of energy levels of magnetic ions in crystalline electric fields},\ }\href {https://doi.org/10.1016/s0081-1947(08)60517-2} {\bibfield  {journal} {\bibinfo  {journal} {Solid State Physics}\ }\textbf {\bibinfo {volume} {16}},\ \bibinfo {pages} {227} (\bibinfo {year} {1964})}\BibitemShut {NoStop}%
\bibitem [{\citenamefont {Danielsen}\ and\ \citenamefont {Lindg{\aa}rd}(1972)}]{Danielsen1972}%
  \BibitemOpen
  \bibfield  {author} {\bibinfo {author} {\bibfnamefont {O.}~\bibnamefont {Danielsen}}\ and\ \bibinfo {author} {\bibfnamefont {P.-A.}\ \bibnamefont {Lindg{\aa}rd}},\ }\href@noop {} {\emph {\bibinfo {title} {Quantum mechanical operator equivalents used in the theory of magnetism}}}\ (\bibinfo  {publisher} {Ris{\o} National Laboratory, Roskilde, Denmark},\ \bibinfo {year} {1972})\BibitemShut {NoStop}%
\bibitem [{\citenamefont {Constable}\ \emph {et~al.}(2017)\citenamefont {Constable}, \citenamefont {Ballou}, \citenamefont {Robert}, \citenamefont {Decorse}, \citenamefont {Brubach}, \citenamefont {Roy}, \citenamefont {Lhotel}, \citenamefont {Del-Rey}, \citenamefont {Simonet}, \citenamefont {Petit},\ and\ \citenamefont {deBrion}}]{Constable2017}%
  \BibitemOpen
  \bibfield  {author} {\bibinfo {author} {\bibfnamefont {E.}~\bibnamefont {Constable}}, \bibinfo {author} {\bibfnamefont {R.}~\bibnamefont {Ballou}}, \bibinfo {author} {\bibfnamefont {J.}~\bibnamefont {Robert}}, \bibinfo {author} {\bibfnamefont {C.}~\bibnamefont {Decorse}}, \bibinfo {author} {\bibfnamefont {J.-B.}\ \bibnamefont {Brubach}}, \bibinfo {author} {\bibfnamefont {P.}~\bibnamefont {Roy}}, \bibinfo {author} {\bibfnamefont {E.}~\bibnamefont {Lhotel}}, \bibinfo {author} {\bibfnamefont {L.}~\bibnamefont {Del-Rey}}, \bibinfo {author} {\bibfnamefont {V.}~\bibnamefont {Simonet}}, \bibinfo {author} {\bibfnamefont {S.}~\bibnamefont {Petit}},\ and\ \bibinfo {author} {\bibfnamefont {S.}~\bibnamefont {deBrion}},\ }\bibfield  {title} {\bibinfo {title} {Double vibronic process in the quantum spin ice candidate {${\mathrm{Tb}}_{2}{\mathrm{Ti}}_{2}{\mathrm{O}}_{7}$} revealed by terahertz spectroscopy},\ }\href {https://doi.org/10.1103/PhysRevB.95.020415} {\bibfield  {journal} {\bibinfo  {journal} {Phys. Rev. B}\ }\textbf {\bibinfo {volume} {95}},\ \bibinfo {pages} {020415(R)} (\bibinfo {year} {2017})}\BibitemShut {NoStop}%
\bibitem [{\citenamefont {Gaudet}\ \emph {et~al.}(2018)\citenamefont {Gaudet}, \citenamefont {Hallas}, \citenamefont {Buhariwalla}, \citenamefont {Sala}, \citenamefont {Stone}, \citenamefont {Tachibana}, \citenamefont {Baroudi}, \citenamefont {Cava},\ and\ \citenamefont {Gaulin}}]{Gaudet2018}%
  \BibitemOpen
  \bibfield  {author} {\bibinfo {author} {\bibfnamefont {J.}~\bibnamefont {Gaudet}}, \bibinfo {author} {\bibfnamefont {A.~M.}\ \bibnamefont {Hallas}}, \bibinfo {author} {\bibfnamefont {C.~R.~C.}\ \bibnamefont {Buhariwalla}}, \bibinfo {author} {\bibfnamefont {G.}~\bibnamefont {Sala}}, \bibinfo {author} {\bibfnamefont {M.~B.}\ \bibnamefont {Stone}}, \bibinfo {author} {\bibfnamefont {M.}~\bibnamefont {Tachibana}}, \bibinfo {author} {\bibfnamefont {K.}~\bibnamefont {Baroudi}}, \bibinfo {author} {\bibfnamefont {R.~J.}\ \bibnamefont {Cava}},\ and\ \bibinfo {author} {\bibfnamefont {B.~D.}\ \bibnamefont {Gaulin}},\ }\bibfield  {title} {\bibinfo {title} {Magnetoelastically induced vibronic bound state in the spin-ice pyrochlore {${\mathrm{Ho}}_{2}{\mathrm{Ti}}_{2}{\mathrm{O}}_{7}$}},\ }\href {https://doi.org/10.1103/PhysRevB.98.014419} {\bibfield  {journal} {\bibinfo  {journal} {Phys. Rev. B}\ }\textbf {\bibinfo {volume} {98}},\ \bibinfo {pages} {014419} (\bibinfo {year} {2018})}\BibitemShut {NoStop}%
\bibitem [{\citenamefont {Santini}\ \emph {et~al.}(2009)\citenamefont {Santini}, \citenamefont {Carretta}, \citenamefont {Amoretti}, \citenamefont {Caciuffo}, \citenamefont {Magnani},\ and\ \citenamefont {Lander}}]{Santini2009}%
  \BibitemOpen
  \bibfield  {author} {\bibinfo {author} {\bibfnamefont {P.}~\bibnamefont {Santini}}, \bibinfo {author} {\bibfnamefont {S.}~\bibnamefont {Carretta}}, \bibinfo {author} {\bibfnamefont {G.}~\bibnamefont {Amoretti}}, \bibinfo {author} {\bibfnamefont {R.}~\bibnamefont {Caciuffo}}, \bibinfo {author} {\bibfnamefont {N.}~\bibnamefont {Magnani}},\ and\ \bibinfo {author} {\bibfnamefont {G.~H.}\ \bibnamefont {Lander}},\ }\bibfield  {title} {\bibinfo {title} {Multipolar interactions in \textit{f}-electron systems: The paradigm of actinide dioxides},\ }\href {https://doi.org/10.1103/revmodphys.81.807} {\bibfield  {journal} {\bibinfo  {journal} {Reviews of Modern Physics}\ }\textbf {\bibinfo {volume} {81}},\ \bibinfo {pages} {807} (\bibinfo {year} {2009})}\BibitemShut {NoStop}%
\bibitem [{\citenamefont {Faure}\ \emph {et~al.}(2024)\citenamefont {Faure}, \citenamefont {Toschi}, \citenamefont {Soh}, \citenamefont {Lhotel}, \citenamefont {Detlefs}, \citenamefont {Prabhakaran}, \citenamefont {McMorrow},\ and\ \citenamefont {Sahle}}]{Faure2024}%
  \BibitemOpen
  \bibfield  {author} {\bibinfo {author} {\bibfnamefont {Q.}~\bibnamefont {Faure}}, \bibinfo {author} {\bibfnamefont {A.}~\bibnamefont {Toschi}}, \bibinfo {author} {\bibfnamefont {J.~R.}\ \bibnamefont {Soh}}, \bibinfo {author} {\bibfnamefont {E.}~\bibnamefont {Lhotel}}, \bibinfo {author} {\bibfnamefont {B.}~\bibnamefont {Detlefs}}, \bibinfo {author} {\bibfnamefont {D.}~\bibnamefont {Prabhakaran}}, \bibinfo {author} {\bibfnamefont {D.~F.}\ \bibnamefont {McMorrow}},\ and\ \bibinfo {author} {\bibfnamefont {C.~J.}\ \bibnamefont {Sahle}},\ }\bibfield  {title} {\bibinfo {title} {Spin dynamics and possible topological magnons in the nonstoichiometric pyrochlore iridate {Tb$_{2}$Ir$_{2}$O$_{7}$} studied by rixs},\ }\href {https://doi.org/10.1103/PhysRevB.110.L140401} {\bibfield  {journal} {\bibinfo  {journal} {Phys. Rev. B}\ }\textbf {\bibinfo {volume} {110}},\ \bibinfo {pages} {L140401} (\bibinfo {year} {2024})}\BibitemShut {NoStop}%
\bibitem [{\citenamefont {Museur}\ \emph {et~al.}(2026)\citenamefont {Museur}, \citenamefont {Robert}, \citenamefont {Morineau}, \citenamefont {Bujault}, \citenamefont {Simonet}, \citenamefont {Pachoud}, \citenamefont {Hadj-Azzem}, \citenamefont {Colin}, \citenamefont {Mangin-Thro}, \citenamefont {Manuel}, \citenamefont {Stewart}, \citenamefont {Holdsworth},\ and\ \citenamefont {Lhotel}}]{Museur2026}%
  \BibitemOpen
  \bibfield  {author} {\bibinfo {author} {\bibfnamefont {F.}~\bibnamefont {Museur}}, \bibinfo {author} {\bibfnamefont {J.}~\bibnamefont {Robert}}, \bibinfo {author} {\bibfnamefont {F.}~\bibnamefont {Morineau}}, \bibinfo {author} {\bibfnamefont {N.}~\bibnamefont {Bujault}}, \bibinfo {author} {\bibfnamefont {V.}~\bibnamefont {Simonet}}, \bibinfo {author} {\bibfnamefont {E.}~\bibnamefont {Pachoud}}, \bibinfo {author} {\bibfnamefont {A.}~\bibnamefont {Hadj-Azzem}}, \bibinfo {author} {\bibfnamefont {C.~V.}\ \bibnamefont {Colin}}, \bibinfo {author} {\bibfnamefont {L.}~\bibnamefont {Mangin-Thro}}, \bibinfo {author} {\bibfnamefont {P.}~\bibnamefont {Manuel}}, \bibinfo {author} {\bibfnamefont {J.~R.}\ \bibnamefont {Stewart}}, \bibinfo {author} {\bibfnamefont {P.~C.~W.}\ \bibnamefont {Holdsworth}},\ and\ \bibinfo {author} {\bibfnamefont {E.}~\bibnamefont {Lhotel}},\ }\bibfield  {title} {\bibinfo {title} {Ferromagnetic fragmented state in the pyrochlore {Ho$_{2}$Ru$_{2}$O$_{7}$}},\ }\href {https://doi.org/10.1103/dzdf-hh4t} {\bibfield  {journal} {\bibinfo  {journal} {Physical Review B}\ }\textbf {\bibinfo {volume} {113}},\ \bibinfo {pages} {L060406} (\bibinfo {year} {2026})}\BibitemShut {NoStop}%
\bibitem [{\citenamefont {Lefran\c{c}ois}\ \emph {et~al.}(2014)\citenamefont {Lefran\c{c}ois}, \citenamefont {Ballou}, \citenamefont {Chapon}, \citenamefont {Lejay}, \citenamefont {Lhotel}, \citenamefont {Ollivier},\ and\ \citenamefont {Simonet}}]{ILLdata1}%
  \BibitemOpen
  \bibfield  {author} {\bibinfo {author} {\bibfnamefont {E.}~\bibnamefont {Lefran\c{c}ois}}, \bibinfo {author} {\bibfnamefont {R.}~\bibnamefont {Ballou}}, \bibinfo {author} {\bibfnamefont {L.-C.}\ \bibnamefont {Chapon}}, \bibinfo {author} {\bibfnamefont {P.}~\bibnamefont {Lejay}}, \bibinfo {author} {\bibfnamefont {E.}~\bibnamefont {Lhotel}}, \bibinfo {author} {\bibfnamefont {J.}~\bibnamefont {Ollivier}},\ and\ \bibinfo {author} {\bibfnamefont {V.}~\bibnamefont {Simonet}},\ }\bibfield  {title} {\bibinfo {title} {Magnetic and crystal field excitation spectra in the iridate pyrochlores {Tb$_{2}$Ir$_{2}$O$_{7}$} and {Er$_{2}$Ir$_{2}$O$_{7}$}},\ }\href {https://doi.ill.fr/10.5291/ILL-DATA.4-01-1380} {\bibfield  {journal} {\bibinfo  {journal} {ILL-DATA.4-01-1380}\ } (\bibinfo {year} {2014})}\BibitemShut {NoStop}%
\bibitem [{\citenamefont {Brown}\ \emph {et~al.}(2006)\citenamefont {Brown}, \citenamefont {Fox}, \citenamefont {Maslen}, \citenamefont {Keefe},\ and\ \citenamefont {Willis}}]{Brown2006}%
  \BibitemOpen
  \bibfield  {author} {\bibinfo {author} {\bibfnamefont {P.~J.}\ \bibnamefont {Brown}}, \bibinfo {author} {\bibfnamefont {A.~G.}\ \bibnamefont {Fox}}, \bibinfo {author} {\bibfnamefont {E.~N.}\ \bibnamefont {Maslen}}, \bibinfo {author} {\bibfnamefont {M.~A.~O.}\ \bibnamefont {Keefe}},\ and\ \bibinfo {author} {\bibfnamefont {B.~T.~M.}\ \bibnamefont {Willis}},\ }\bibfield  {title} {\bibinfo {title} {Intensity of diffracted intensities},\ }in\ \href@noop {} {\emph {\bibinfo {booktitle} {International Tables for Crystallography}}},\ Vol.\ \bibinfo {volume} {C: \textit{Mathematical, physical and chemical tables}},\ \bibinfo {editor} {edited by\ \bibinfo {editor} {\bibfnamefont {E.}~\bibnamefont {Prince}}}\ (\bibinfo  {publisher} {International Union of Crystallography, Chester, England},\ \bibinfo {year} {2006})\ p.\ \bibinfo {pages} {554}\BibitemShut {NoStop}%
\end{thebibliography}
\end{document}